\documentclass[aps,prb,epsf,floatfix]{revtex4}
\usepackage{color}
\usepackage{epsfig}
\usepackage{amssymb}
\usepackage{subfigure}

\newcommand{\be}{\begin{equation}}
\newcommand{\ee}{\end{equation}}
\newcommand{\bea}{\begin{eqnarray}}
\newcommand{\eea}{\end{eqnarray}}
\begin{document}

\title{Magnetic impurities in a superconductor: Effect of domain
walls and interference
}

\author{P. D. Sacramento$^1$, V. K. Dugaev$^{1,2}$ and V. R. Vieira$^1$}
\address{$^1$Departamento de F\'isica and  CFIF, Instituto Superior
T\'ecnico, Universidade T\'ecnica de Lisboa, Av. Rovisco Pais, 1049-001 Lisboa, Portugal\\
$^2$Department of Mathematics and Applied Physics, Rzesz\'ow University of Technology,
Al. Powsta\'nc\'ow Warszawy 6, 35-959 Rzesz\'ow, Poland
}

\date{\today}

\begin{abstract}
We consider the effect of magnetic impurities, modeled by classical
spins, in a conventional superconductor. We study their effect on
the quasiparticles, specifically on the spin density and local
density of states (LDOS). As previously emphasized, the impurities
induce multiple scatterings of the quasiparticle wave functions
leading to complex interference phenomena. Also, the impurities
induce quantum phase transitions in the many-body system.
Previous authors studied the effect of either a small number of
impurities (from one to three) or a finite concentration of
impurities, typically in a disordered distribution. In this work
we assume a regular set of spins distributed inside the superconductor
in such a way that the spins are oriented,
forming different types of domain walls, assumed stable.
This situation may be particularly interesting in the context
of spin transfer due to polarized currents traversing the
material.
\end{abstract}
\vspace{0.3cm}
\pacs{PACS numbers: }
\maketitle

\section{Introduction}

The effect of perturbing the superconducting state is a topic of
interest since often it provides information about the
superconducting state, its nature and origin, but also it provides
information about its parent normal state. This has gained particular
interest due to the elusive nature of pairing in high-temperature
superconductors. In particular, the effect of magnetic fields in
its various forms has attracted interest for a long time, such
as the effect of vortices in type-II superconductors and in
general the interplay between superconductivity and magnetic field.

Also, the effect of impurities has been studied considering both non-magnetic and
magnetic impurities in conventional and unconventional superconductors \cite{review}.
In the case of non-magnetic impurities and $s$-wave
pairing, Anderson's theorem states that, at least for low concentrations, they
have little effect since the impurities are not pair-breaking \cite{AndersonT}. In $d$-wave
superconductors however, non-magnetic impurities cause a strong pair breaking
effect \cite{Ueda}. In the limit of strong scattering it was found that the lowest energy
quasiparticles become localized below the mobility gap, even in a regime where
the single-electron wave functions are still extended \cite{Lee}. This result has been
confirmed solving the Bogoliubov-de Gennes equations with a finite concentration
of non-magnetic impurities \cite{Franzet}. However, allowing for angular dependent impurity
scattering potentials it has been found that the scattering processes close to
the gap nodes may give rise to extended gapless regions \cite{Haas}. 

Several conflicting predictions have appeared
in the literature regarding
the effect of the presence of impurities.
Some progress towards understanding the
disparity of theoretical results has been achieved
realizing that the details of
the type of disorder significantly affect the density of states \cite{report}.
Particularly in the case of $d$-wave superconductors, in contrast to conventional
gaped $s$-wave superconductors, the presence of gapless
nodes is expected to affect the transport properties.
Using a field theoretic description and linearizing the spectrum around the
four Dirac-like nodes it has been suggested
that the system is critical. It was obtained that the density of states is
of the type $\rho(\epsilon) \sim |\epsilon |^{\alpha}$, where $\alpha$
is a non-universal
exponent dependent on the disorder, and that the low
energy modes are extended states
(critical metal) \cite{NTW}. Taking into account the effects of inter-nodal
scattering (hard-scattering) it has been shown that an
insulating state is obtained
instead, where the density of states still vanishes at low energy but with an
exponent $\alpha=1$ independent of disorder \cite{Senthil}.
Using the Bogoliubov-de Gennes (BdG) equations it was found
that the
$d$-wave superconductivity is mainly destroyed locally near a strong scatterer.
The superfluid density is strongly suppressed near the impurities but
only mildly
affected elsewhere \cite{Atkinson}. No evidence for localization of the
low energy
states was found. The superfluid density is suppressed
but less than expected \cite{Franzet,Randeria1} and, accordingly, the decrease
of the
critical temperature with disorder is much slower than previously expected, in
accordance with experiments \cite{Ulm}. Similar results of an inhomogeneous
order parameter
were also obtained for $s$-wave superconductors \cite{Randeria2}.
The results show that the order parameter is only significantly affected
close to the impurity locations.

On the other hand, magnetic impurities induce in-gap bound states in
conventional superconductors, while for instance in $d$-wave
superconductors, due to their gapless nature, they just induce
virtual bound states. The local nature of these boundstates can now
be studied in detail due to the progress in experimental techniques
like Scanning Tunneling Microscopy (STM).
In their pioneer work,\cite{abrikosov60} Abrikosov and Gor'kov considered
the properties of a superconductor with magnetic impurities. They
demonstrated that the noninteracting magnetic impurities suppress
the superconductivity, so that at some critical impurity density
the superconducting gap $\Delta $ shrinks to zero, which was later
identified as a quantum critical point \cite{ramazashvili97}.

It was also shown long ago that the presence of few magnetic impurities 
(vanishing concentration) is enough to lead to
rather interesting quantum phase transitions, in particular in the
total magnetic moment of the condensate \cite{sakurai}. In the simplest case the magnetic
impurities can be treated as classical spins inserted in the superconductor,
acting as local magnetic fields. Indeed it was shown that if the coupling
is weak enough, the Kondo coupling can be overlooked and a much simpler treatment
of the classical case provides a good description \cite{satori}. Actually, even though there
are similarities between the interaction of the electrons with an impurity both
in the standard Kondo effect in metals and the coupling to the magnetic impurity
in the superconductor, while the Kondo effect is merely
a crossover between a free spin at high energies or temperatures and the Kondo
singlet (for a $S=1/2$ impurity in a single $s$ band) at low energies \cite{hewson}, in the
case of the superconductor there is a true first order quantum phase transition \cite{salkola}.
The phase transition occurs through a level crossing between two states as the
coupling between the spin density of the electrons and the impurity spin grows.
The level crossing occurs between one state that 
describes an uncompensated local spin (at smaller coupling)
and a state where the impurity spin is compensated (partially since for the classical
description to be valid the spin has to be large).

The case of two (or three) impurities has also been studied \cite{morr1,morr2}
and quantum phase transitions
have been identified both as a function of coupling but also as a function of other
parameters such as the distance between the impurities, or the angle between their
spin orientations. Varying judiciously these parameters one may cross various phase
transition points. The richness of these alternations was explained in terms
of interference effects between the locally induced states at the impurity sites,
in a related way to the study of mirages and other interference effects in other
systems \cite{morr2,manoharan,bruno,pereg,corrals}.

In the case of interacting magnetic moments in a superconductor, the
nature of the superconducting transition changes, and the critical temperature
slightly increases \cite{larkin71}. Recently, Galitski and Larkin
studied \cite{galitski02} the effect of a spin-glass ordering of
magnetic impurities on the superconductivity. They showed that the
superconducting properties depend on the state of the magnetic
system and found a shift of the superconducting quantum critical
point.
A related problem is the mechanism of the exchange interaction
of magnetic moments in the superconductor. In the case of normal
metal, this is the RKKY interaction due to Friedel oscillations of the
magnetic density in the electron gas, induced by a single moment.
In the case of a superconductor, the RKKY interaction is affected
by the gap at the Fermi surface \cite{abrikosov88}. This effect
was revisited in some recent works \cite{aristov97,galitski02}. The
main result is that the form of the RKKY interaction in
superconductors is mainly preserved but the interaction contains a
decaying exponential factor which vanishes when $\Delta \to 0$.

As is well known, in the case of dominating ferromagnetic long-range ordering
of the moments, the competition between superconductivity and
ferromagnetism leads to the absence of superconductivity. In other
words, ferromagnetism suppresses the superconductivity acting like
an external magnetic field. However, it is possible to reach a
coexistence of the magnetization in magnetic domains with the
superconducting state \cite{bulaevskii85}.
On the other hand, the coexistence of superconductivity and antiferromagnetism
is much more favorable, since on average a Cooper pair, if the coherence
length is large enough, will feel a zero magnetic field. In general,
the two phases compete with each other but in some cases there is a coexistence.
Competition between antiferromagnetic and superconducting orders is an
important characteristic of  heavy-fermion systems \cite{Lonzarich},
which is also shared by high-$T_c$ materials \cite{hightc}
and low-dimensional systems \cite{Jerome}.
Heavy fermion systems that exhibit both superconductivity and antiferromagnetism
exhibit ratios between the N\'eel temperature $T_N$
and the superconducting critical temperature $T_c$  that can vary substantially
(of the order of $T_N/T_c\sim 1-100$),
with  coexistence of both types of order below $T_c$.
The coexistence of both types of order can be tuned  by external
parameters such as externally applied pressure or chemical pressure
(involving changes in the stoichiometry) \cite{Lonzarich,ishida99}.
It has recently been found that $UPd_2Al_3$
($T_N=14.3$ K and $T_c=2$ K) and $UNi_2Al_3$  ($T_N=4.5$ K and $T_c=1.2$ K)
show coexistence of superconductivity and local moment antiferromagnetism
\cite{Lonzarich,steglich93,feyerherm94,bernhoeft98,also00}.
However, in the Ce-based heavy-fermions magnetism typically competes
with superconductivity.
It has been found in the context of the Anderson model that, both in
the problem of local moment formation in the superconductor \cite{miguel}
and in the context of the Anderson lattice model, that, in a certain regime, 
a quantum phase transition is found \cite{sacramento}
which is characterized by an abrupt expulsion of magnetic order by  $d$-wave
superconductivity,
as an externally applied pressure increases. This transition
takes place when the $d$-wave superconducting critical temperature,
$T_c$, intercepts the magnetic critical temperature, $T_m$, under increasing pressure.

In our theoretical model, the magnetic moments are considered as a
certain magnetic structure embedded into a superconductor. There
are different ways to realize it practically by using modern
nanotechnology. One of them is related to the formation of a
magnetic structure on top of the superconducting layer. As another 
example, it can be also a laterally organized magnetic superlattice or
a hybrid ferromagnet/superconductor
structure \cite{lange03,stamopoulos05,yang04,gillijns05}.

Various types of heterostructures of superconductors and ferromagnets
have been considered in the recent literature \cite{revbuzdin,revlyuk,revbergeret}.
Also, the influence of magnetic dots (randomly or regularly distributed)
coupled to a superconductor has received attention \cite{peeters}.
In the first case the proximity effect due to the vicinity of the various
systems, where both magnetic fluctuations penetrate the superconductor
and superconducting fluctuations penetrate the magnetic system \cite{halterman},
has received particular attention due to potential device applications.
The penetration of the magnetic field in the superconductor splits the up and
down spin electron bands, due to the Zeeman effect, in addition to the orbital effect
through the vector potential. The Cooper pairs have a finite momentum due to the
band splitting and the order parameter oscillates in the superconducting phase.
If the size of the superconducting region is small enough, these oscillations
have noticeable consequences, such as for instance oscillations in the critical
temperature of the superconducting region in F/S/F structures \cite{revbuzdin} or
varying relative phases of the superconducting wave functions in S/F/S structures
if the thickness of the superconducting region is changed. This has been
confirmed looking at the Josephson current through the heterostructure \cite{revbuzdin}.
In particular, the so-called $\pi$-junctions, where the Josephson current vanishes,
have received attention in the literature \cite{shelukhin,kharitonov}.
In the second case of a distribution of magnetic dots in the close vicinity of the
superconductor it has been shown that if the magnetic
moments coupled are oriented via an external field, they act as
very effective pinning centers for vortices present in the superconductor
\cite{lyuksyutov1} originating so called frozen flux superconductors \cite{lyuksyutov2}.
The insertion of magnetic rods in the superconductor is also interesting
\cite{lyuksyutov3,vorticity,confinement}.

Now the interest to the superconductivity on a magnetic profile
arises on a quite different ground. It was found that electronic
properties of magnetic nanostructures can be used in various
magnetoelectronic devices, in which the magnetic state can be
effectively controlled by the magnetic field or electric current
and, in turn, the variation of the magnetic state changes strongly the
electronic characteristics of devices (and {\it vice
versa}) \cite{wolf01,zutic04}. More recently, the semiconducting
magnetic materials are included into
consideration \cite{dietl00,ohno00,dietl01,ruster03}, and the
superconductors are also used as some elements of the hybrid
structures for magnetoelectronic applications \cite{maekawa02}.
It seems therefore
worthwhile to consider the same type of phenomena in a superconductor with
magnetic impurities disposed in some form of ordering that may
be controllable from the outside \cite{takahashi,zhu,berciu}. 
In particular, we have in mind finite systems
to which we may attach leads through which we may insert currents that go through
a superconducting material with classical spins immersed with their spins
oriented in such a way that they form domain walls. These may be achieved
for instance imposing different boundary conditions between two sides of the
material. Therefore in this work we have in mind systems that are finite
and with non-periodic boundary conditions.

While it is interesting to consider the effect the superconducting state may have
on the magnetic profile in view of possible spintronics applications, it is
also interesting to see the effect of the 
patterned magnetization profile on the superconducting properties. 
In this work we focus on the latter aspect of the problem imposing a fixed
magnetic pattern. 
Usually people consider, say, a semiconductor in a potential
quantum well or a metal in a magnetic profile but not a superconductor
in a magnetic profile.  
We find that the impurities affect the properties
in a very local way and the pattern of interferences between the impurity induced
states is rather complex. We also find quantum phase transitions in these situations.
Even though we present our results for a fixed magnetic profile, we also study
the stability of the magnetic profile taking as stabilizing factor a possible
RKKY interaction between the impurities mediated by the quasiparticles of the
superconductor.

In section II we introduce the model that describes the magnetic impurities inserted
in the BCS $s$-wave superconductor. In section III we consider the quantum phase
transitions originated by the change of the spin coupling between the classical impurity
spins and the conduction electron spin density, revealed in the structure of
the energy levels, local spin density, local gap function and global spin density.
In section IV we study the nature of the quasiparticle states revealed in the local
density of states and the local kinetic energy. In section V we consider the stability
of the domain wall, taking into account an effective interaction between the impurity
spins that may originate in a RKKY interaction, and in section VI we study the
effect of a finite temperature, particularly on the quantum phase transition. We conclude
with section VII.

\begin{figure}
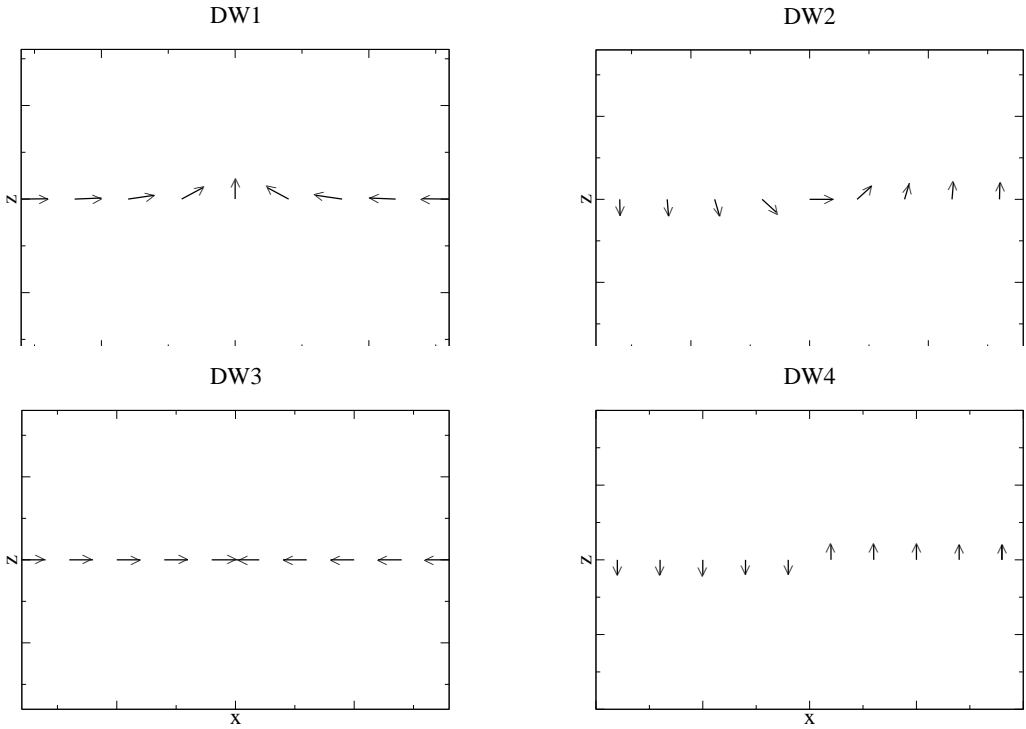

\includegraphics[width=0.33\textwidth]{fig1a.eps}
\hspace{1.5cm}
\includegraphics[width=0.33\textwidth]{fig1b.eps}
\vspace{1.5cm}
\includegraphics[width=0.33\textwidth]{fig1c.eps}
\hspace{1.5cm}
\includegraphics[width=0.33\textwidth]{fig1d.eps}
\caption{\label{dws}
Various N\'eel type domain walls considered in this paper.
}
\end{figure}

\section{Model}
Consider a set of classical spins immersed in a two-dimensional $s$-wave conventional
superconductor. 
We consider a two-dimensional system for computational simplicity and because
it is easier to experimentally control either the location of the magnetic
impurities or the local magnetic fields induced by vicinity of, for instance,
magnetic dots. 
We use a lattice description of the system.
In some sites we place classical spins parametrized like
\be
\frac{\vec{S}_l}{S} = \cos \varphi_l \vec{e}_x + \sin \varphi_l \vec{e}_z
\ee
where $S$ is the modulus of the spin. Thus, we assume that the spins lie
in the $x-z$ plane. 
The Hamiltonian of the system is given by
\bea
H &=& - \sum_{<i,j>,\sigma} t_{i,j} c_{i\sigma}^{\dagger} c_{j\sigma} - \mu \sum_{i\sigma}
c_{i\sigma}^{\dagger} c_{i\sigma} \nonumber \\
&+& \sum_i \left( \Delta_i c_{i\uparrow}^{\dagger} c_{i\downarrow}^{\dagger}
+ \Delta_i^* c_{i\downarrow} c_{i\uparrow} \right) \nonumber \\
&-& \sum_{i,,l,\sigma,\sigma'} J_{i,l} [ \cos \varphi_l c_{i\sigma}^{\dagger}
\sigma_{\sigma,\sigma'}^x c_{i\sigma'} \nonumber \\
&+& \sin \varphi_l c_{i\sigma}^{\dagger} \sigma_{\sigma,\sigma'}^z c_{i\sigma'} ] ,
\eea
where the first term describes the hopping of electrons between different
sites on the lattice, the second term includes the chemical potential $\mu $, the third 
one corresponds to the superconducting $s$-pairing with the site-dependent order
parameter $\Delta _i$, and the last term is the exchange interaction of an electron at
site $i$ with the magnetic impurity located at site $l$.
The hopping matrix is given by
$t_{i,j}=t \delta_{j,i+\delta} + t' \delta_{j,i+\delta'}$ where $\delta$ is a vector
to a nearest-neighbor site and $\delta'$ to a next-nearest site.
Most of our calculations will be performed taking $t=1$, $t'=0$ and $\mu=-1$.
For this value of the chemical potential the band is between quarter and half-filling.
The effects of introducing a next-nearest-neighbor hopping or varying
the chemical potential are discussed ahead.
Note that both the indices $l$ and $i,j$ specify sites on a two-dimensional
system.
The indices $i,j=1,...,N$, where $N$ is the number of lattice sites.
We take $J_{i,l}=J \delta_{i,l}$ and therefore the last sum is
over the sites, $l$, where a spin is located.
We assume that the spin configuration is fixed and
static. Later on we will study the stability of the spin configuration.

The diagonalization of this Hamiltonian is performed using the Bogoliubov transformation in the form
\bea
c_{i\uparrow} &=& \sum_n \left[ u_n(i,\uparrow) \gamma_n - v_n^*(i,\uparrow) \gamma_n^{\dagger}
\right] \nonumber \\
c_{i\downarrow} &=& \sum_n \left[ u_n(i,\downarrow) \gamma_n + v_n^*(i,\downarrow)
\gamma_n^{\dagger} \right]
\eea
Here $n$ is a complete set of states, $u_n$ and $v_n$ are related to the eigenfunctions of Hamiltonian (2),
and the new fermionic operators $\gamma_n$ are the quasiparticle operators. These are chosen
such that in term of new operators
\be
H = E_g + \sum_n \epsilon_n \gamma_n^{\dagger} \gamma_n
\ee
where $E_g$ is the groundstate energy and $\epsilon_n$ are the excitation
energies. As a consequence
\bea
\left[H , \gamma_n \right] &=& -\epsilon_n \gamma_n \nonumber \\
\left[H , \gamma_n^{\dagger} \right] &=& \epsilon_n \gamma_n^{\dagger}
\eea
The coefficients $u_n(i,\sigma )$, $v_n(n,\sigma )$ can be obtained solving
the Bogoliubov-de Gennes equations \cite{degennes}.
Defining the vector
\[ \psi_n(i) = \left( \begin{array}{c}
u_n(i,\uparrow) \\
v_n(i,\downarrow) \\
u_n(i,\downarrow) \\
v_n(i,\uparrow) \end{array} \right)
\]
the BdG equations can be written as
\be {\cal H} \psi_n = \epsilon_n \psi_n
\label{BdG}
\ee
where the matrix ${\cal H}$ at site $i$ is given by
\[ {\cal H} = \left( \begin{array}{cccc}
-h
-\mu -J_{i,l} \sin \varphi_l & \Delta_i &
-J_{i,l} \cos \varphi_l & 0
 \\
\Delta_i^* & h
+\mu -J_{i,l}  \sin \varphi_l  & 0 &
-J_{i,l} \cos \varphi_l
 \\
-J_{i,l} \cos \varphi_l & 0  & -h
 -\mu +J_{i,l}
 \sin \varphi_l
 & \Delta_i \\
0 & -J_{i,l} \cos \varphi_l & \Delta_i^*  & h +\mu +
J_{i,l} \sin \varphi_l
 \end{array} \right)
\]
\noindent where $h=(t\hat{s}_{\delta}+t'\hat{s}_{\delta'})$ with 
$\hat{s}_{\delta} f(i)=f(i+\delta)$
and $\hat{s}_{\delta'} f(i)=f(i+\delta')$.
The solution of these equations gives both the energy eigenvalues and eigenstates.
The problem involves the diagonalization of a $(4N)\times(4N)$ matrix.
The solution of the BdG equations
is performed self-consistently imposing at each iteration that
\be
\Delta_i = \frac{V}{2} \left[ <c_{i \uparrow} c_{i\downarrow}> -
<c_{i \downarrow} c_{i\uparrow}> \right]
\ee
where $V$ is the effective attractive interaction between the electrons.
Using the canonical transformation this can be written as
\bea
\Delta_i &=& -V \sum_{n,0<\epsilon_n<\hbar \omega_D} \{ f_n \left( u_n(i,\uparrow) v_n^*(i,\downarrow) +
u_n(i,\downarrow) v_n^*(i,\uparrow) \right)  \nonumber \\
&-& \frac{1}{2} \left[ u_n(i,\uparrow) v_n^*(i,\downarrow) +
u_n(i,\downarrow) v_n^*(i,\uparrow) \right] \}
\eea
where $\omega_D$ is the Debye frequency,
and $f_n$ is the Fermi function defined as usual like
\[
f_n=\frac{1}{e^{\epsilon_n/T}+1}
\]
where $T$ is the temperature.
We note that the Bogoliubov-de Gennes equations are invariant under the substitutions
$\epsilon_n \rightarrow -\epsilon_n$, $u(\uparrow) \rightarrow v(\uparrow)$,
$v(\uparrow) \rightarrow u(\uparrow)$, $v(\downarrow) \rightarrow -u(\downarrow)$,
$u(\downarrow) \rightarrow -v(\downarrow)$.

We are interested in calculating various quantities.
In particular we calculate the quasiparticle spin densities
\bea
s_z(i) &=& \frac{1}{2} <c_{i,\sigma}^{\dagger} \sigma_{\sigma,\sigma'}^z c_{i,\sigma'}>
\nonumber \\
&=& \frac{1}{2}
\sum_n \{ f_n \left[ |u_n(i,\uparrow)|^2 - |u_n(i,\downarrow)|^2 \right] \nonumber \\
&+& (1-f_n) \left[ |v_n(i,\uparrow)|^2 - |v_n(i,\downarrow)|^2 \right] \}
\nonumber \\
s_x(i) &=& \frac{1}{2} <c_{i,\sigma}^{\dagger} \sigma_{\sigma,\sigma'}^x c_{i,\sigma'}>
\nonumber \\
&=& \frac{1}{2}
\sum_n \{ f_n \left[ u_n^*(i,\uparrow) u_n(i,\downarrow)+
u_n^*(i,\downarrow) u_n(i,\uparrow) \right] \nonumber \\
&-& (1-f_n) \left[ v_n^*(i,\uparrow) v_n(i,\downarrow)+
v_n^*(i,\downarrow) v_n(i,\uparrow) \right] \}
\eea
where the sums are taken over the positive excitation energies.
Also, we are interested in calculating the local density of states
\bea
\rho(\epsilon,i) &=&
\sum_{n,\sigma} \{ |u_n(i,\sigma)|^2 \delta(\epsilon-\epsilon_n) \nonumber \\
&+& |v_n(i,\sigma)|^2 \delta(\epsilon+\epsilon_n)  \}
\eea
Since the system is finite the states are discrete. This can be written as
\be
\rho(\epsilon, i) = \sum_{n,\sigma, \alpha} \rho_{\alpha}(\epsilon_n,i,\sigma)
\ee
where $\alpha=+$ runs over the positive energy eigenvalues of
Eq. (\ref{BdG}) $\epsilon_n^+$ and $\alpha=-$ runs
over the negative energies $\epsilon_n^-$. Therefore 
\be
\rho_+(\epsilon_n,i,\sigma) = |u_n(i,\sigma)|^2 \delta_{\epsilon,\epsilon_n^+} +
|v_n(i,\sigma)|^2 \delta_{\epsilon,\epsilon_n^-}
\ee
and
\be
\rho_-(\epsilon_n,i,\sigma) = |v_n(i,\sigma)|^2 \delta_{\epsilon,\epsilon_n^+}+
|u_n(i,\sigma)|^2 \delta_{\epsilon,\epsilon_n^-}
\ee
where $\epsilon_n^{\alpha}=-\epsilon_n^{-\alpha}$.

\begin{figure}
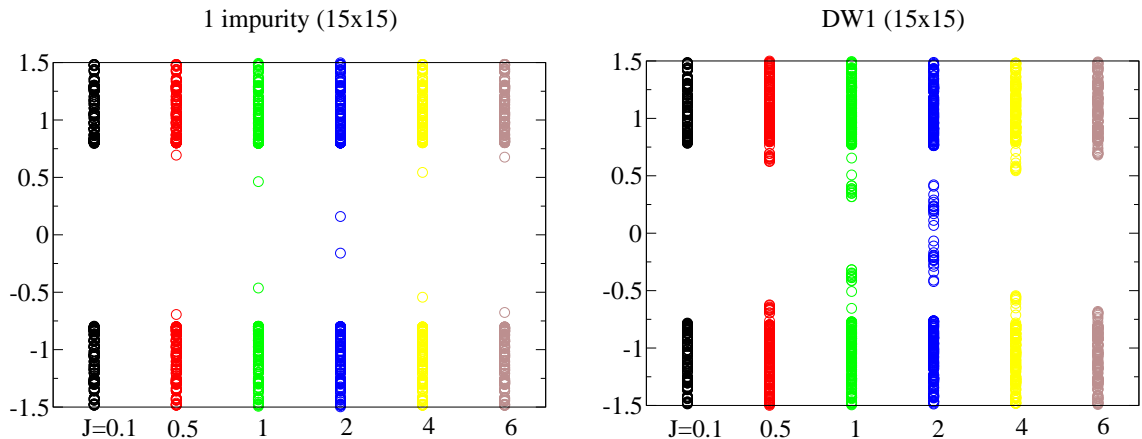

\includegraphics[width=0.4\textwidth]{fig2a.eps}
\hspace{0.5cm}
\includegraphics[width=0.4\textwidth]{fig2b.eps}
\caption{\label{fig1}
a) Energy levels for 1 impurity. There is one pair of boundstates originated by the impurity.
As $J$ grows the level appears to "cross" zero energy. The crossing is between $J=1$
and $J=2$ (see ahead). b) Energy levels for DW1 domain wall. 
}
\end{figure}

Consider now the possibility that there is a line of impurities, $\vec{S}_l$,
such that their spin orientations are correlated. For simplicity
we consider a one-dimensional array of spins that traverse the two-dimensional
material from one border to the other along the $x$ direction.
A dense impurity distribution will destroy superconductivity if the spin
coupling is strong enough. Therefore, to study the effects of interference
we limit ourselves to lines of spins in the two-dimensional electronic system.
One may impose a magnetic field at each side of the sample such
that it orients the first and last spins. Choosing different boundary
conditions we may create different domain walls assuming that there
are interactions between the spins that tend to orient them. These interactions
may either be ferromagnetic or antiferromagnetic and may have its origin
in RKKY type interactions via the superconducting substrate. As we mentioned before,
the form of the interactions mediated by the quasiparticles has a form similar
to the one of a standard metal. 

Clearly if the number of spins increases considerably and/or their coupling
to the electron density increases enough superconductivity will be destroyed.
We will be focusing on situations where superconductivity prevails, as evidenced
by the self-consistent solution of the BdG equations. In particular, if we consider
a fully two-dimensional distribution of classical spins, unless the coupling is small,
in general superconductivity will be destroyed. Since the Zeeman term of each classical
spin acts like a local magnetic field we expect at the very least that if the applied
field exceeds the critical field superconductivity will be destroyed.

The various cases considered are displayed in Fig. \ref{dws}. Besides the cases of
one impurity and two impurities previously considered, and that we briefly
consider here to compare with the new results, we consider various situations
where for instance, we have a domain wall of the N\'eel type (here limited to a line of spins
to simplify) where the leftmost spin is either oriented along the chain
(defined as the $x$ direction) or perpendicularly to it (these are the cases
$DW1$ and $DW2$, respectively). 
Specifically in the case of domain wall $DW1$ we choose
\be
\varphi_l = \frac{\pi}{2} + \frac{\pi}{2} \tanh \frac{x-x_c}{\lambda}
\label{lambda1}
\ee
for the domain wall $DW2$
\be
\varphi_l = \frac{\pi}{2} \tanh \frac{x-x_c}{\lambda} 
\label{lambda2}
\ee
The other configurations are characterized by
$\varphi_l=\frac{\pi}{2} \theta(-x) -\frac{\pi}{2} \theta(x)$ for the domain wall $DW3$,
$\varphi_l =\pi \theta(-x)$ and $\varphi_l=0$ for $x>0$, for $DW4$
and $\varphi_l = \frac{\pi}{2}(1+(-1)^x)$,
for the antiferromagnetic case (AF), $DW5$,

Most our results were obtained for a system of $15 \times 15$ sites. Increasing
the system size does not affect qualitatively the results. For instance, we have
considered a system of $25 \times 25$ sites and the results are very similar.
The only visible difference is the reduction of the finite size effects near the
borders of the system. We used parameters such that superconductivity is stable
and the superconducting gap is relatively large so that the in-gap states are easily
identified. Choosing units where the hopping $t=1$, we get a gap of the order of
$0.4$ choosing $V=4t$. Typically we choose $\lambda=3$. 

\section{Quantum phase transitions}

\subsection{Energy levels}

We begin by revisiting some results for a situation where we have one or two
impurities immersed in the material.
As is typical of a $s$-wave pairing the clean system has a gap. Introducing
one magnetic impurity generates two bound states in the gap. One is at positive
energies and the other is at a symmetric negative energy. 
This is clearly shown
in Fig. \ref{fig1}. The nature of the wavefunctions corresponding to the two eigenstates
will be discussed later.
If we introduce two impurities the number of states in the gap doubles. In general
if there are $N_i$ impurities there are $2N_i$ states in the gap, $N_i$ at
positive energies and the same at negative energies.
As the coupling between the electron spin density and the impurity
spin grows, the energy of the boundstate lowers and approaches the Fermi level.
Increasing further the coupling
the level is repelled from the chemical potential. The nature of the groundstate
has changed, as we will see ahead.

In Fig. \ref{fig1} we also show the energy spectra for the domain wall DW1 for different values
of the coupling. There are now as many boundstates as impurity spins (for positive
energies).
For small coupling the energy levels are close to the top of the gap, but as the
coupling grows the trend is similar to the single impurity case. For a large
system the gap will be virtually filled close to the transition where the level
crossing(s) take place. As we will see there are several level crossings
as the coupling varies. 
Note that $J$ actually means $JS$
where $S$ is the magnitude of the local spins. Therefore we may consider
$S=1/2$ and change the Zeeman term coupling
value. But we can as well consider that the coupling is not changing
much but we are inserting impurity spins with different values of
$S$. This is truly the classical limit.
Previously the effect of the change in the relative angles or distances between
two impurities was considered \cite{morr1}. 
These changes also induce changes in the energy spectra
and affect the various level crossings.
The corresponding of the change in relative angles and/or
distances between two impurities in the case of domain walls is the change in the width of the
domain wall, $\lambda$. However, the analysis of these interference effects on the
states is more complex due to the multiple scatterings, as we will show ahead.

\begin{figure}
\includegraphics[width=0.4\textwidth]{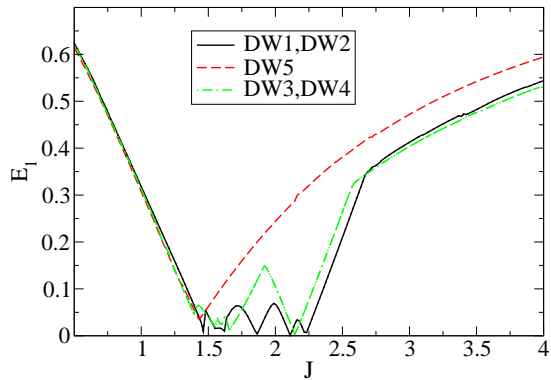}
\caption{\label{fig3}
Lowest energy level as a function of $J$ for the various domain walls.
Note the various level crossings (near zeros) for the various cases.
The crossings are concentrated for $1<J<2$. Note that for the antiferromagnetic
chain there is a single level crossing. Domain walls DW1 and DW2 have the same
lowest state energy and the same happens with domain walls DW3 and DW4.
}
\end{figure}

In contrast to the case of a small number of isolated magnetic moments creating a number of levels in the gap, an ordered array of 
impurities creates some minibands. It can be clearly seen in Fig.~\ref{fig1}, even though we performed our calculations for a finite system.  

The detail of the level crossings is better studied considering the evolution of the
lowest level (with positive energy) as a function of the coupling, $J$. This is shown
in Fig. \ref{fig3} for various systems. In the case of a single impurity there is
a single level crossing, which for our parameters occurs at a value between $J=1$ 
and $J=2$. In the case of the domain walls there are several level crossings, as
illustrated in Fig. \ref{fig3}. For the same set of parameters the level crossings
are contained in an interval of coupling strengths that is of the same order.
We recall that at each level crossing the lowest energy does not strictly reach zero. At a finite
value of the coupling the crossing occurs without closing the gap indicating that the quantum phase
transition is actually a first order one. The same occurs in the cases of the domain
walls. However, since the number of boundstates inside the gap increases, the mini-gaps
near the various transitions are quite small and in the thermodynamic limit should become
vanishingly small. Even for a small system, considering, for instance, a two-dimensional
distribution of impurity spins, one expects that the gap will be filled with states.
Actually, for a moderate coupling of the order of $J \sim 1$, 
the superconductivity can be destroyed by
the local magnetic fields created by the impurities. The states in the gap
are localized as will be shown later.

\begin{figure}
\includegraphics[width=0.3\textwidth]{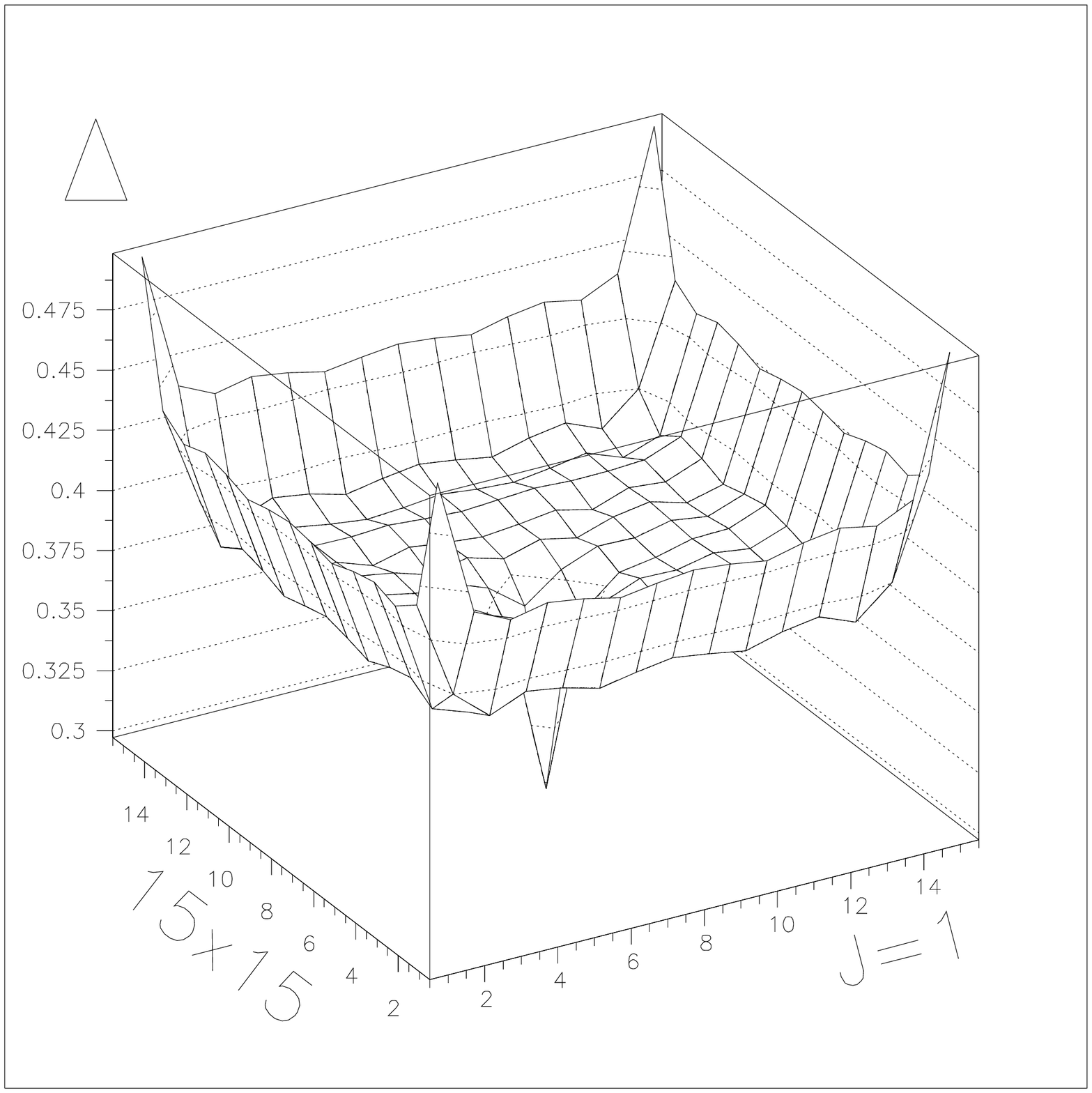}
\includegraphics[width=0.3\textwidth]{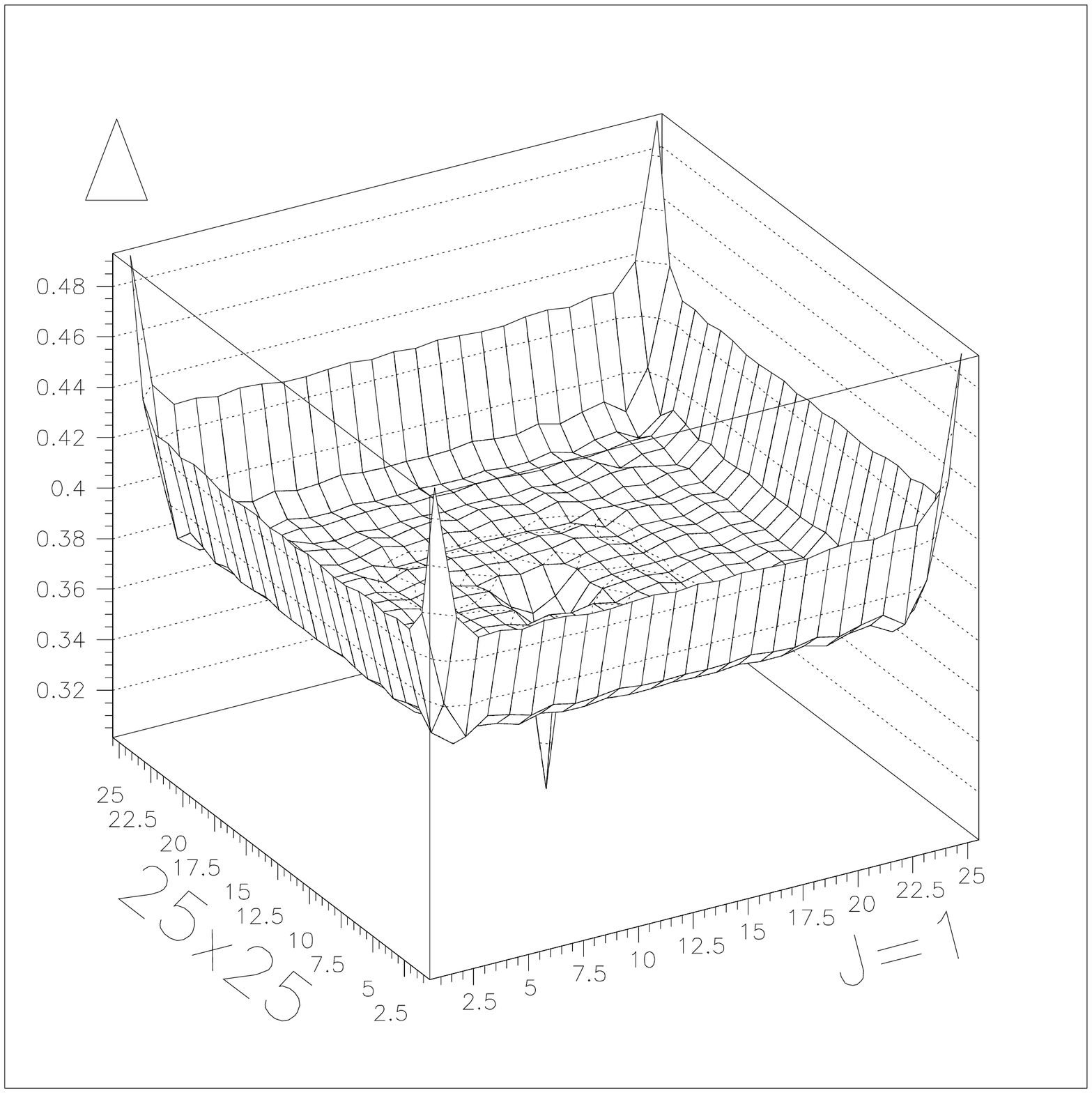}
\includegraphics[width=0.3\textwidth]{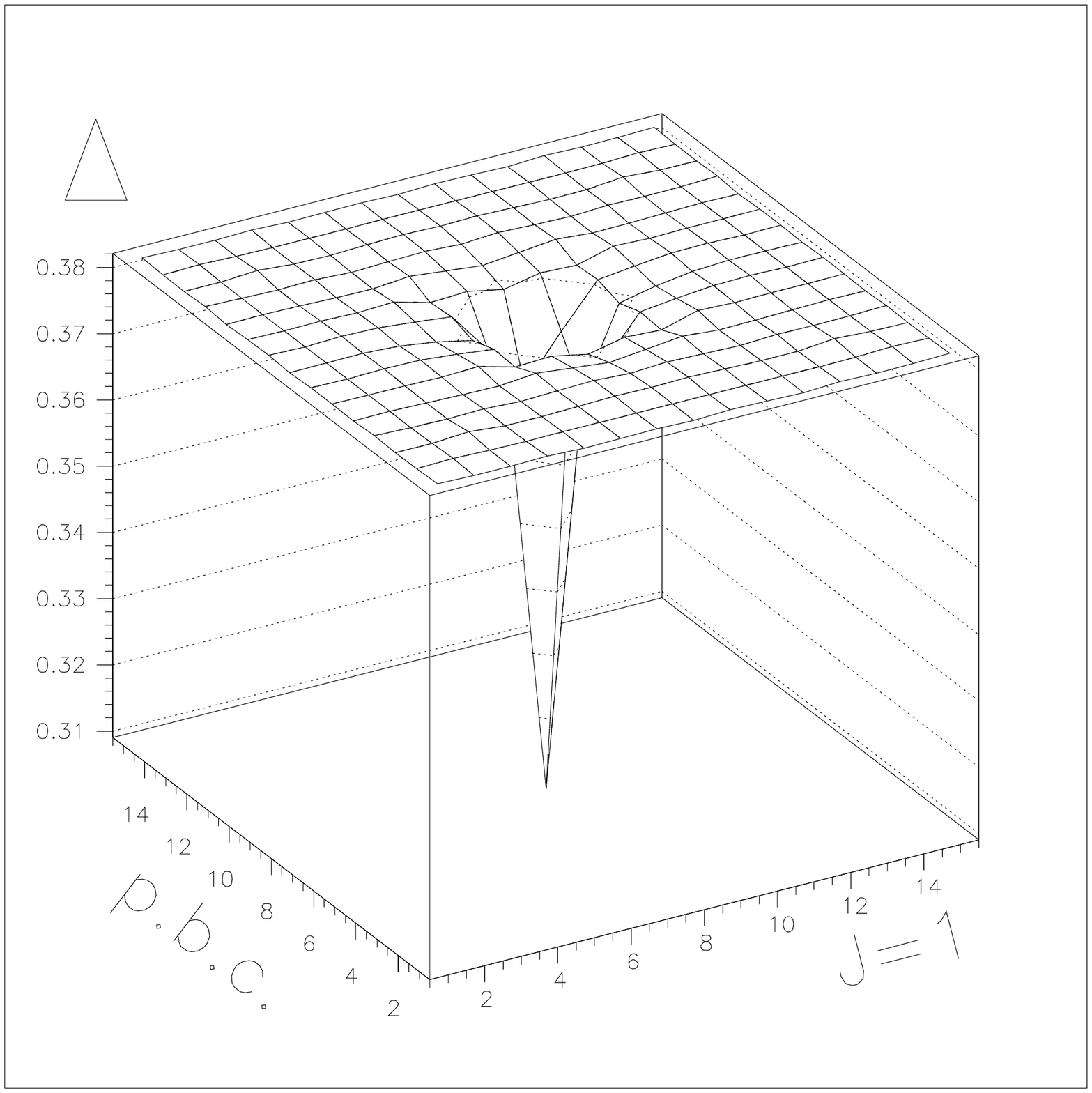}
\caption{\label{figcompare}
Comparison of the spatial dependence of the gap function for the
case of a single impurity, for the cases a) $15 \times 15$ and b) $25 \times 25$
for open boundary conditions and c) periodic boundary conditions, for $J=1$. 
}
\end{figure}

\begin{figure}
\includegraphics[width=0.4\textwidth]{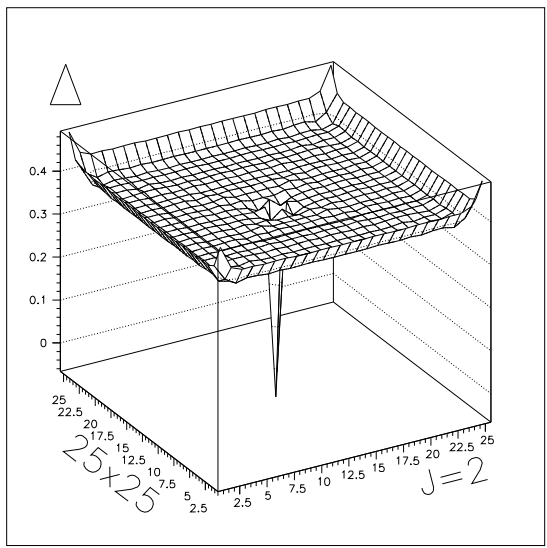}
\caption{\label{fig7}
$\Delta$ for $J=2$ for one impurity.
$\Delta$ decreases at impurity site (this is the same as for
a non-magnetic impurity). Note that there is a $\pi$ shift
in $\Delta$ (it becomes negative) at the impurity site in contrast to the case
for $J=1$, shown in Fig. (\ref{figcompare}b).
}
\end{figure}

\begin{figure*}
\includegraphics[width=0.4\textwidth]{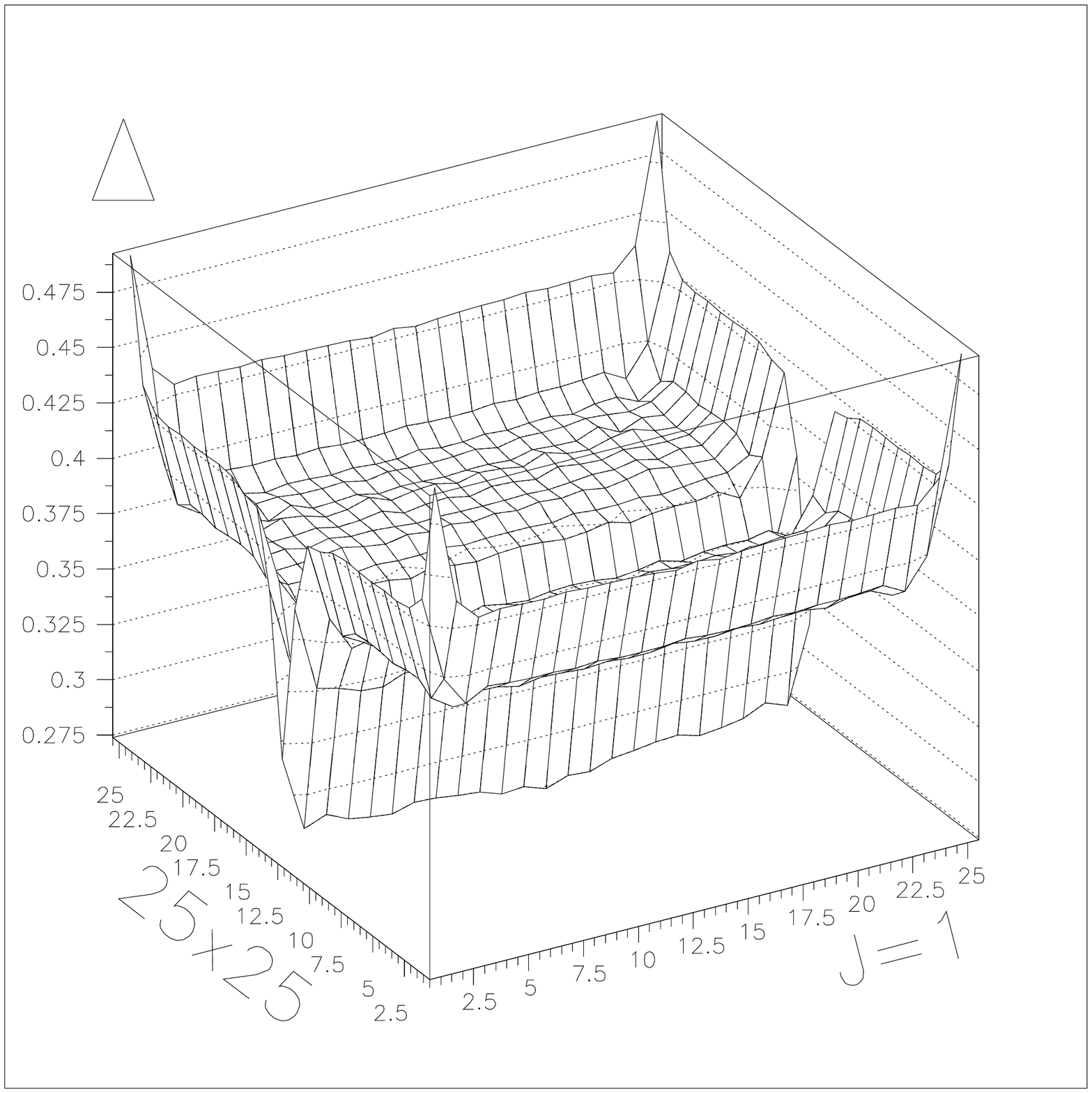}
\includegraphics[width=0.4\textwidth]{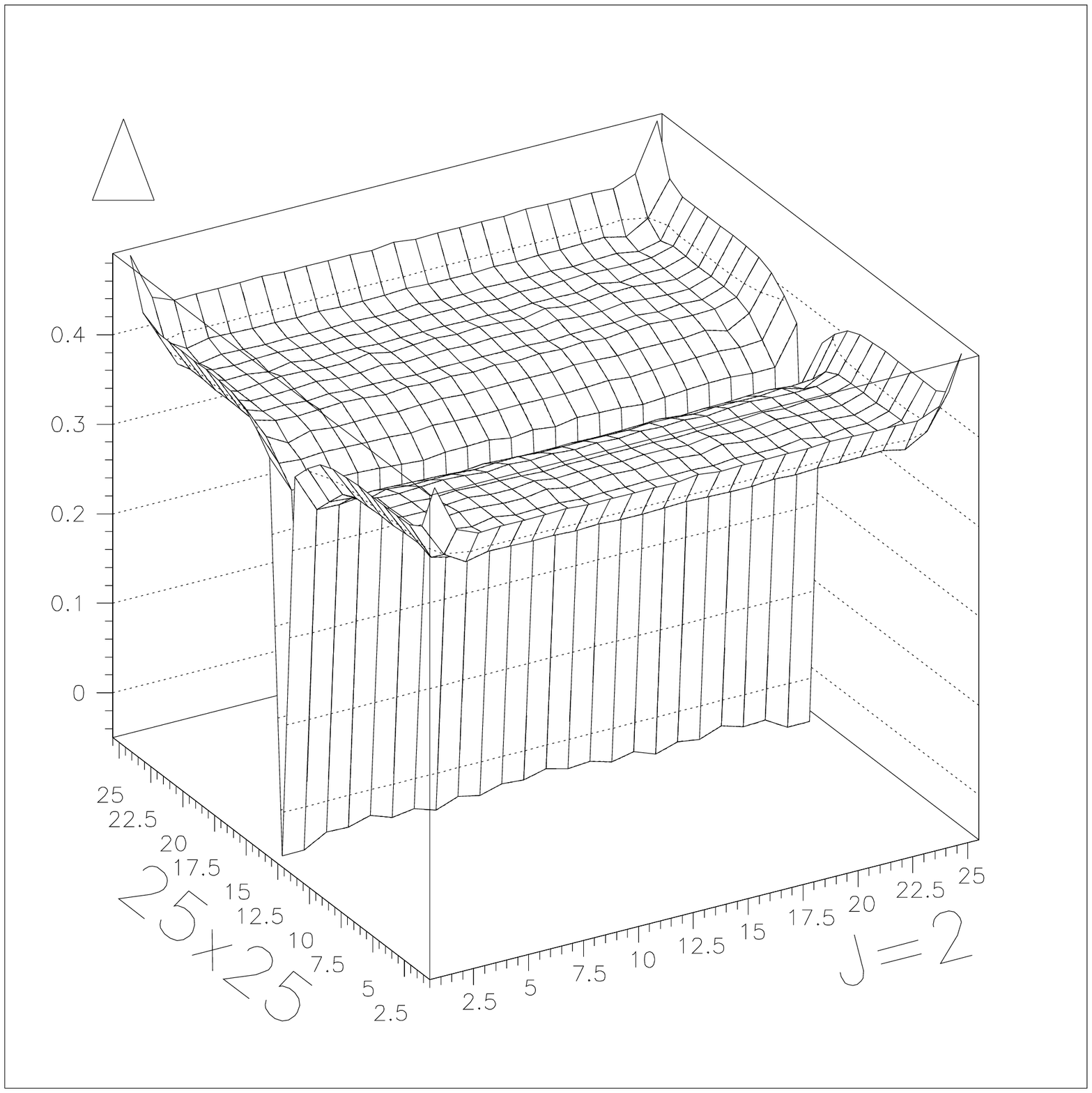}
\caption{\label{fig8}
$\Delta$ for DW1 for $J=1,2$. 
Note that for $J=2$ there is again a $\pi$ shift.
}
\end{figure*}

\begin{figure*}
\includegraphics[width=0.4\textwidth]{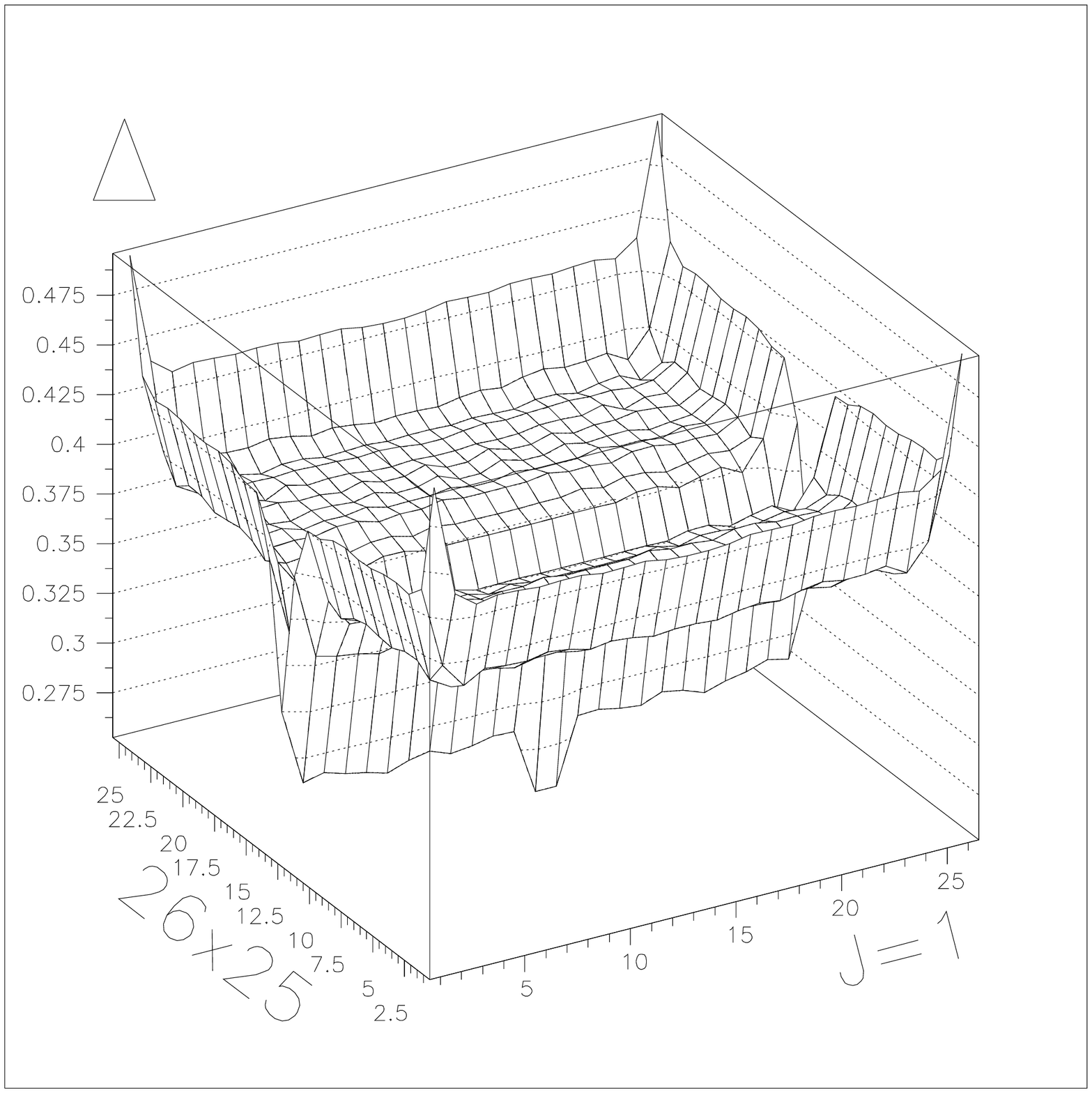}
\includegraphics[width=0.4\textwidth]{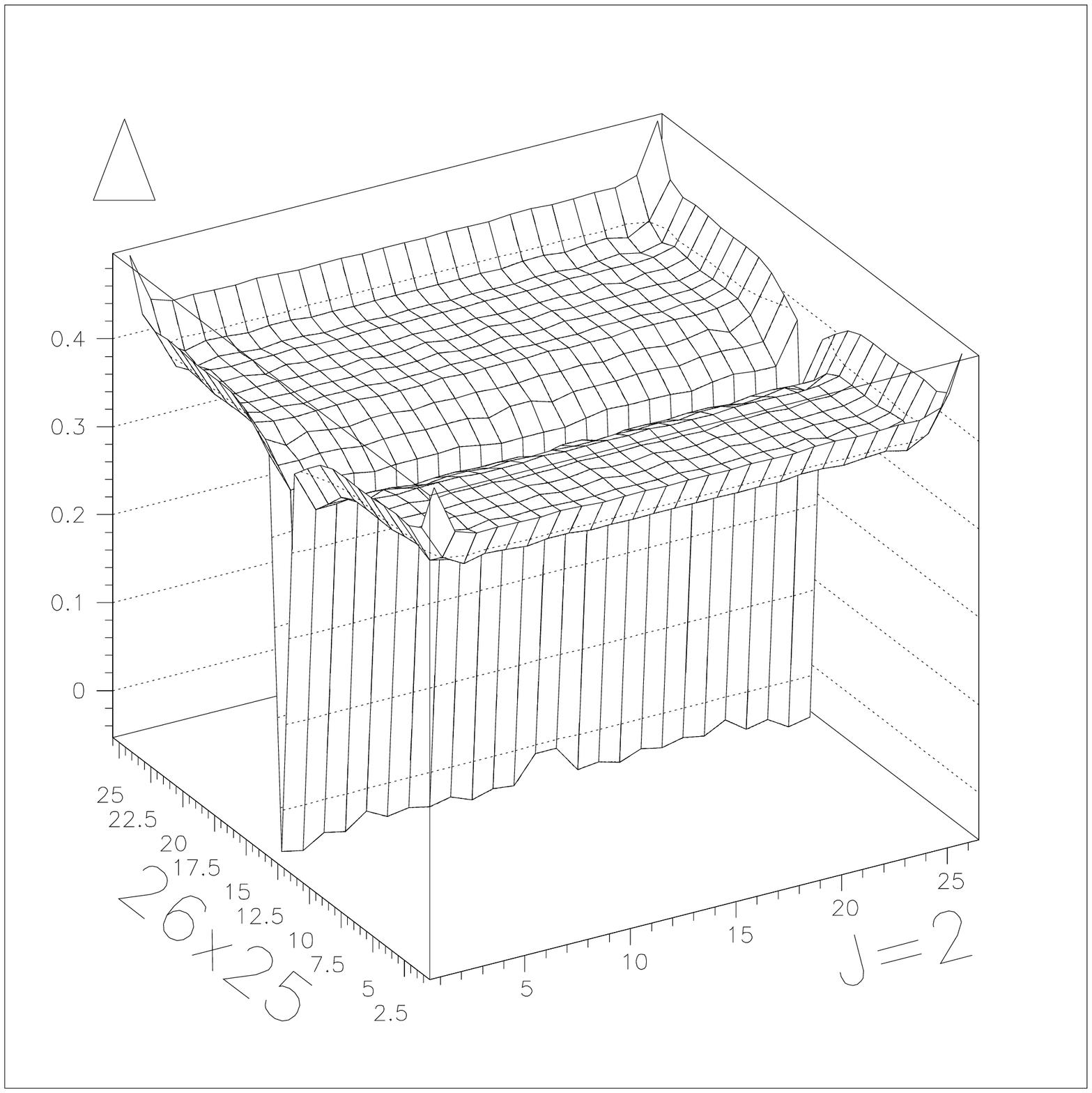}
\caption{\label{fig9}
$\Delta$ for DW4 for $J=1,2$. Once again there is a $\pi$ shift.
}
\end{figure*}

\subsection{Local gap function}

A characteristic of the order in the system is given by the behavior of the superconducting order
parameter, as a function of space \cite{schlottmann1}. 
In Fig. \ref{figcompare} we study the effects of system size and boundary conditions on
the gap function in the case of a single impurity, located at the center of the system.
We see that for open boundary conditions the finite size effects are important
near the border of the system. Increasing the system size these diminish. 
In the case of periodic boundary conditions 
the finite size effects are virtually absent. Note that away from the borders
the results for the cases $15\times 15$ and $25 \times 25$ are qualitatively the same.
Therefore in the rest of the paper we will illustrate the results considering either
system sizes preferring however the smaller system size since this decreases the 
computational time. Also, as discussed above,
we are aiming at the equilibrium properties of a system which is finite, and therefore
we will consider the open boundary conditions instead of the more standard periodic
boundary conditions.

In Fig. \ref{fig7} we compare $\Delta$ for the typical
couplings of $J=1$ and  $J=2$ for the case of one impurity. 
The first thing to notice is that the order parameter is
only affected very close to the impurity site. Since we are using open boundary conditions
the behavior of the order parameter, and other quantities is also affected near
the border. If we were to use periodic boundary conditions these effects would be
strongly reduced and for a relatively large system they would be almost vanishing.
We note however that near the impurity the boundary conditions have no effect as expected.
We note the previously observed $\pi$ shift of the order parameter when $J$ is large enough.
This effect prevails when we have several impurities.
We see from Figs. \ref{fig8}, \ref{fig9} that the
$\pi$ shift is observed for the more complex structures, indicating once
again that some properties are of a local nature.
Also we see that the orientation of the spins does not affect significantly
the order parameter. This is to be expected since we are considering
singlet pairing, and therefore rotationally invariant. While the spin
density is obviously strongly dependent on the impurity spin orientation,
the order parameter is only mildly affected particularly at the center of
the solitonic like spin configuration. If the coupling is strong enough
the order parameter is basically constant along the impurity line.

As the level crossings occur there is a change of phase of $\pi$ locally,
as evidenced by the single impurity results. Therefore, as the various
level crossings occur in the cases of several impurities there may be inhomogeneities
since the capture of the electrons by the impurities is not necessarily
uniform, as evidenced by the non-homogeneous changes of local spin density. Therefore it
is natural to expect some inhomogeneities along the chain. This is also seen
in the LDOS shown ahead.

A simple explanation for the occurrence of the $\pi$ shift is not available \cite{review}.
It is argued that it is related to the $\pi$-junctions referred above, but these occur
for specially commensurate widths of the ferromagnetic slab in /S/F/S heterostructures.
Here the effect is quite local and therefore a relation is not clear.

\subsection{Local spin density}

\begin{figure}
\includegraphics[width=0.35\textwidth]{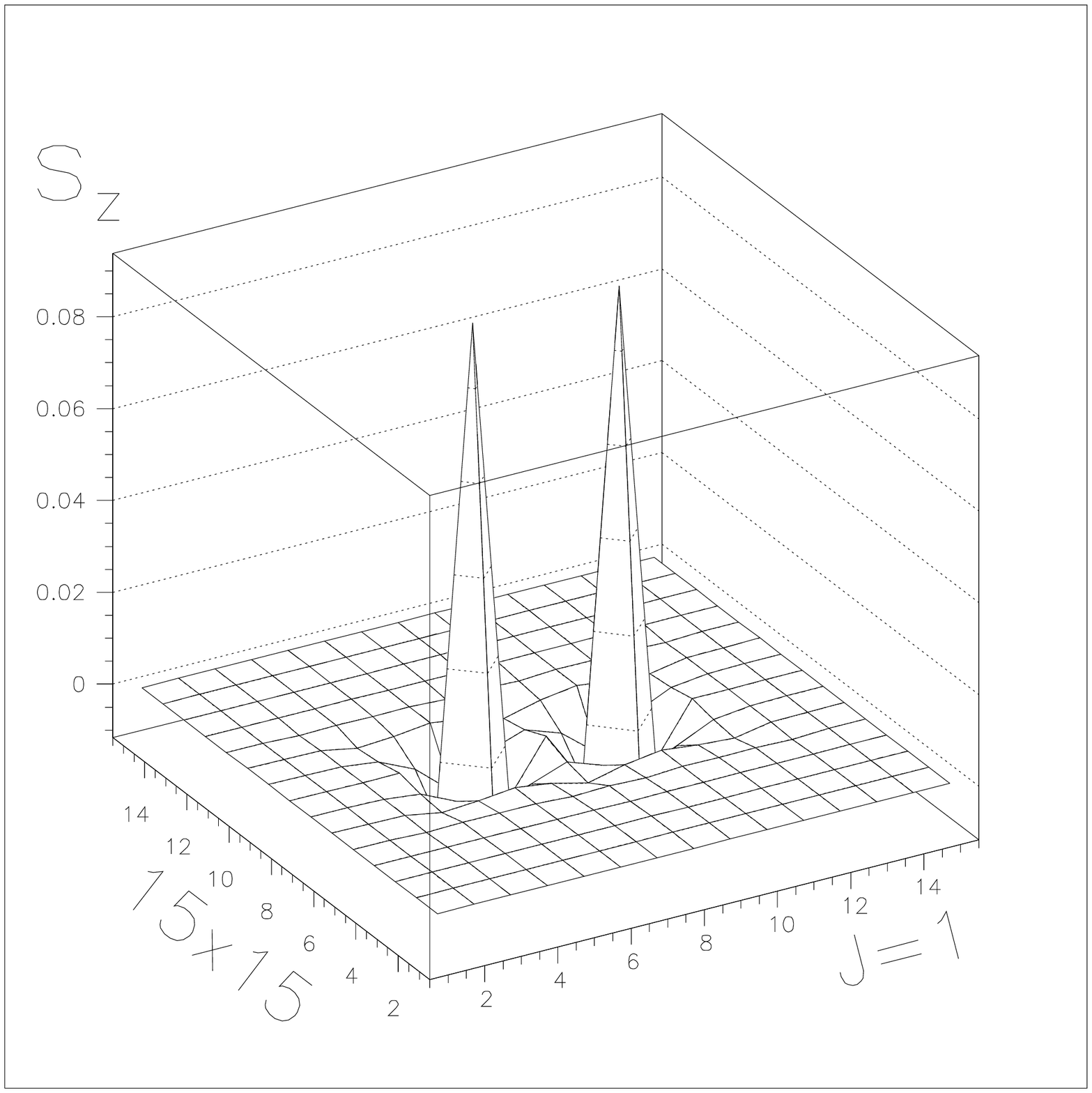}
\includegraphics[width=0.35\textwidth]{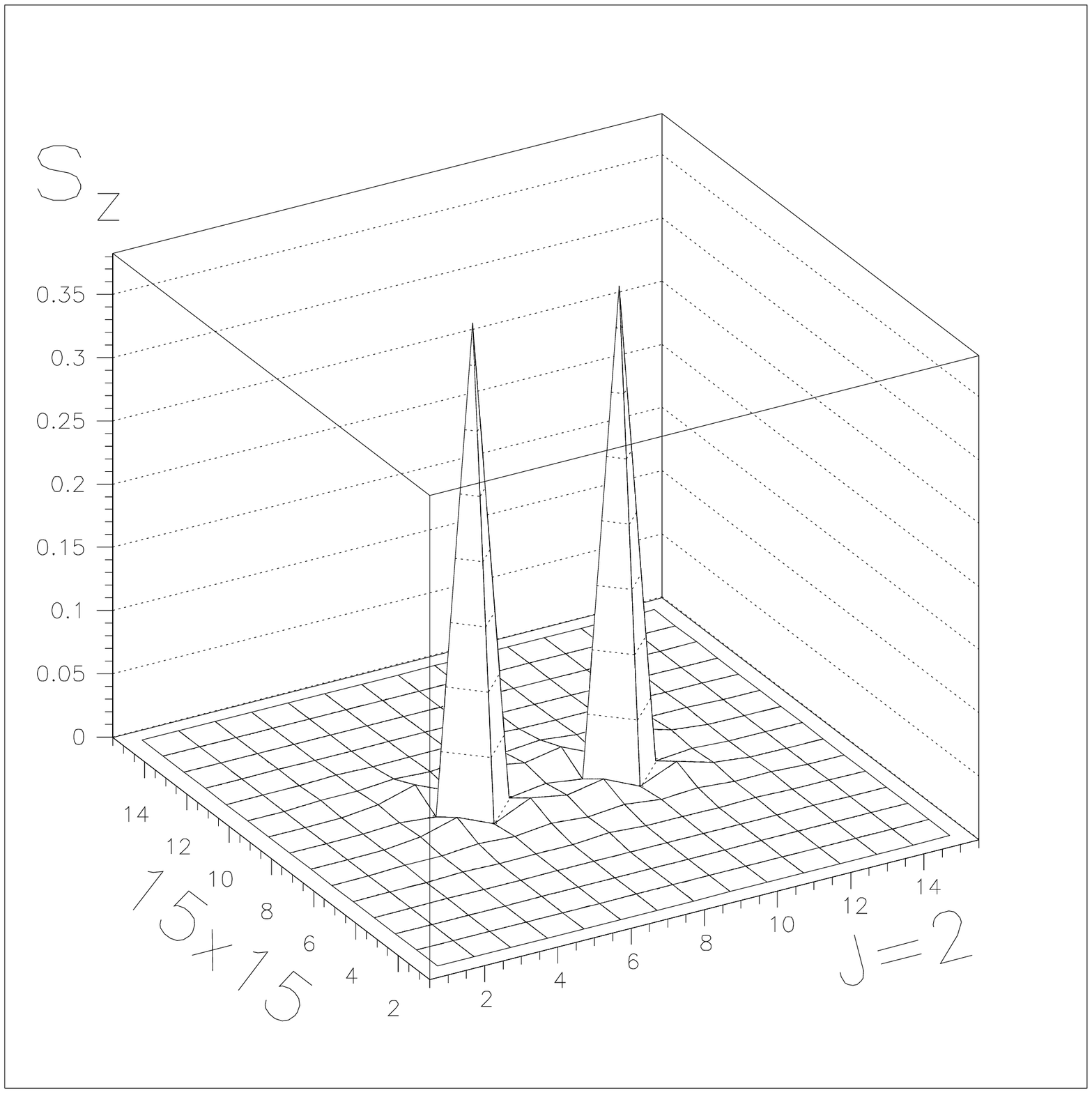}
\includegraphics[width=0.35\textwidth]{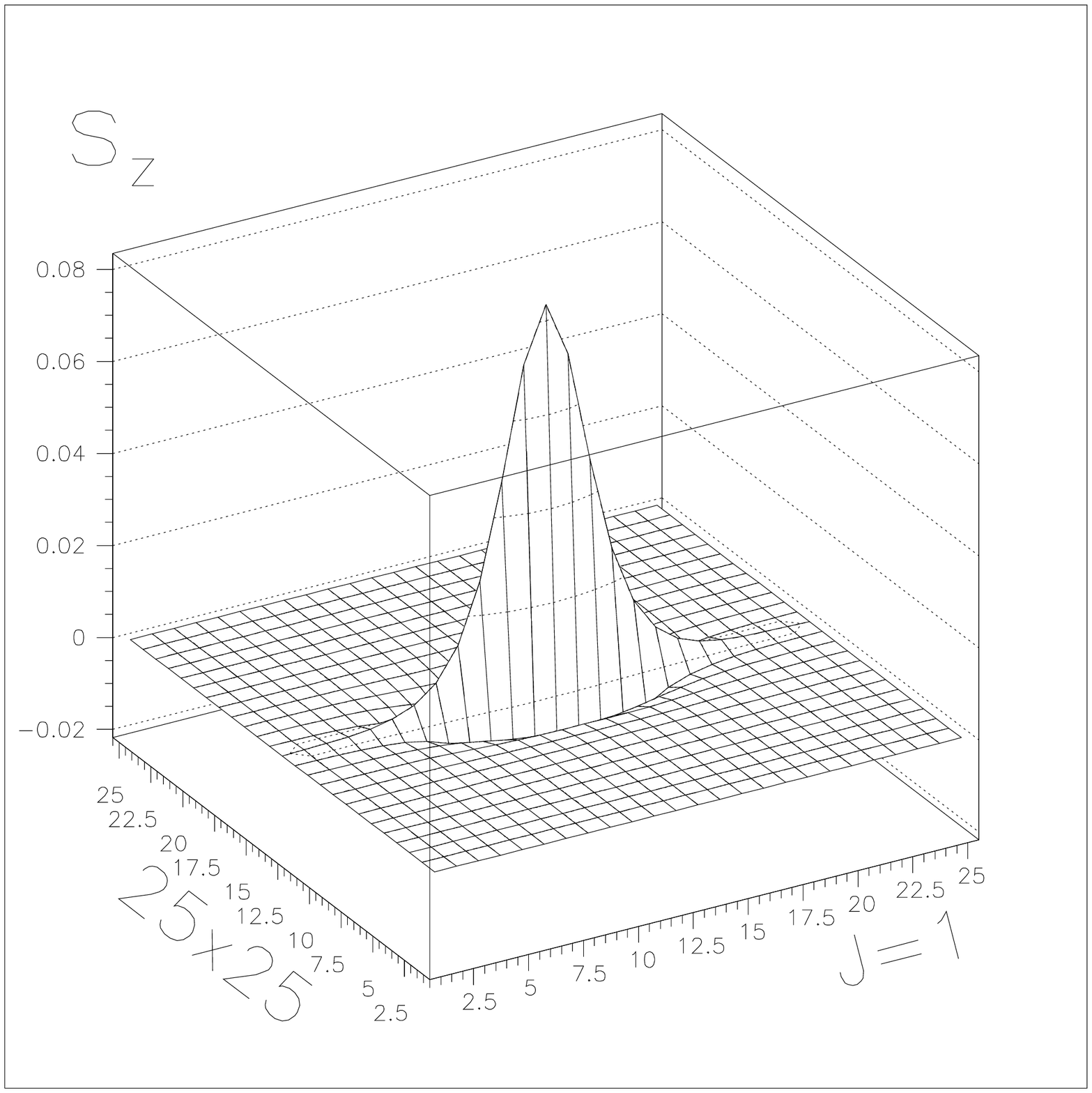}
\includegraphics[width=0.35\textwidth]{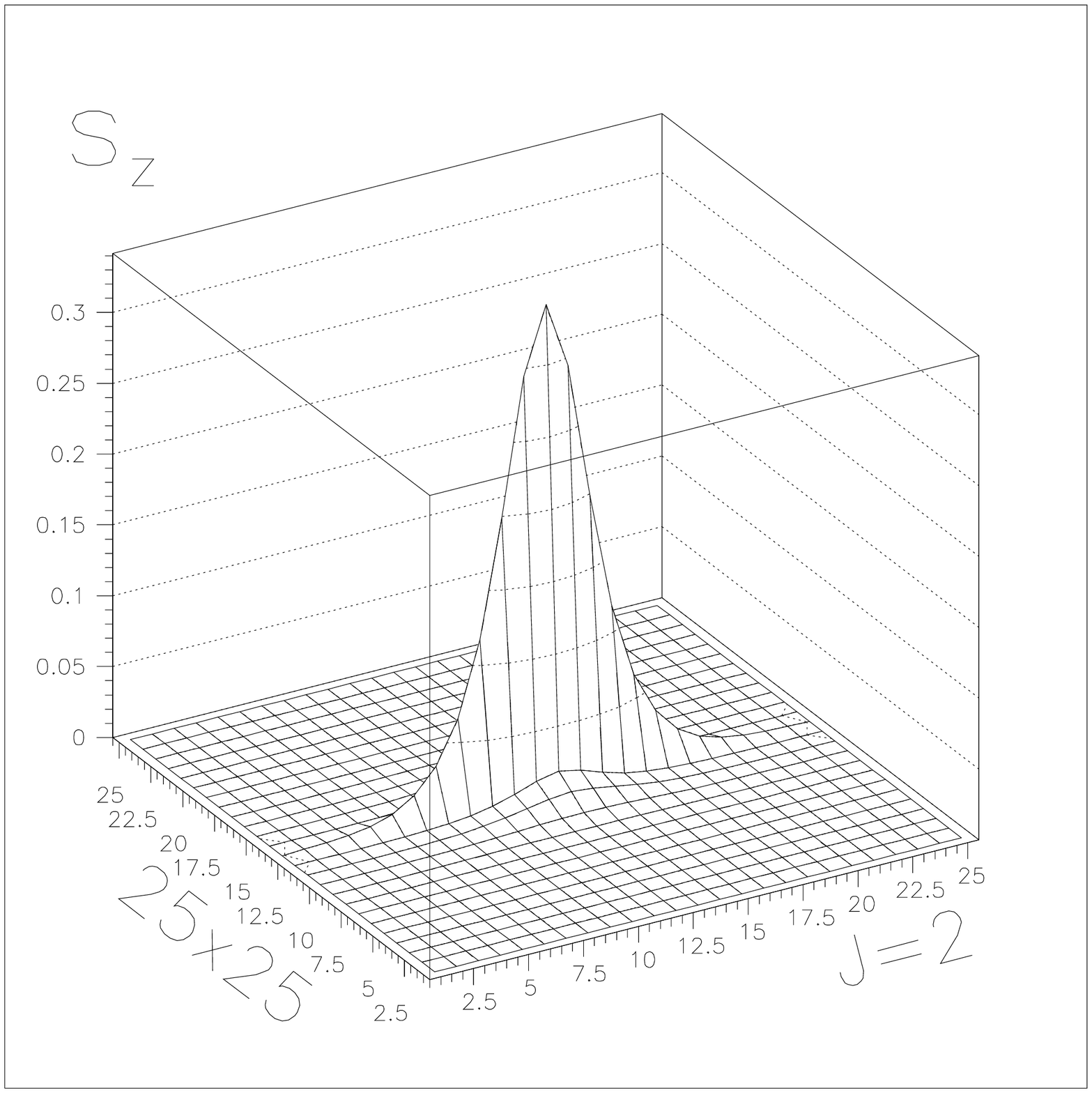}
\includegraphics[width=0.35\textwidth]{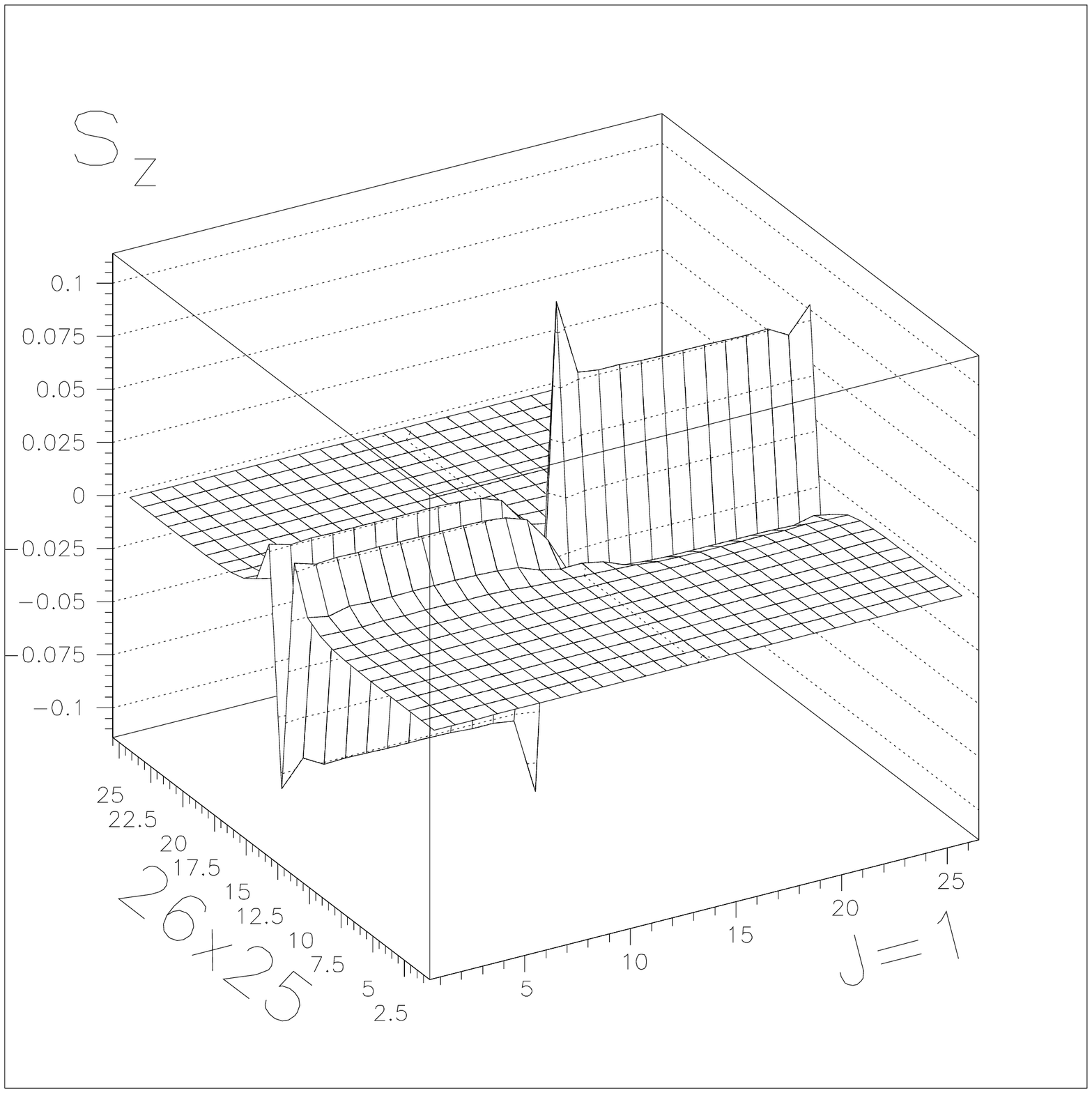}
\includegraphics[width=0.35\textwidth]{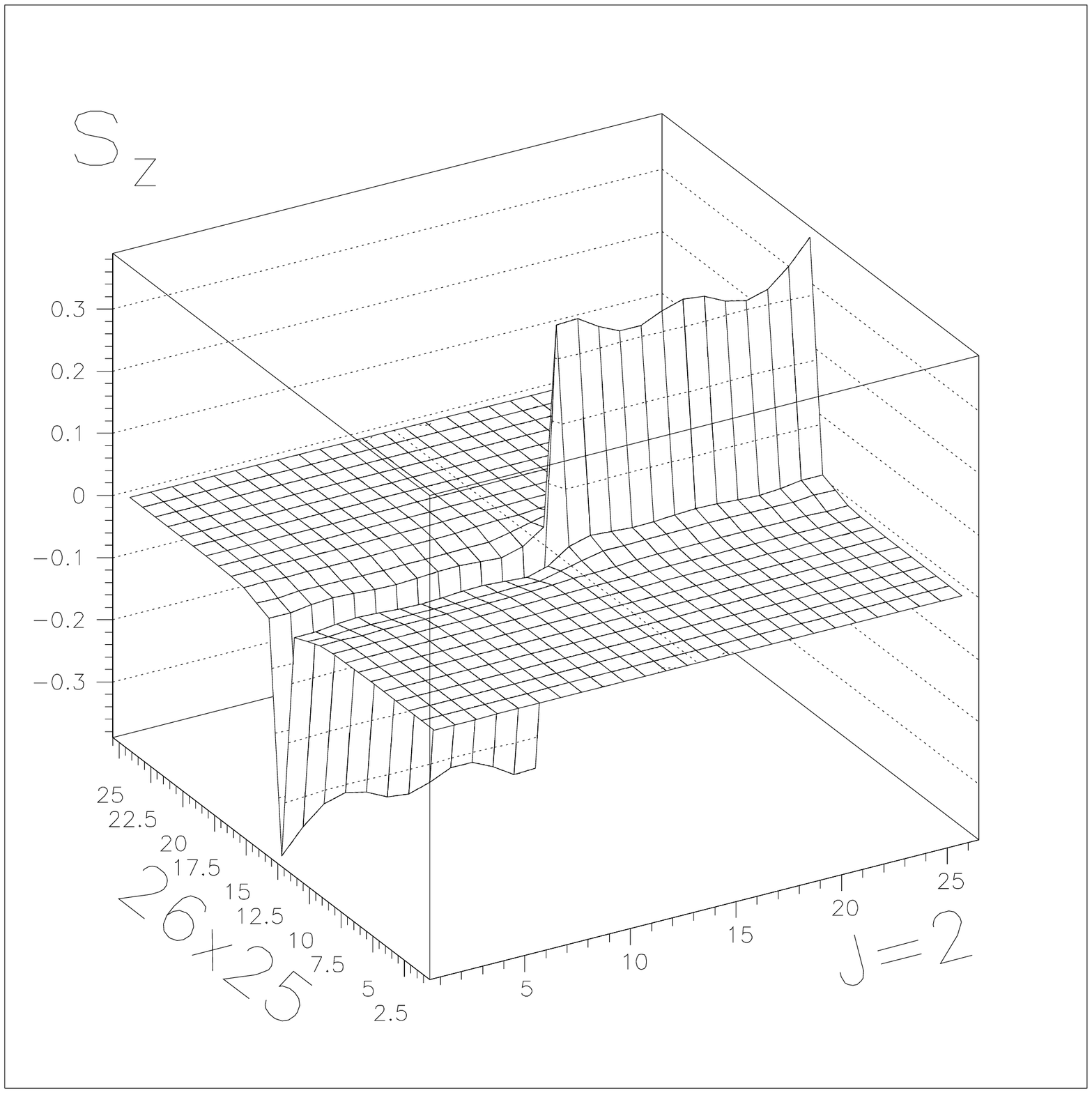}
\caption{\label{fig4}
(a) and (b) Plots of $s_z$ the local $z$ component of the electronic spin density
for the case of two impurities for
$J=1,2$.
(c) and (d) Plot of $s_z$ for $J=1,2$ for DW1. Note that along the
line of spins constituting the domain wall the spin density
is large and positive and that for $J=1$ in the neighbors
it is negative and for the other case $J=2$ it is positive.
As for the cases of one and two impurities this transition
is associated with a transition from $s_z^T=0$ to a $s_z^T\neq 0$.
(e) and (f) Plot of $s_z$ for $J=1,2$ for DW4. Note that now
the total magnetizations are zero but if we divide the system in two
halves similar considerations apply. 
}
\end{figure}

In Fig. \ref{fig4} (a) and (b) we show the behavior of the electron spin density in the presence
of two impurities. Since the impurity spin acts like a local magnetic field we 
expect that the spin density will align along the local field. At the
impurity site the spin density is aligned as shown in the figure. In Fig. \ref{fig4}
we compare two values of the coupling.
For $J=1$ note the negative spin density around the impurity site.
At the impurity site it is positive, as expected. For larger $J$ such as $J=2$ notice
that the spin density in the vicinity of the impurity site is now
positive. We will see that for $J=1$ the total $s_z=0$, while for larger
values of $J$ it is positive (this is one of the hallmarks of the
phase transition). One interpretation is that if $J$ is strong enough
the impurity captures one electron breaking a Cooper pair leaving
the other electron unpaired and the overall electronic
system becomes polarized.
At zero temperature, where the quantum phase transition occurs, the magnetization
reduces to
\bea
s_z(i) &=& \frac{1}{2} \sum_{n,\epsilon_n>0} \left( |v_n(i,\uparrow)|^2 -
|v_n(i,\downarrow)|^2 \right) \nonumber \\
 &=& \frac{1}{2} \sum_{n,\epsilon_n<0} \left( |u_n(i,\uparrow)|^2 -
|u_n(i,\downarrow)|^2 \right)
\eea
and therefore may be calculated either from the hole states at positive energies
(given by $v_n$) or by the particle states at negative energies (given by $u_n$).
At zero temperature there are no quasiparticles and therefore all the states
of positive energies are empty. We will see later the nature of the states
as a function of coupling and their relation to the spin orientation.

In Fig. \ref{fig4} (c)-(f) we show the results for the spin density for two
typical cases, $DW1$ and $DW4$, for the same couplings. For $J=1$ the
first level crossing has not yet occurred and the total magnetization is zero,
as for the single impurity case. This implies that the system responds to the
local polarization of the electronic spin density at the impurity site 
by an overall negative polarization, to yield a total magnetization
that vanishes. Since the effect of the impurities is quite local, it is around the
impurity line that the magnetization is negative. This happens for both
domain walls. Increasing the coupling, for instance $J=2$ the situation changes.
It is clear from the figures that around the impurity line
the magnetization is no longer negative. The influence of the impurity spins is
now strong enough to polarize the electronic system, not only at the impurity
sites but also in their vicinity. The total magnetization no longer vanishes
but has a finite value. This is further discussed ahead.

As the coupling changes, in general the magnetization varies in a continuous way.
However, as the system goes through the various phase transitions the
spin density changes discontinuously consistently with a first order quantum phase transition. 
Since there are now several spins that may bind electrons in succession,
there is now a sequence of phase transitions.
These changes are not in general
uniform along the line of impurities. For instance in the case of the
domain wall $DW1$ at the first transition (which, for the parameters
chosen, occurs near $J=1.55$) there is an increase centered around
the middle point of the chain, for the next transition (around $J=1.62$)
the increase occurs both at the middle point and at both ends of the chain,
at the next transition (around $J=1.8475$) there is a sharp peak at the
middle point and at the transition around $J=2.215$ the increase has a broad
maximum centered around the middle point. This inhomogeneity is characteristic
of the various domain walls where the space distribution of the wave functions
is complex due to the multiple interferences of the quasiparticles off
the various impurities.

\begin{figure}
\includegraphics[width=0.4\textwidth]{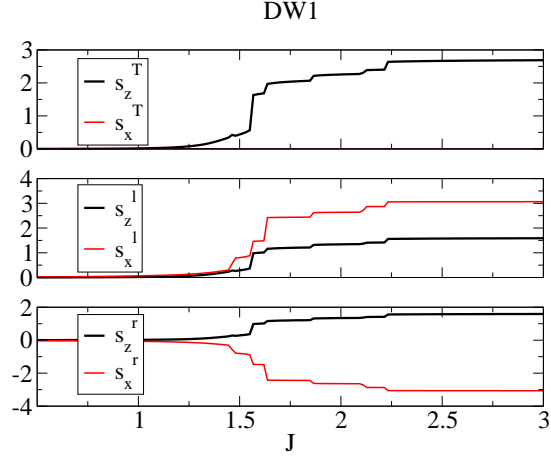}
\caption{\label{fig10}
Various quantities for DW1 as a function of $J$. We plot
$s_z^T$, $s_x^T$, $s_z^l$, $s_x^l$, $s_z^r$ and $s_x^r$. Here $l$ and $r$
stand for left half of the system and right half of the system. Due to the symmetry
of the domain wall chosen $s_x^T=0$. Note however that if we consider the magnetization
of the left or right halves there are interesting phase transitions.
}
\end{figure}

\begin{figure}
\includegraphics[width=0.4\textwidth]{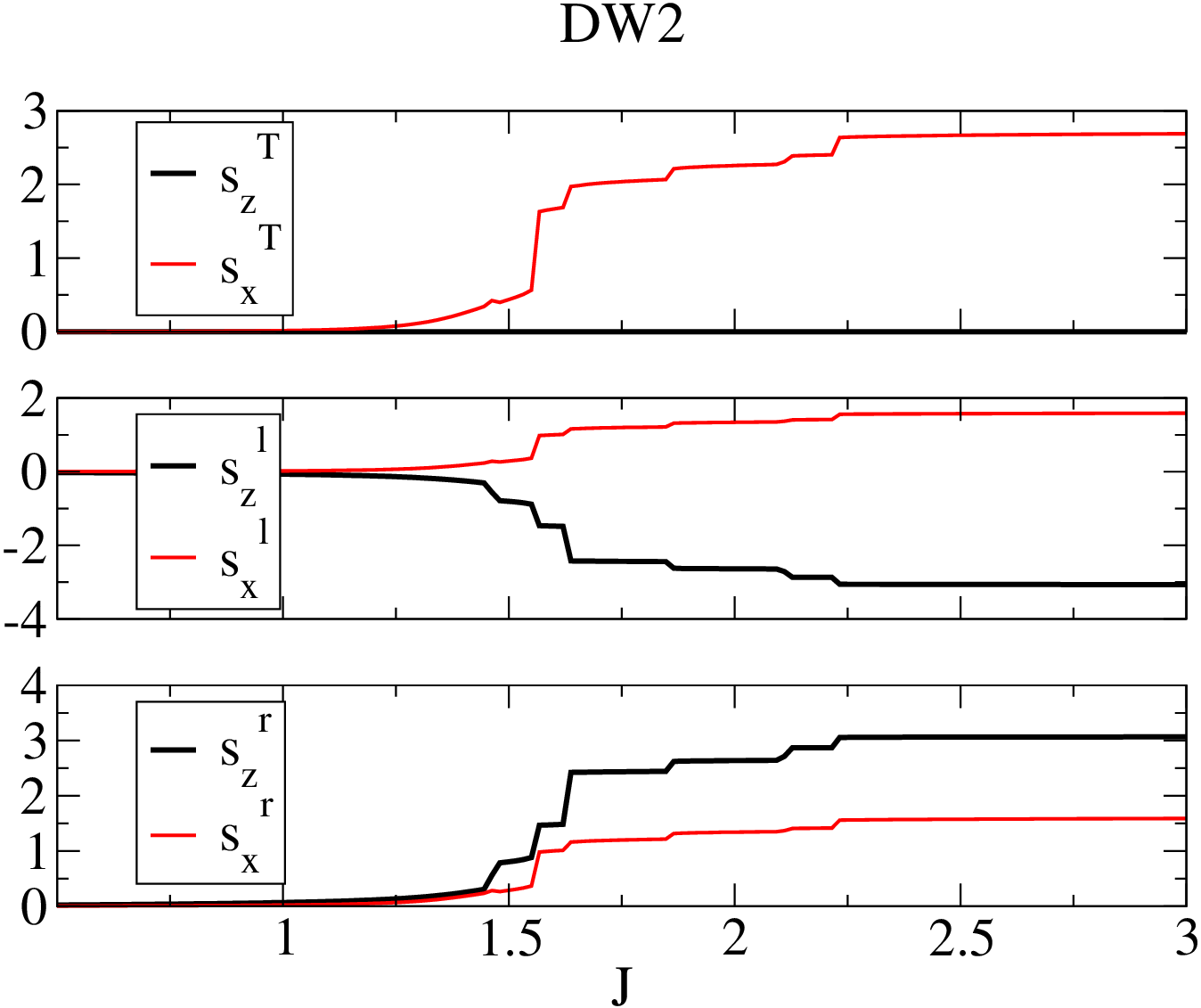}
\caption{\label{fig11}
Various quantities for DW2 as a function of $J$. We plot
$s_z^T$, $s_x^T$, $s_z^l$, $s_x^l$, $s_z^r$ and $s_x^r$.
}
\end{figure}

\begin{figure}
\includegraphics[width=0.4\textwidth]{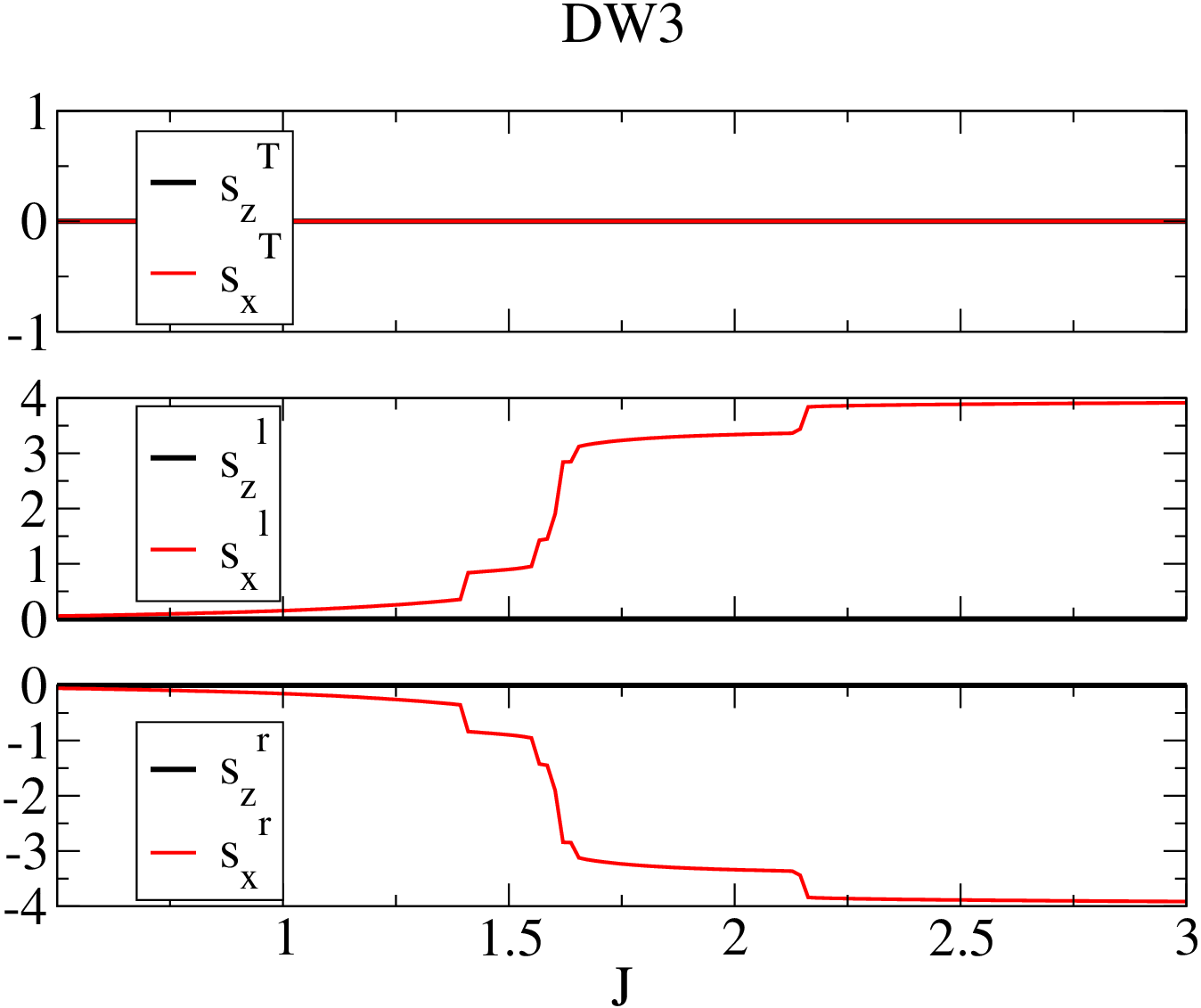}
\caption{\label{fig13}
Various quantities for DW3 as a function of $J$. We plot
$s_z^T$, $s_x^T$, $s_z^l$, $s_x^l$, $s_z^r$ and $s_x^r$.
}
\end{figure}

\begin{figure}
\includegraphics[width=0.4\textwidth]{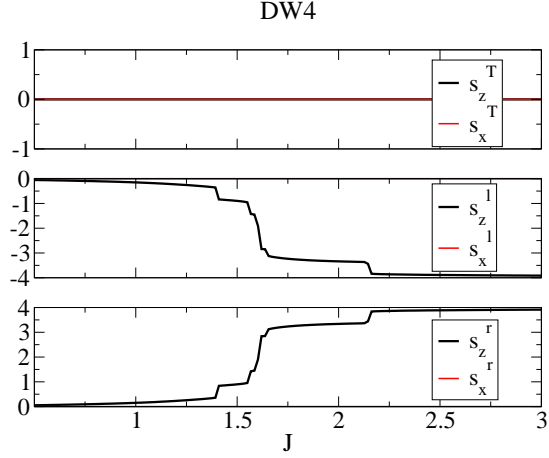}
\caption{\label{fig14}
Various quantities for DW4 as a function of $J$. We plot
$s_z^T$, $s_x^T$, $s_z^l$, $s_x^l$, $s_z^r$ and $s_x^r$.
}
\end{figure}

\begin{figure}
\includegraphics[width=0.4\textwidth]{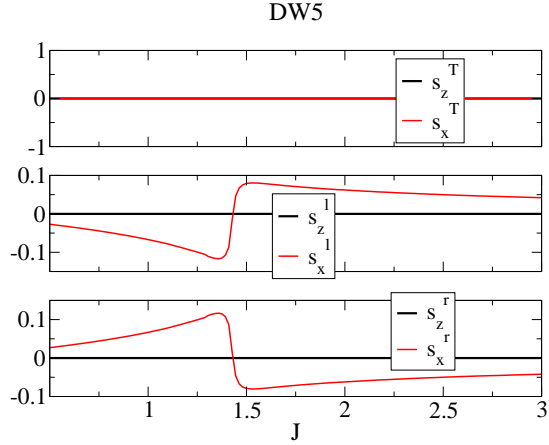}
\caption{\label{fig12}
Various quantities for the AF spin configuration (DW5) as a function of $J$. We plot
$s_z^T$, $s_x^T$, $s_z^l$, $s_x^l$, $s_z^r$ and $s_x^r$.
}
\end{figure}

\subsection{Global spin density}

The sequence of the quantum phase transitions is also clearly displayed
if we consider the total magnetization of the system as a function of
the coupling. In the cases of the domain walls $DW1$ and $DW2$ the total
value of $s_z^T$ and $s_x^T$ changes as a function of the coupling in
basically the same way since one may obtain one case from the other by
a rotation in spin space. However, the other domain walls considered
have no overall magnetization in any direction. We may however consider
only the left or right magnetizations and these will display the same
type of quantum phase transitions, even though the total magnetizations are
in principle zero.

The detail of the dependence on the coupling value is better shown in 
Figs. \ref{fig10},\ref{fig11},\ref{fig12},\ref{fig13},\ref{fig14} 
where we present various average spin densities for the various domain walls.
Due to the shape of the domain walls, in some cases the total spin densities,
summed over all electron sites, vanish. However, the quantum phase transitions
are clearly shown if we take averages over, say, half of the system. Associated
with the various level crossings there are various phase transitions
between plateaus corresponding to an increasing spin density component as the
electron spins bind to the impurity spins.

The antiferromagnetic case is peculiar. There are no true discontinuities but
the left and right $s_x$ do not vanish and show an interesting change of sign 
at a value of $J \sim 1.4$. This is probably a finite size effect that should
disappear as the system size increases.

\subsection{Effect of changing the Fermi level and effect of next-nearest-neighbor
hopping}

We have chosen to work with a chemical potential of $\mu=-1$. The electron
density is therefore not fixed and has to be determined self-consistently.
We do not present here the results for the electron density but will consider them in future
publications, particularly in the context of transport properties
where the density of charge carriers will play an important role.
The results do not depend in a significant way on the value of the
chemical potential. 
The value of $\mu=-2$ has been chosen before since the system in this
case is quarter-filled and the
analysis of the interference pattern for the two impurities case is simplified \cite{morr2},
since along the $x$ axis the Fermi momentum is $\pi/2$.
In the case of this paper $\mu=-1$ means the band-filling is larger but is
of the order of $0.65$ and therefore is still quite far from half-filling.
The change in the chemical potential slightly affects the location of the
quantum phase transitions but the results are qualitatively the same.
However, changing the chemical potential affects the band-filling.
The band-filling is also affected by the spin coupling.
As the coupling grows the band-filling increases since electrons are trapped
by the impurities. 
We note that if we consider a more realistic situation, where we
take into account next-nearest-neighbor hopping, the results are also
qualitatively the same. The effect of this extra term is to change the
location of the quantum phase transition points. Both decreasing the band-filling
and introducing the next-nearest-neighbor hopping decrease the order
parameter and therefore increase in proportion the importance of the coupling
between the spin density and the impurity spin anticipating the appearance of
the quantum phase transition.

We note that considering the case of a more dense impurity spin distribution
will both tend to destroy superconductivity for smaller values of the coupling
but also to increase the band filling approaching the half-filling situation
for moderate values of the coupling. This effect will be considered elsewhere.

\begin{figure}
\includegraphics[width=0.4\textwidth]{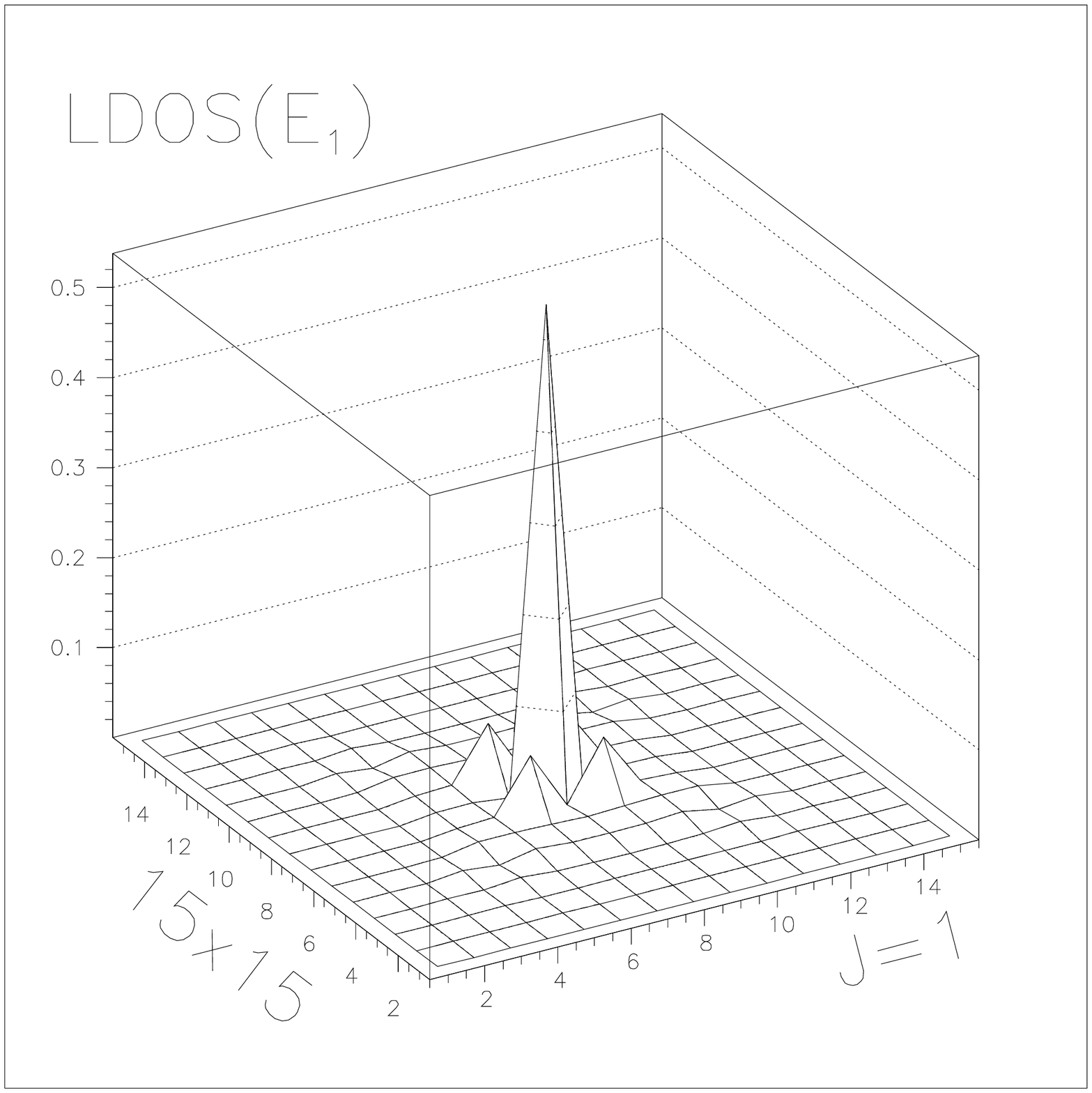}
\includegraphics[width=0.4\textwidth]{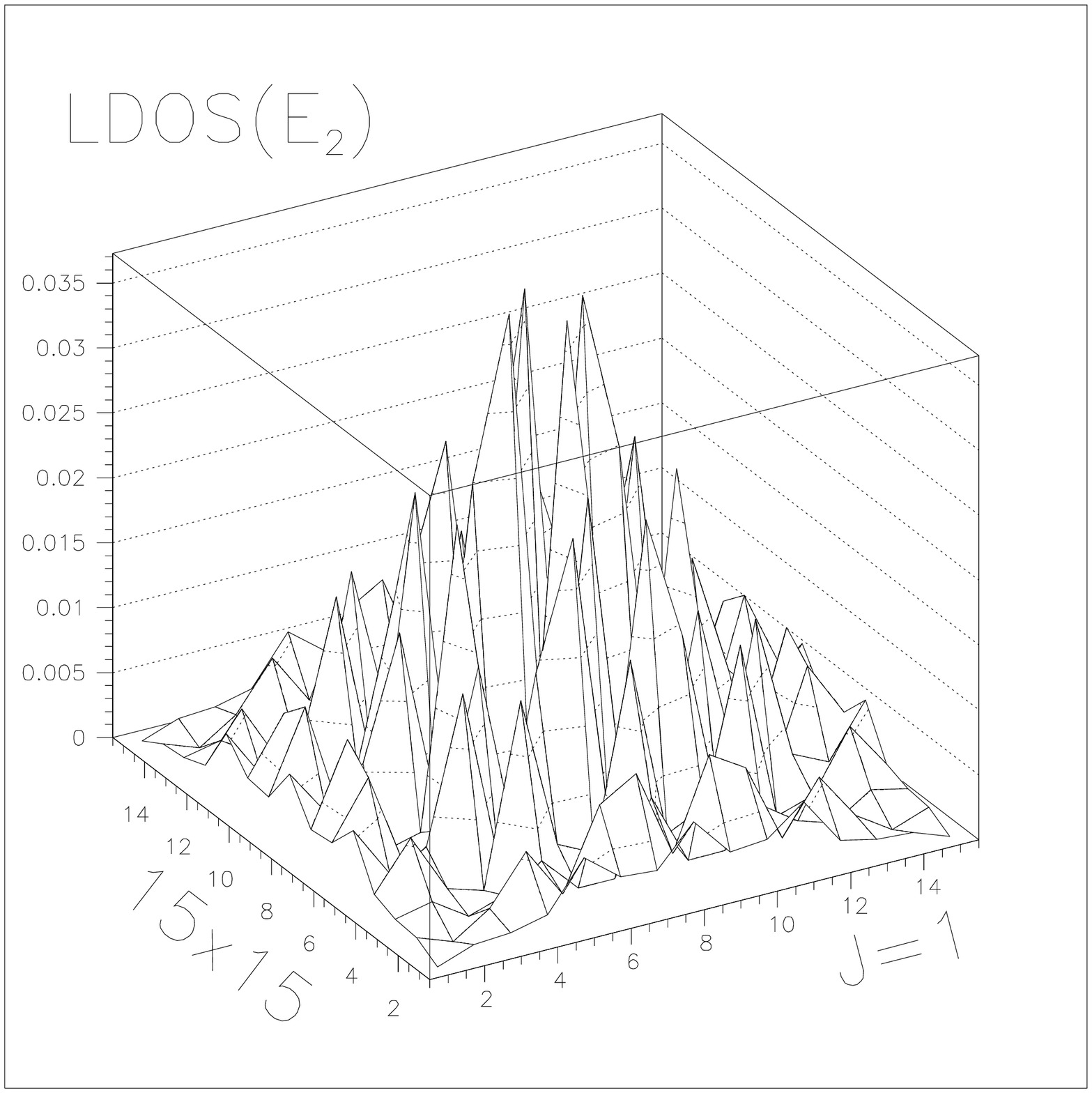}
\caption{\label{fig15}
LDOS for the lowest energy state (with positive energy) and the second
lowest state (also with positive energy) for $J=1$ in the case of a single impurity.
}
\end{figure}

\begin{figure*}
\includegraphics[width=0.4\textwidth]{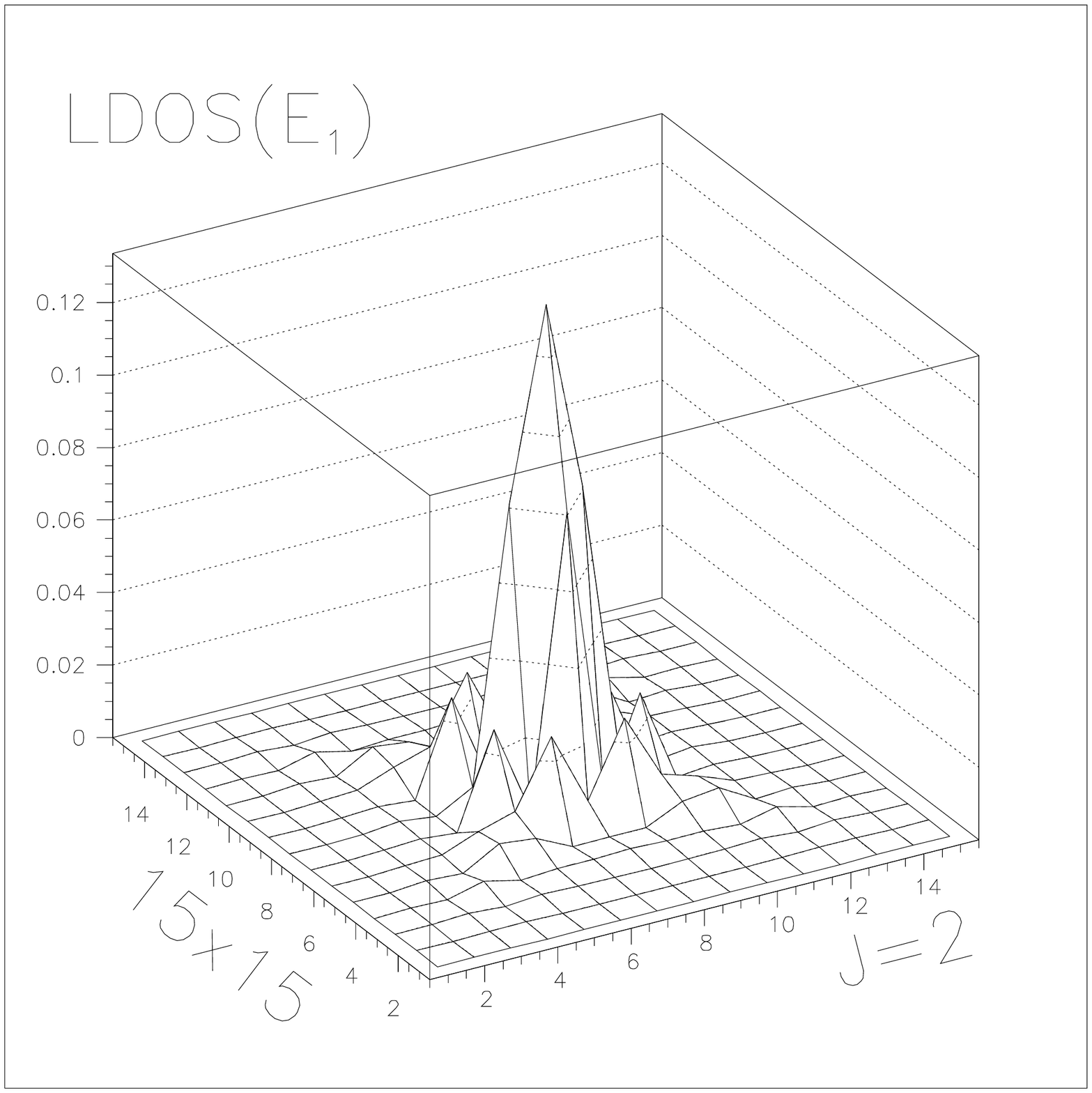}
\includegraphics[width=0.4\textwidth]{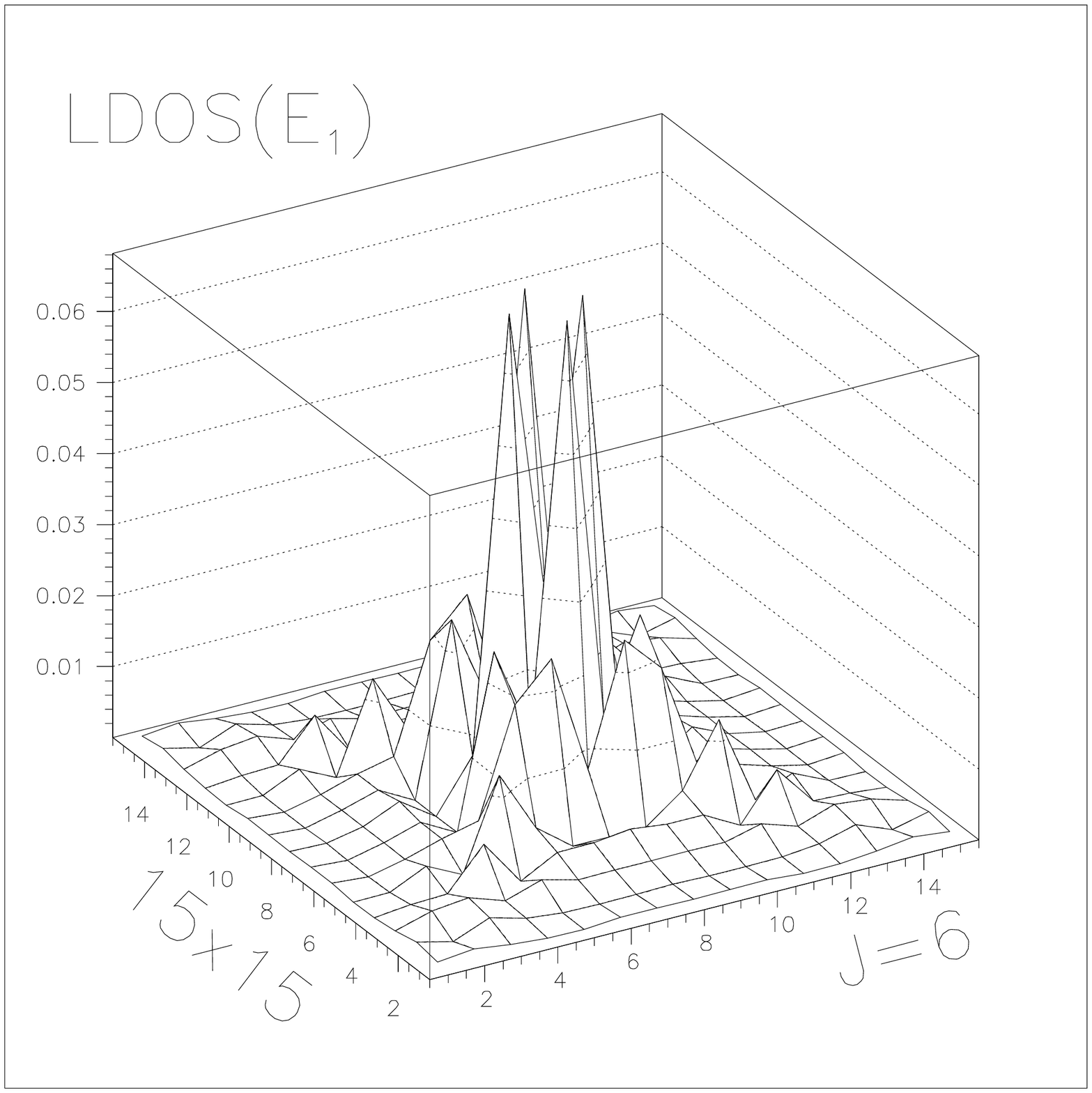}
\caption{\label{fig16}
LDOS for the lowest energy state (with positive energy) 
for $J=2,6$ in the case of a single impurity.
}
\end{figure*}

\section{Nature of states}

\subsection{LDOS: $\rho(\epsilon,i)$}

As mentioned before, since the spectrum is discrete and symmetric,
the density of states is composed by a series of delta function peaks where for
each energy (for instance positive) there are two contributions one
from a level of positive energy, for instance $\epsilon_n$, with weight
$|u_n(i,\sigma)|^2$ and another from the symmetric energy level with energy
$-\epsilon_n$ with weight $|v_{\bar{n}}(i,\sigma)|^2$, where $\bar{n}$ is the
level symmetric to the level $n$. 

Consider first the case of a single impurity.
The true nature of the boundstate of positive energy is better evidenced by the LDOS
at the energy of the lowest level. Specifically, and calling this energy level $n=1$,
the LDOS is given by
\bea
\rho(\epsilon=\epsilon_1,i) &=& \sum_{\sigma} \left( |u_1(i,\sigma)|^2 + |v_{\bar{1}}(i,\sigma)|^2
\right) \nonumber \\
&=& \rho_+(\epsilon_1,i,\uparrow) + \rho_+(\epsilon_1,i,\downarrow)
\eea
This is shown in Fig. \ref{fig15} where
the LDOS of the second level (located in the continuum) is also shown
for comparison.
The LDOS of the second level is
spread over the system. The lowest state is
localized near the impurity site. The LDOS of the
lowest peak is nicely confined around the impurity site and has a large
spectral weight. Here we considered
$J=1$. Notice that the oscillations are strongly damped beyond the nearest-
neighbors.
In Fig. \ref{fig16} we consider higher values of the coupling $J=2,6$.
Notice that for $J=2$
the peak has broadened but is still quite localized.
Also, the oscillations are now less damped particularly along
the diagonals of the square. Note also that the spectral weight at the impurity site is now reduced.
In the case of a large coupling $J=6$ the peak is still localized but has several new features.
It has a broader spectral weight and there is no peak at the central point. At the
impurity site the gap function is still negative even though its magnitude decreases for
this large $J$ value.   
Instead of a central peak there are now four peaks in
the nearest neighbors of the impurity site and a square symmetry, even though the
state is still quite localized. Note the extension of the wavefunctions along
the diagonals of the system. 

\begin{figure*}
\includegraphics[width=0.35\textwidth]{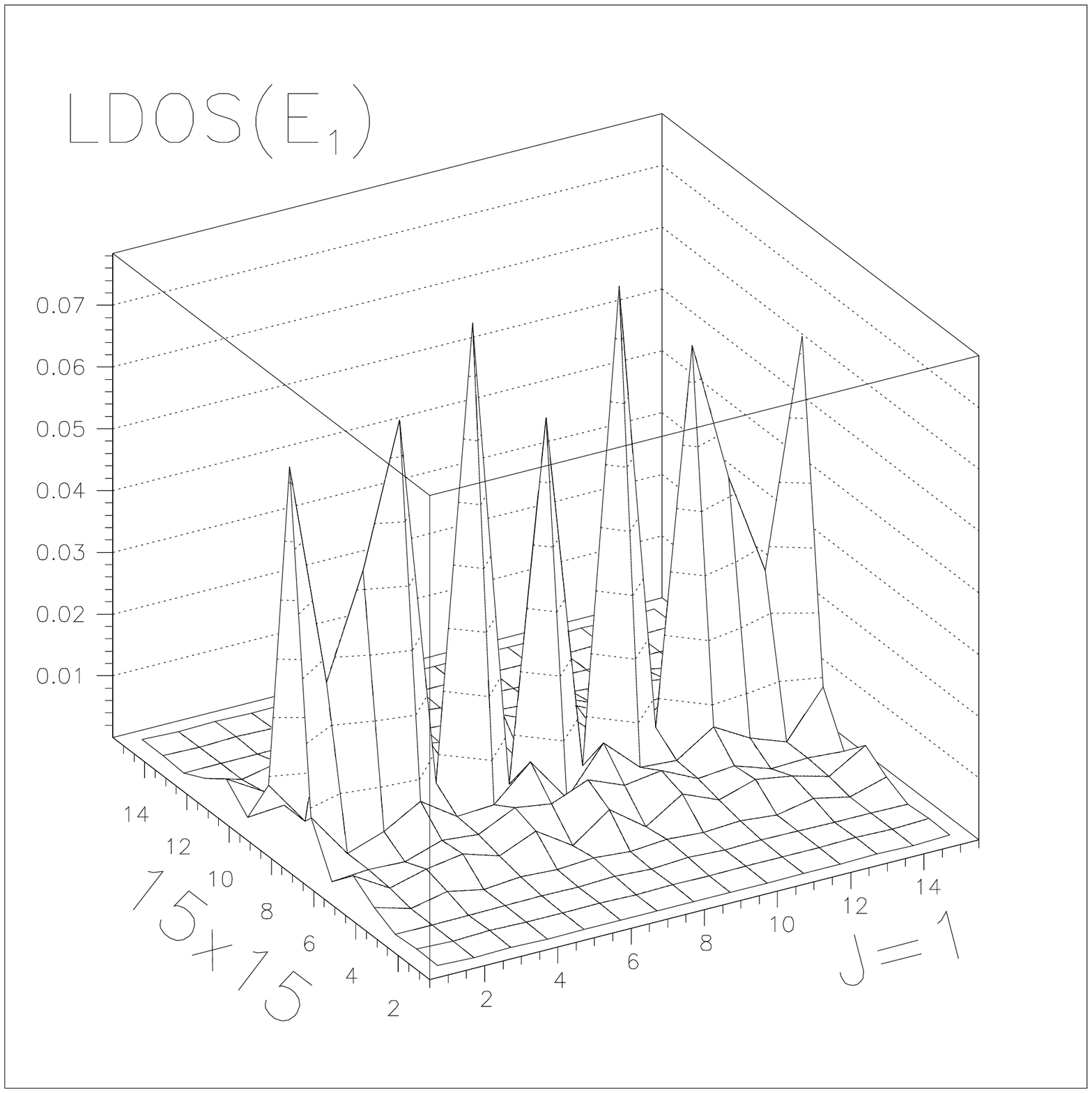}
\includegraphics[width=0.35\textwidth]{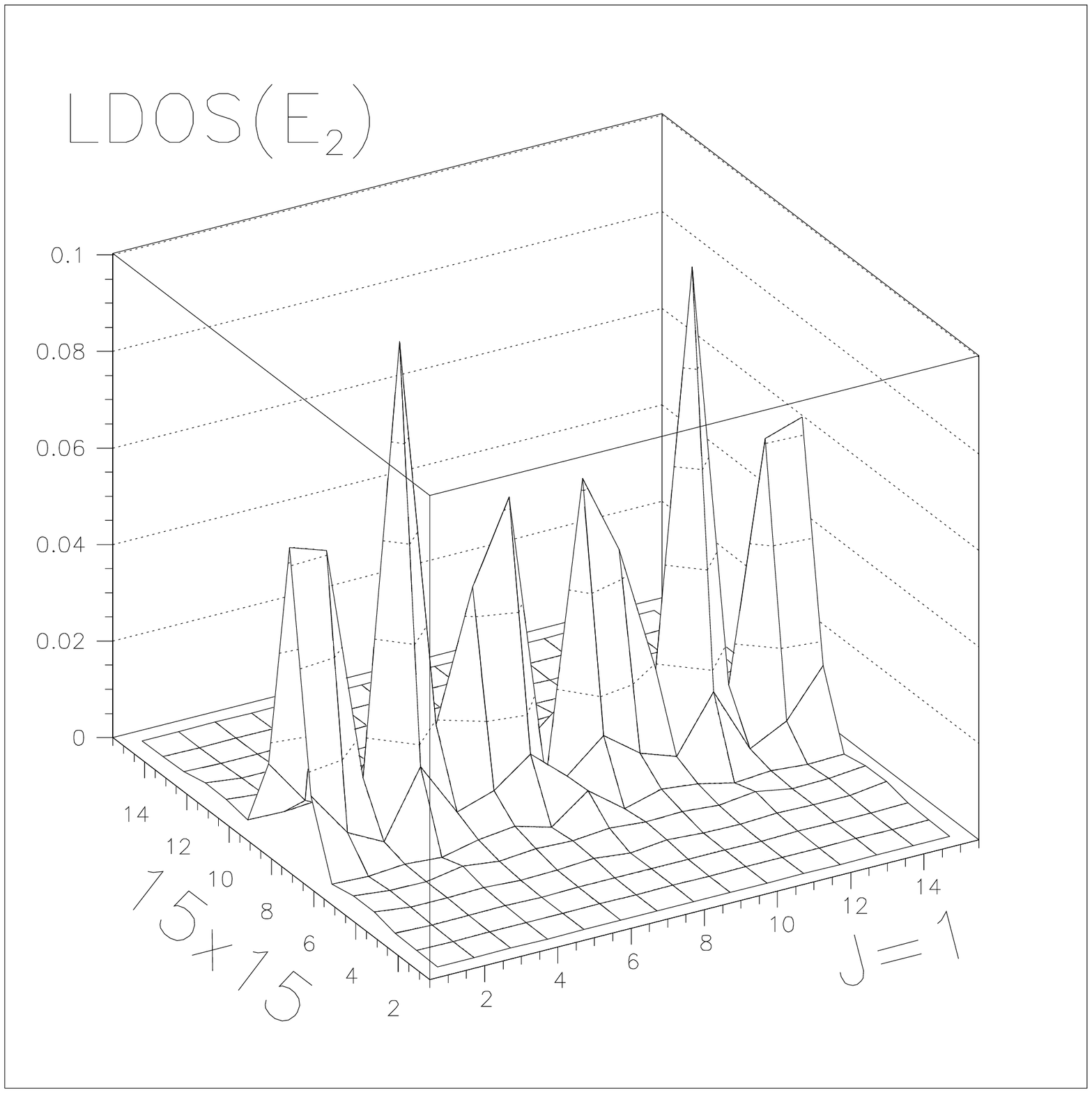}
\includegraphics[width=0.35\textwidth]{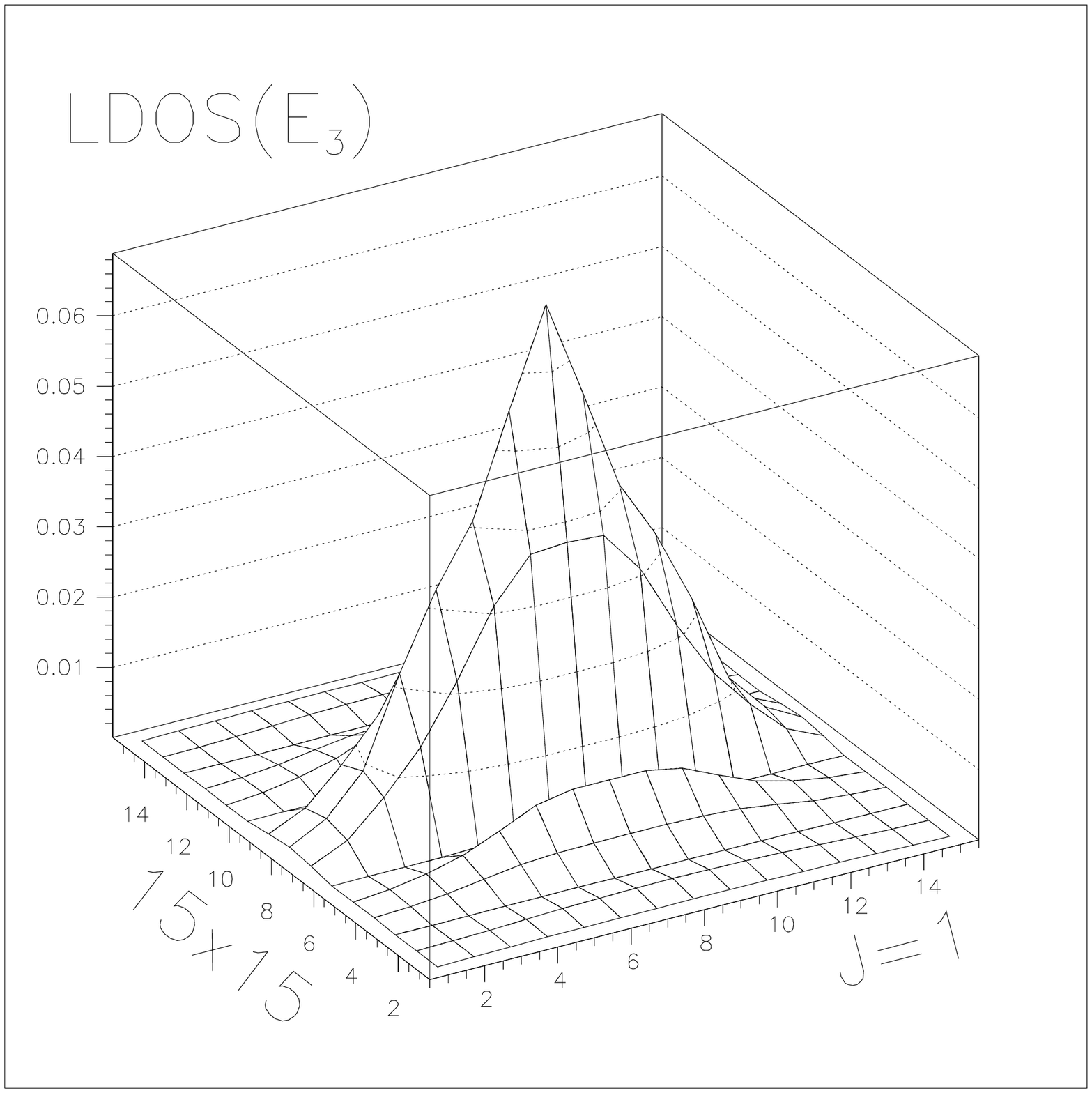}
\includegraphics[width=0.35\textwidth]{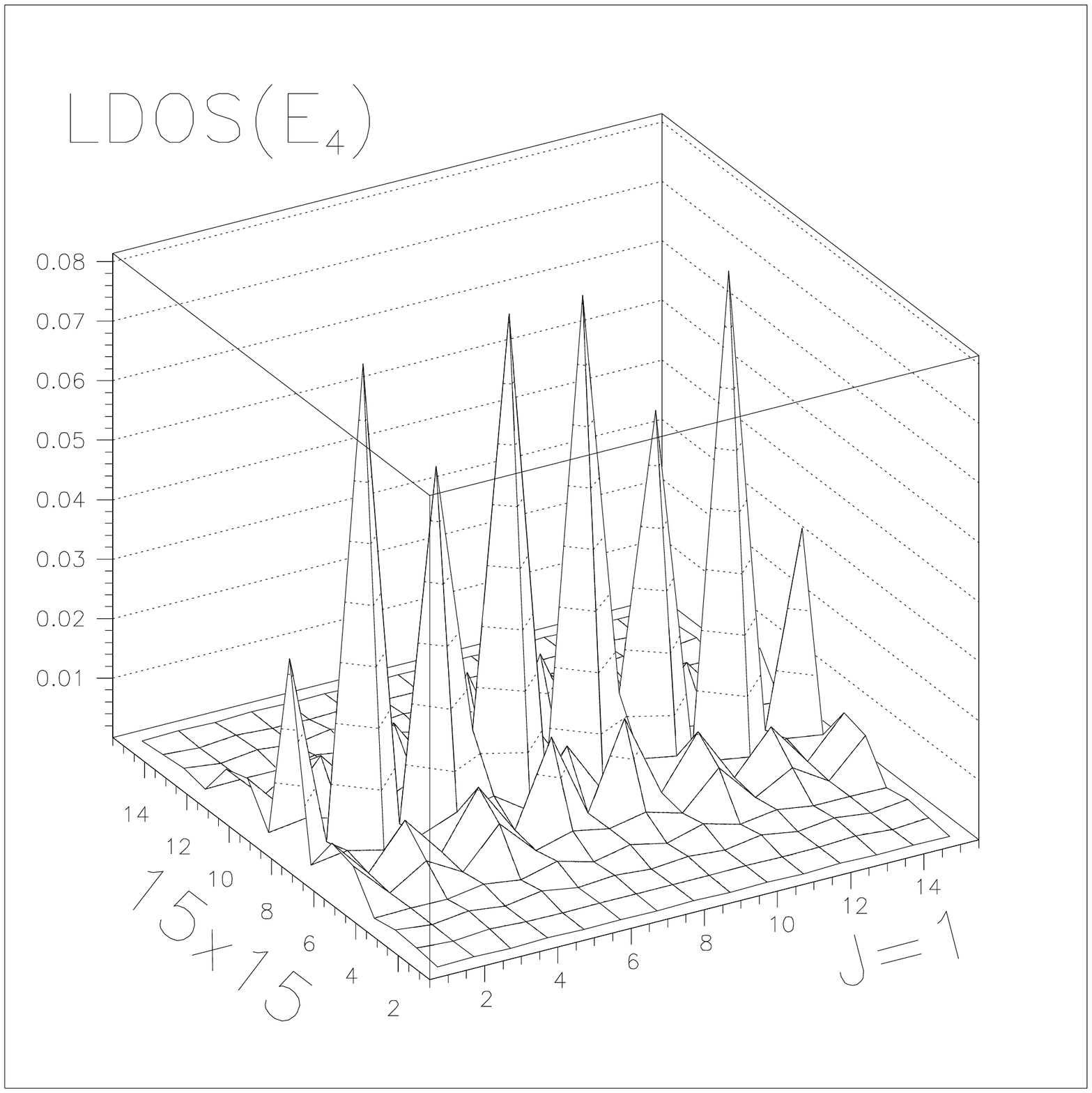}
\includegraphics[width=0.35\textwidth]{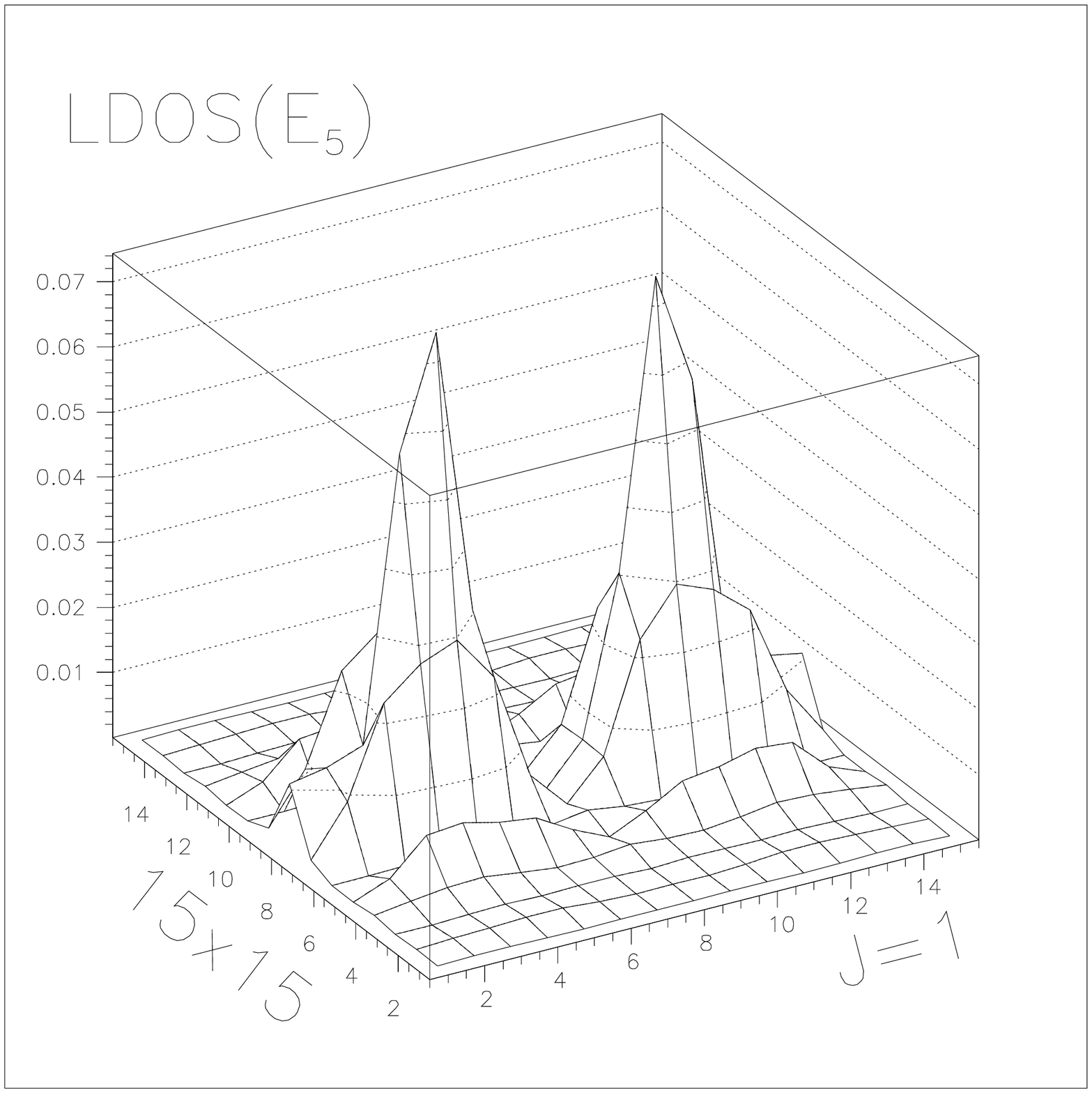}
\includegraphics[width=0.35\textwidth]{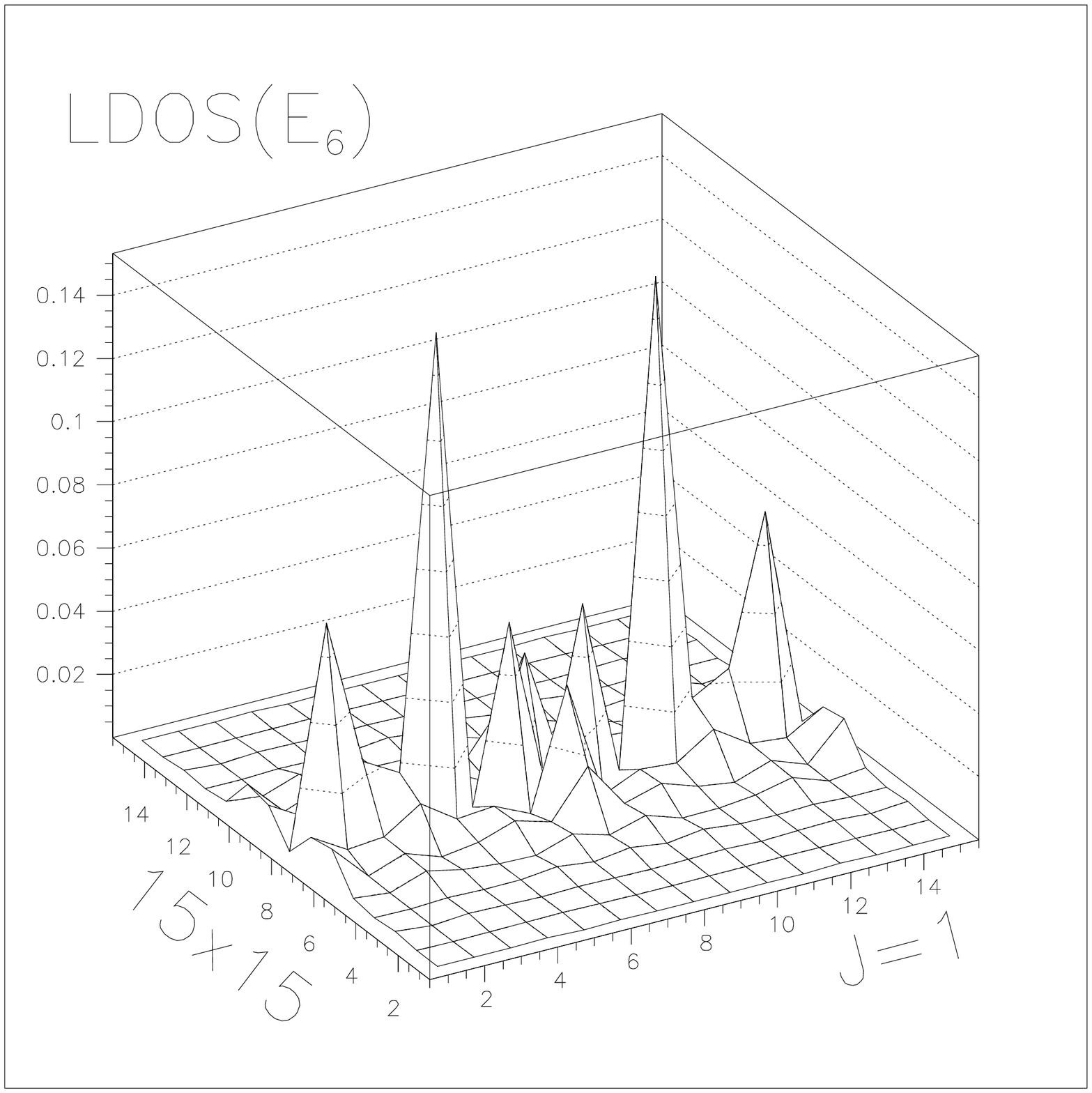}
\includegraphics[width=0.35\textwidth]{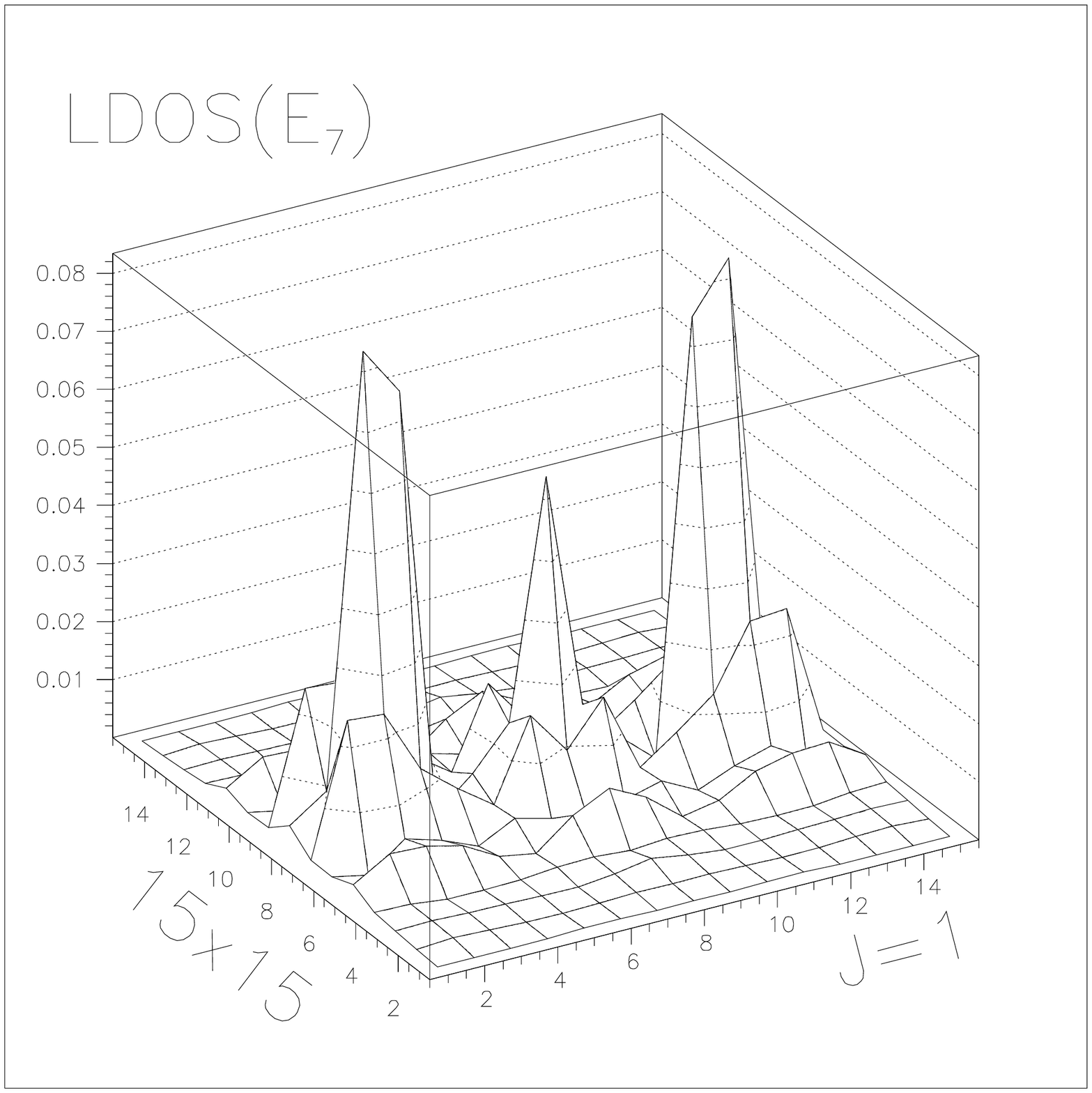}
\includegraphics[width=0.35\textwidth]{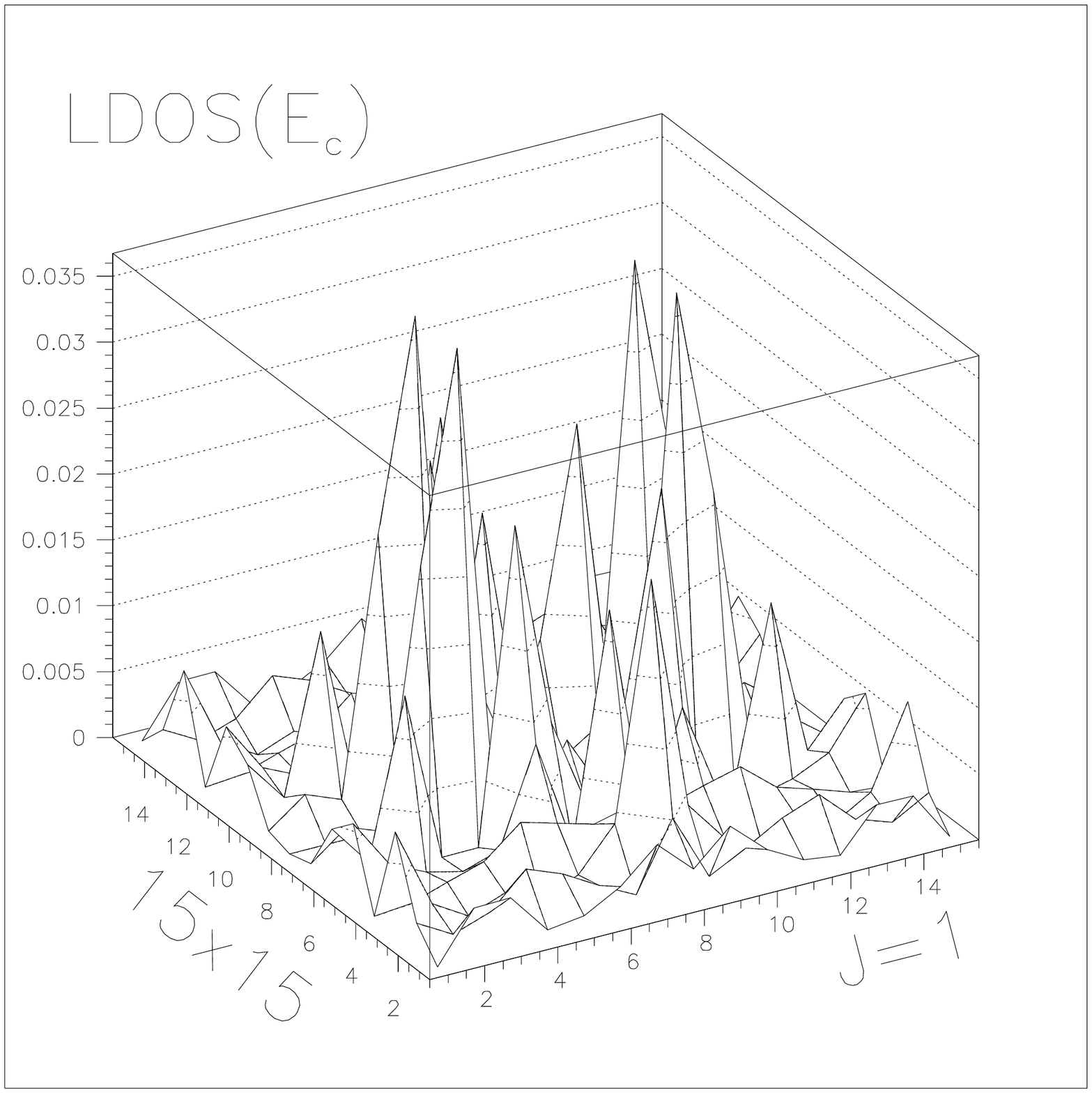}
\caption{\label{fig17}
LDOS for DW1 for the first seven levels for $J=1$ and a state in the continuum.
}
\end{figure*}

\begin{figure*}
\includegraphics[width=0.35\textwidth]{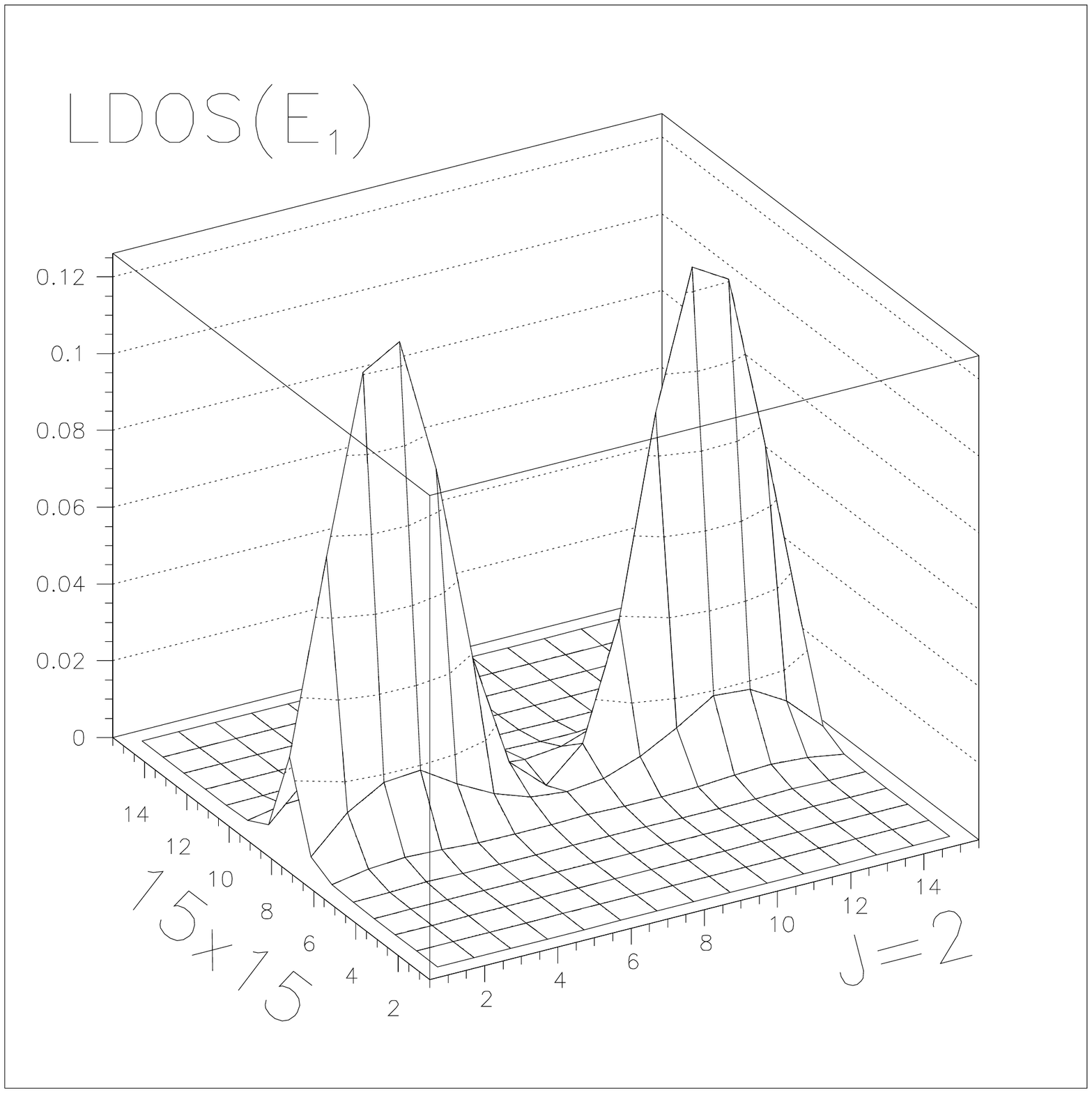}
\includegraphics[width=0.35\textwidth]{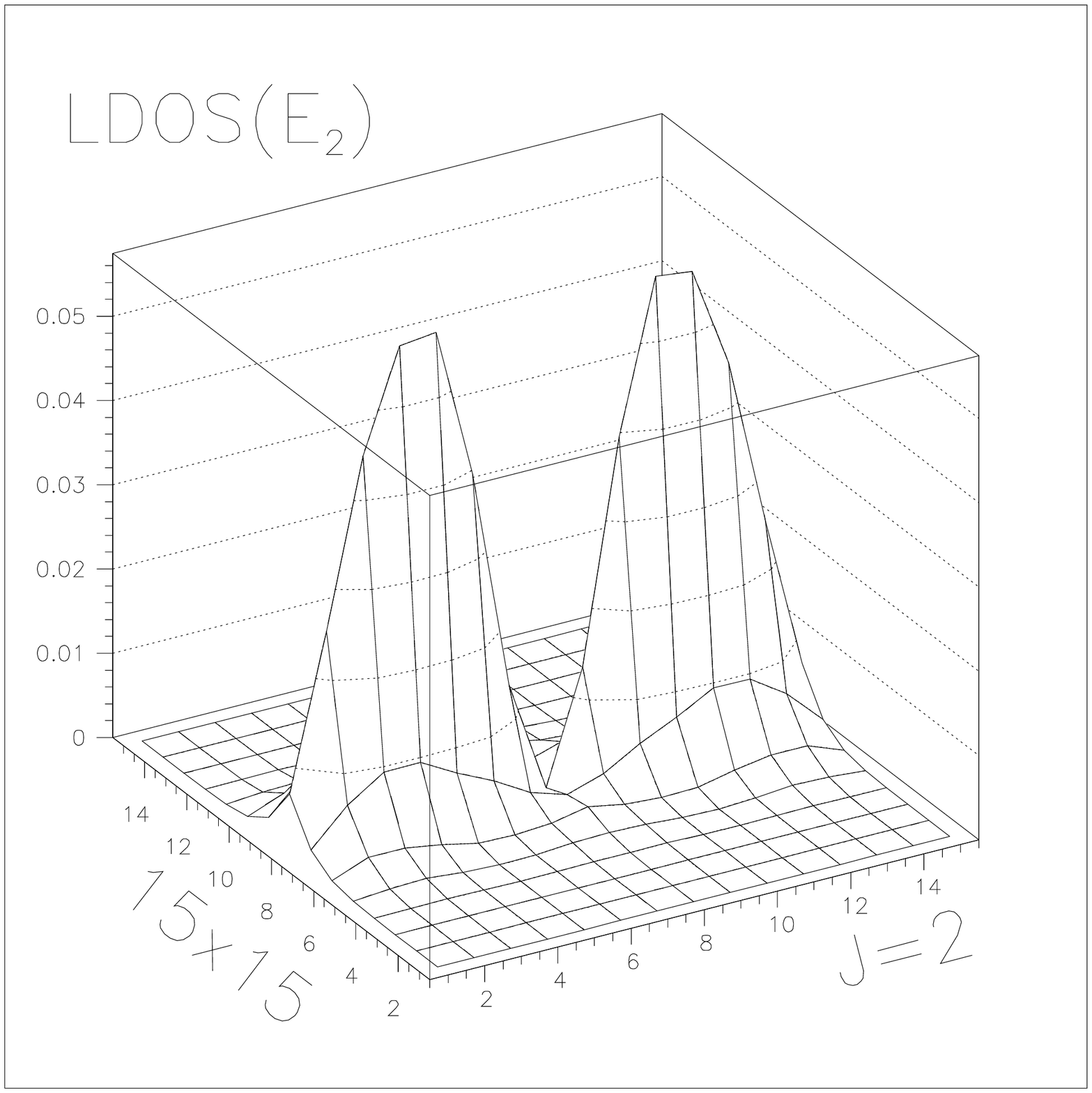}
\includegraphics[width=0.35\textwidth]{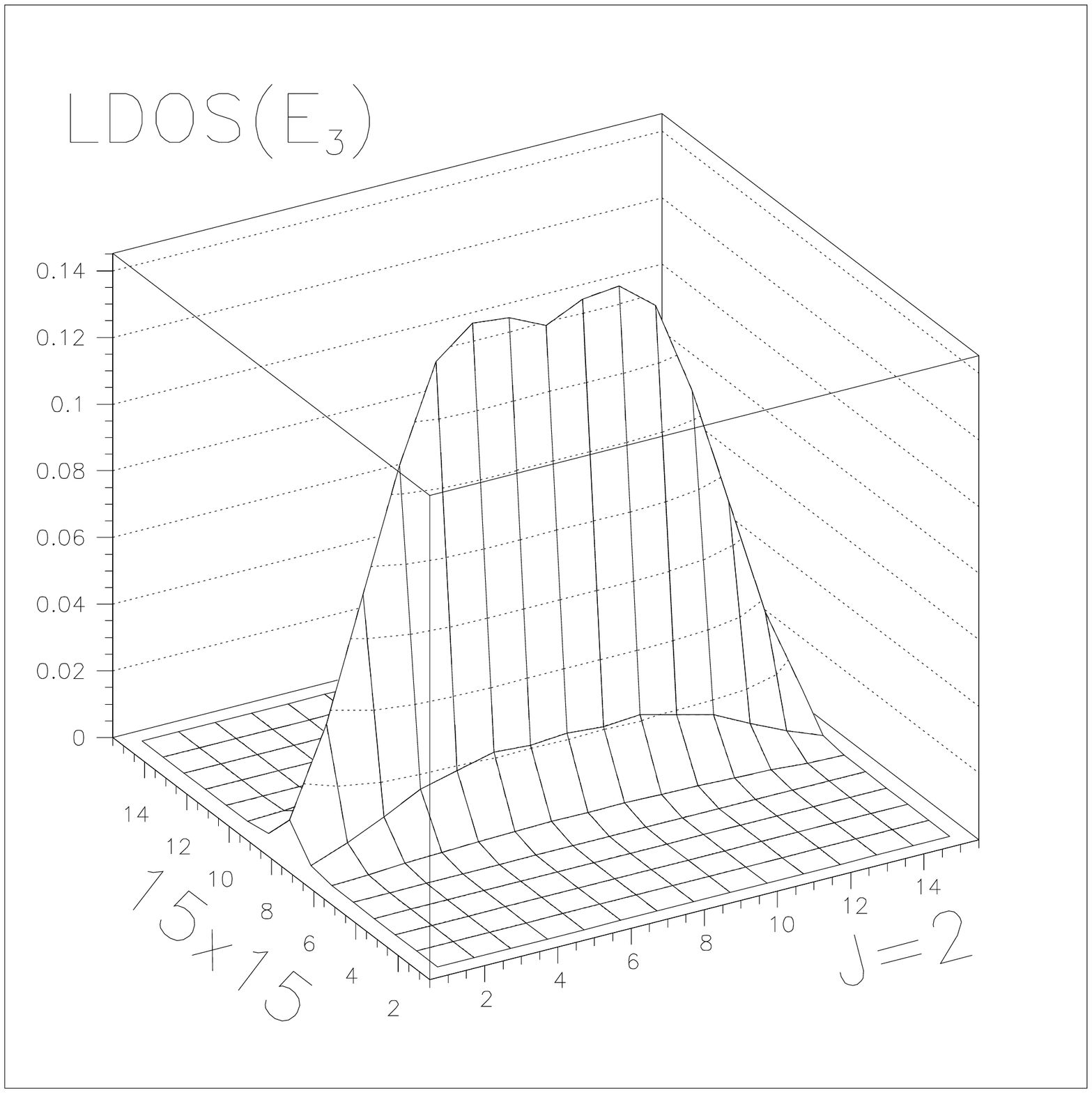}
\includegraphics[width=0.35\textwidth]{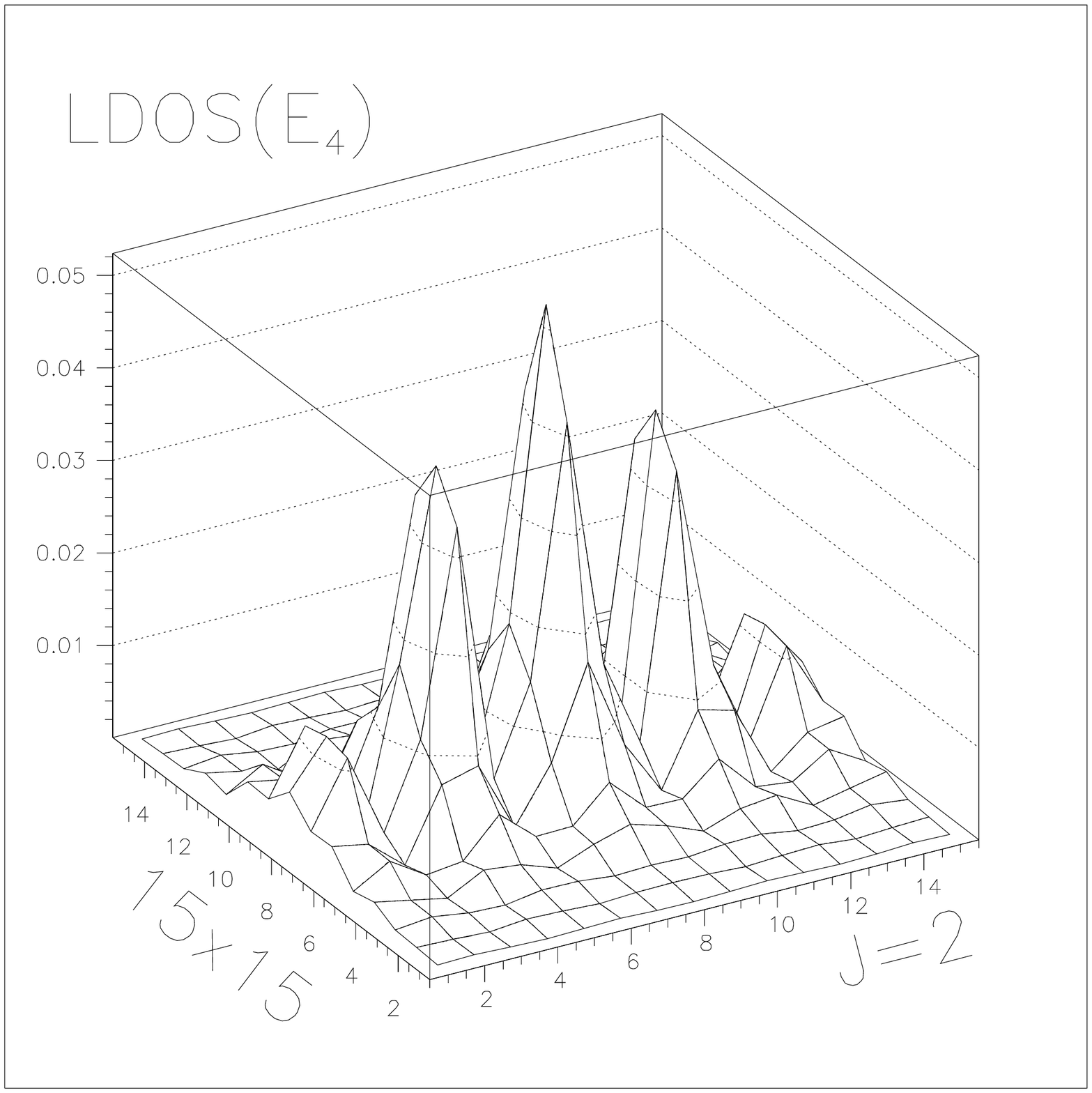}
\includegraphics[width=0.35\textwidth]{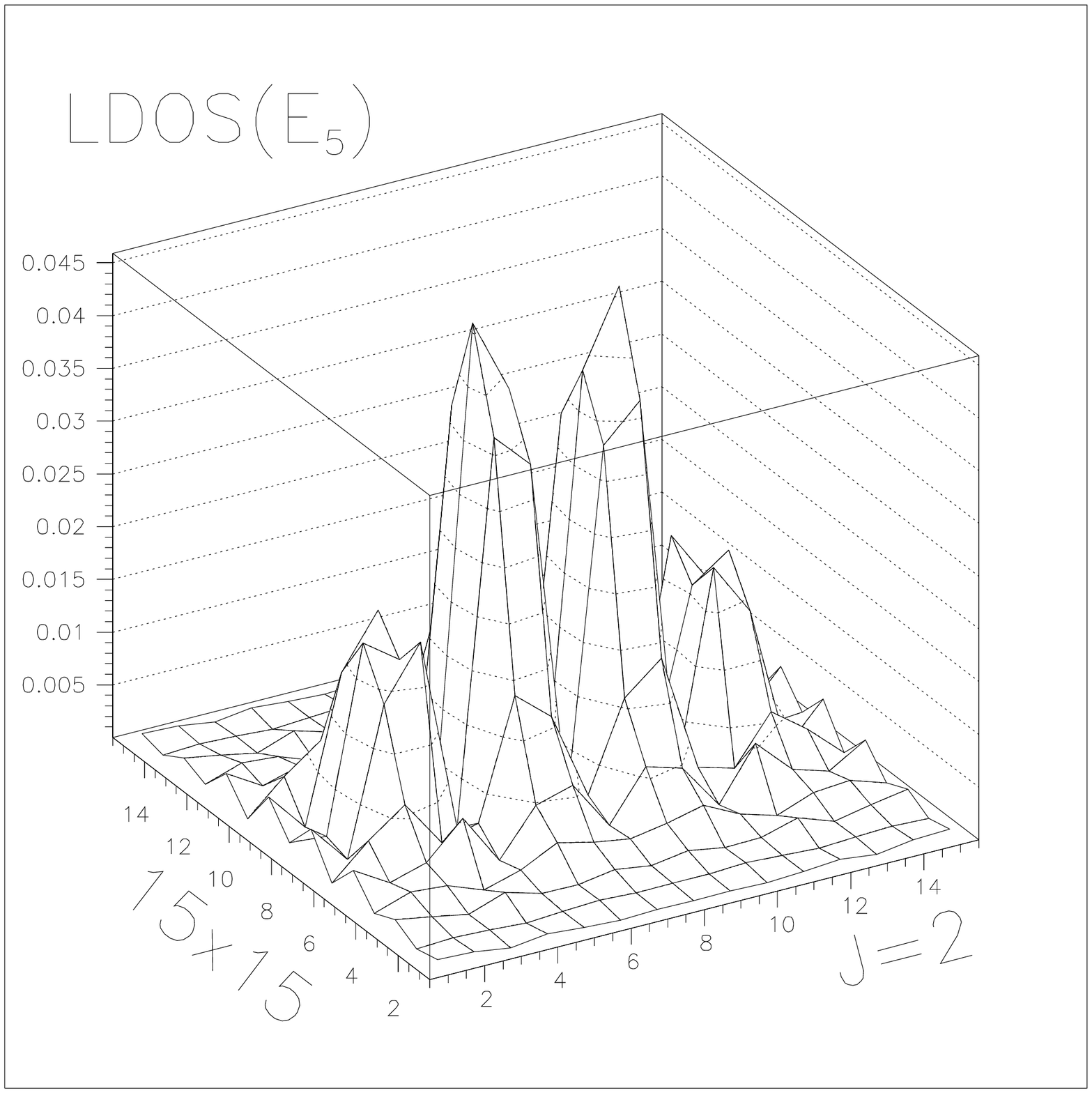}
\includegraphics[width=0.35\textwidth]{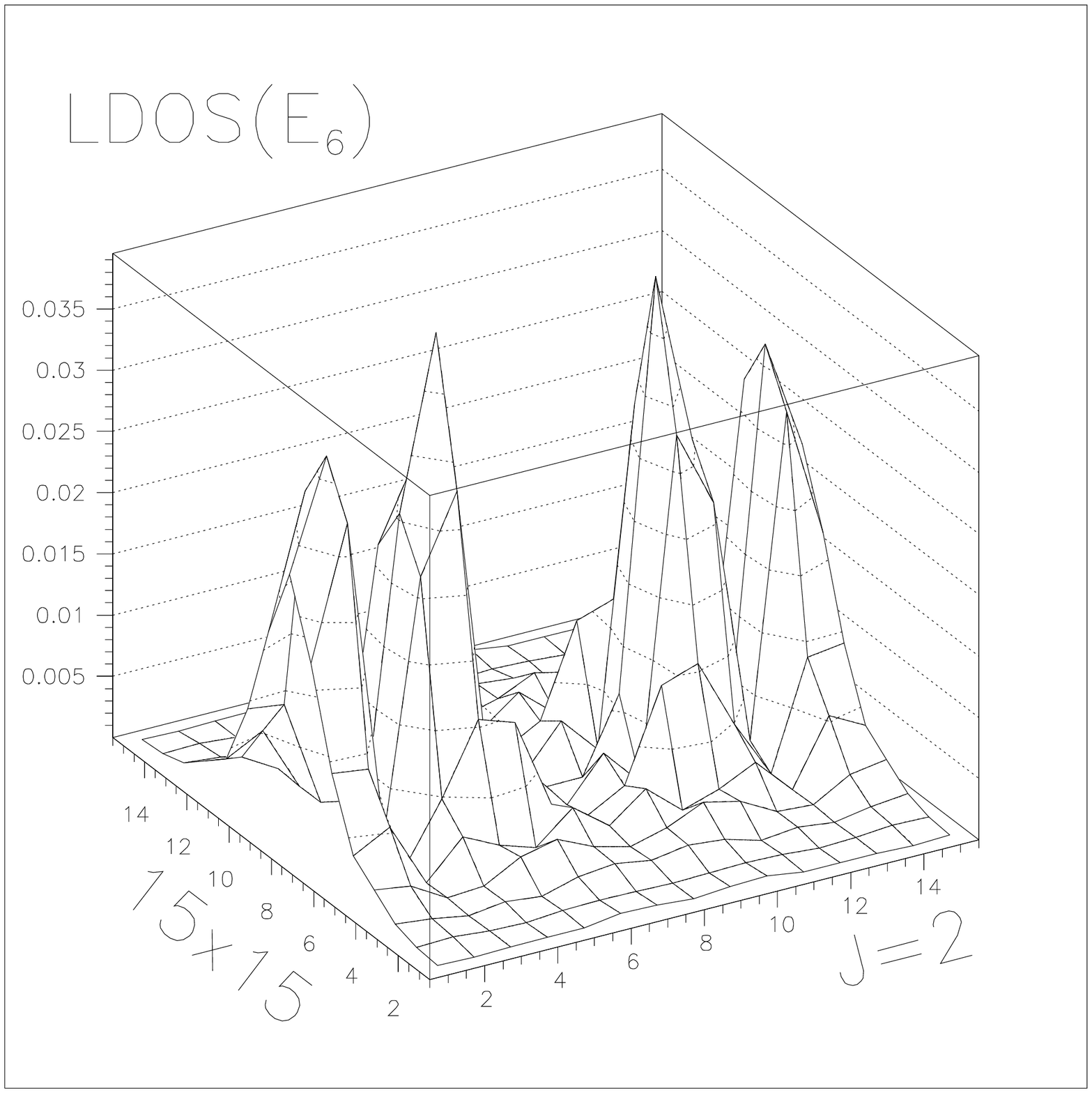}
\includegraphics[width=0.35\textwidth]{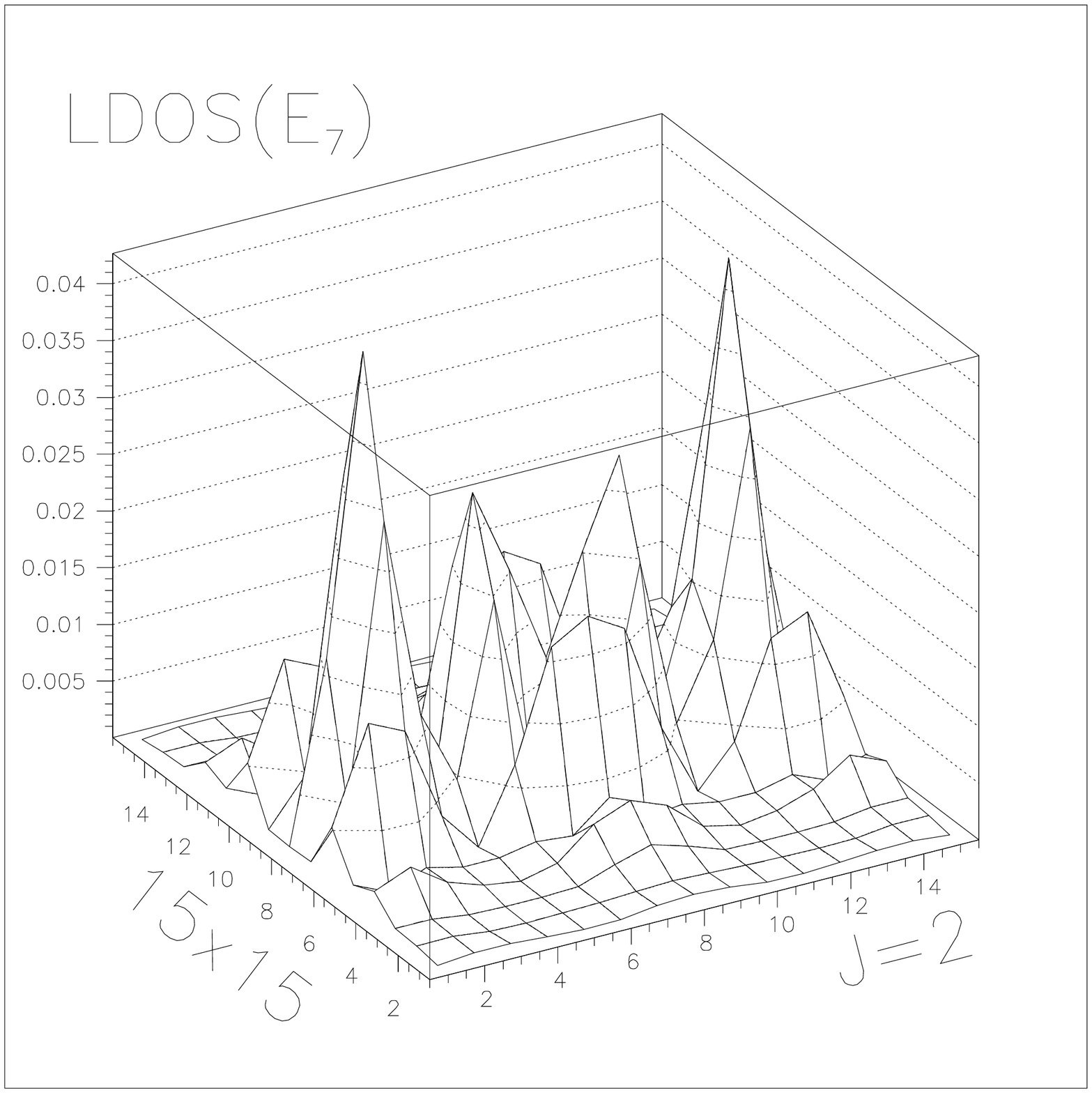}
\includegraphics[width=0.35\textwidth]{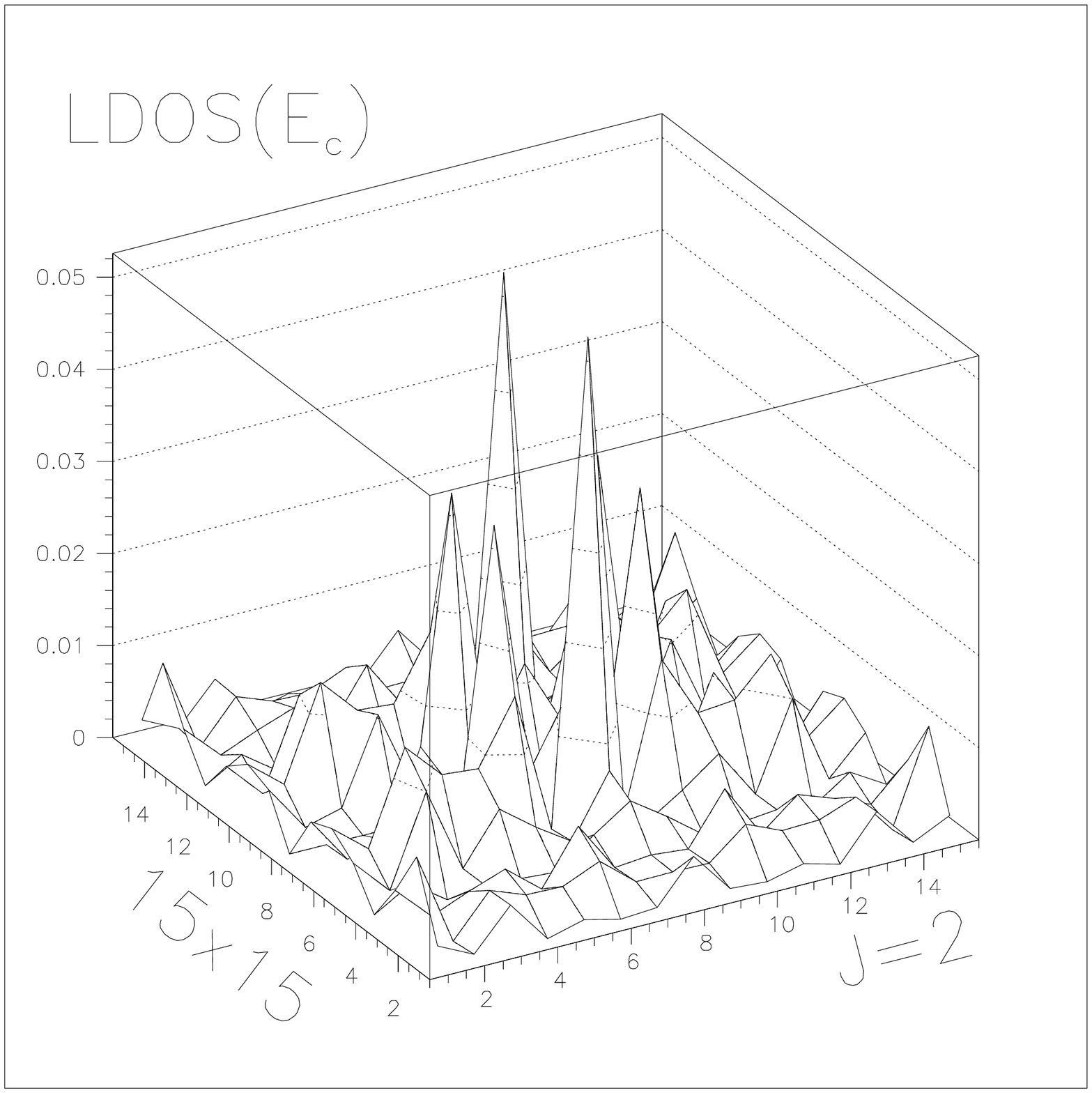}
\caption{\label{fig18}
LDOS for DW1 for the first seven levels for $J=2$ and a state in the continuum.
}
\end{figure*}

If we consider the case of two impurities 
there are now two boundstates. The behavior is quite similar to the
case of a single impurity with two maxima localized in the vicinity of the
impurity sites. The third level and beyond are extended, as expected.

In the cases of a few impurities, like two or three impurities,
a study of the relation between the phase transitions and the
level structure is possible \cite{morr2}. However, increasing
the number of impurity spins this analysis is complicated by the
multiple interferences of the wave functions of the boundstates.
We may however look at the LDOS for the lowest levels and examine
their structure. This is shown in Figs. \ref{fig17},\ref{fig18} for the domain wall $DW3$
for the couplings $J=1,2$. 

The case of $J=1$ is similar to the case of a small coupling,
where no phase transition has occurred, and the local magnetic field is a small
perturbation. Clearly as the coupling increases the shielding of the perturbation
in the vicinity of the impurity site increases but no qualitative change is
observed until the first level crossing. There are now as
many boundstates as impurities and therefore there are several
"localized" states. In Fig. \ref{fig17} we show the LDOS for the 
first seven positive energy levels and for a level in the continuum.
The states in the gap have a structure along the line of impurities and, in that
sense, they are localized. However, since there are several impurities 
there is a series of maxima and minima at the impurity sites. In the single
impurity or two impurities cases the boundstates have maxima at the impurity
locations but we see from Fig. \ref{fig17} that this is not so. For instance,
the lowest level wavefunction has a series of peaks at alternating sites
symmetrically around the central point (note that the number of sites is odd).
On the other hand the second level has a zero at the central point indicative of
an anti-symmetric state. This is reminiscent of the results obtained for two
impurities \cite{morr2} where there is naturally two states, one symmetric and
one anti-symmetric. On the other hand the third peak is quite spread along the domain
wall, and extends further along the perpendicular direction.
As the energy increases the states are still fairly localized with more or less complex
structures. As before, the state in the continuum is of a different nature and spread
through the system, even though not homogeneously. The spectral weights are similar
even though their magnitudes are correlated inversely with their extension, as expected.

Increasing the coupling to $J=2$ changes somewhat the wave-functions. The first levels
are still localized but with different spatial distributions. For instance
the two lowest levels display two broad peaks and deep minima at the
central position of the domain wall. The third level is somewhat similar
to the case of $J=1$. Also, we see an alternancy of symmetric and anti-symmetric
states as the energy increases.

We should note that the results obtained for the $15 \times 15$ system are not general.
For instance, considering the case of a system $25 \times 25$ the details of
the LDOS are different. In particular, due to the increased number of boundstates
the sequence of states is more complex. Also, the increased number of states
in the gap increases the near degeneracies of the states and mixes their
symmetry properties. As discussed above, as the spin coupling grows the impurities
capture electrons. In the case of the extended spin configurations, since the states
are extended along the chain of spins, even though the lowest states
are localized in the perpendicular direction, it is not possible
to disentangle the capture of the electrons by each impurity since
the states are a superposition over many sites.

\begin{figure*}
\includegraphics[width=0.33\textwidth]{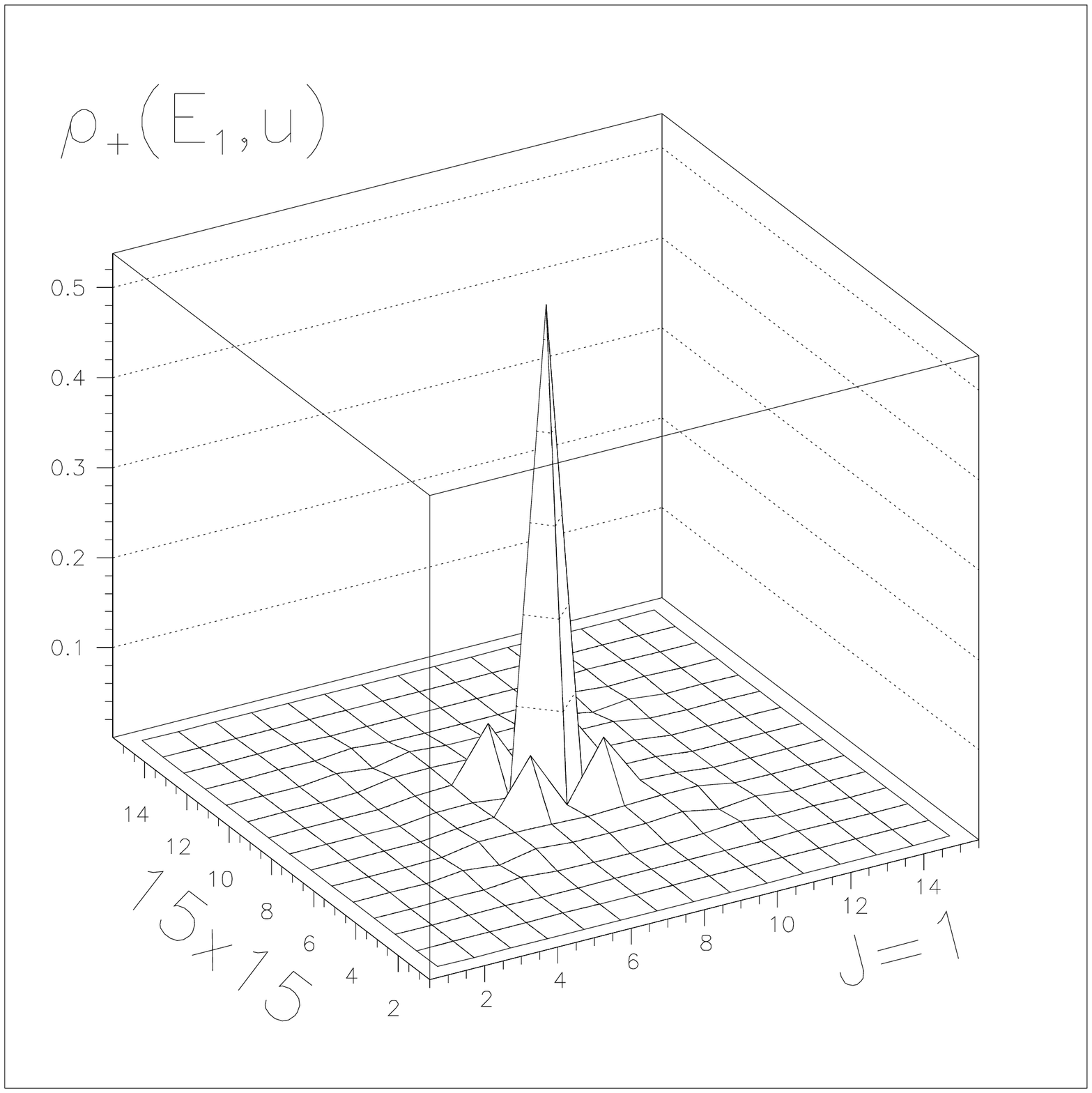}
\includegraphics[width=0.33\textwidth]{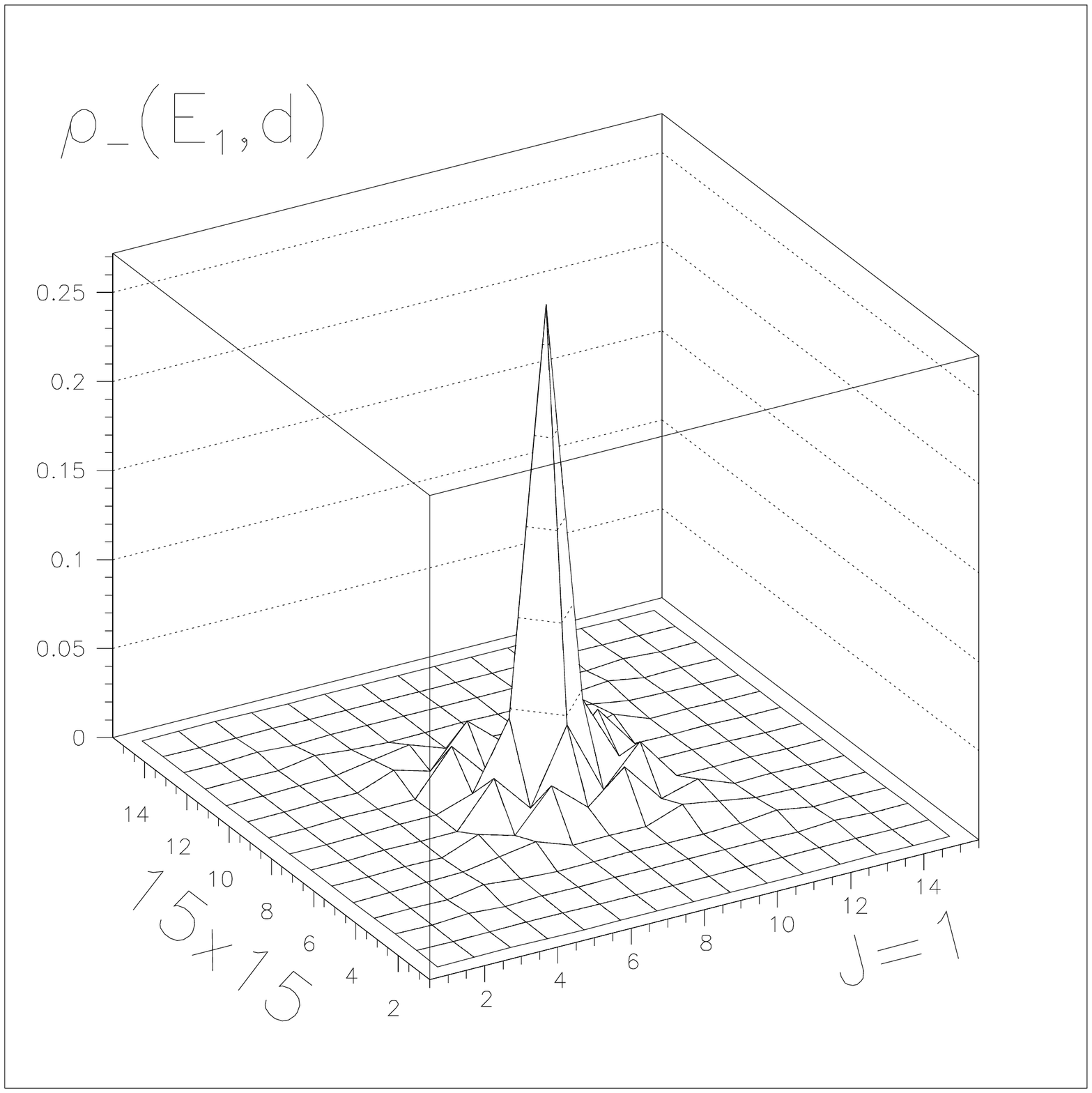}
\includegraphics[width=0.33\textwidth]{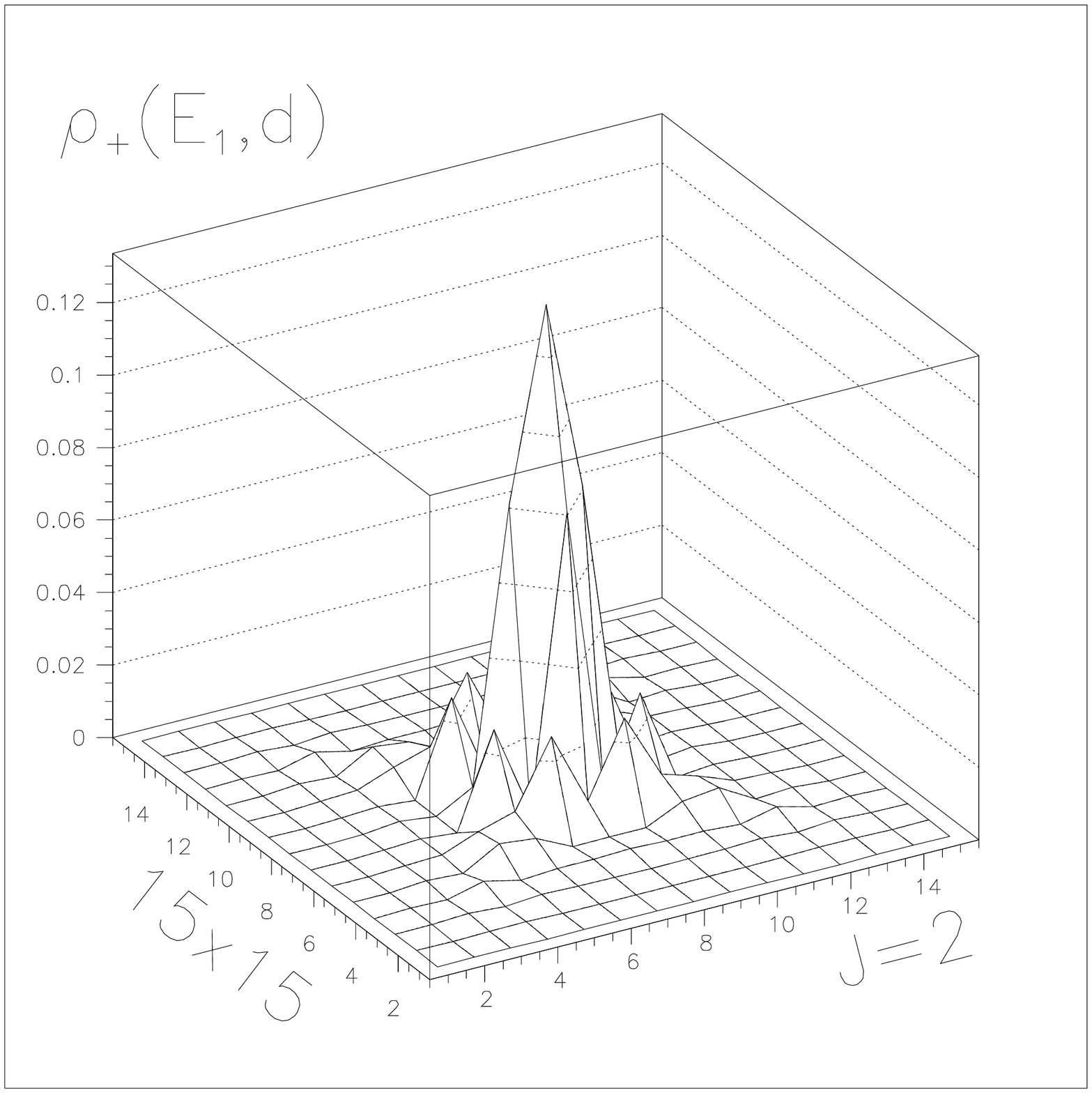}
\includegraphics[width=0.33\textwidth]{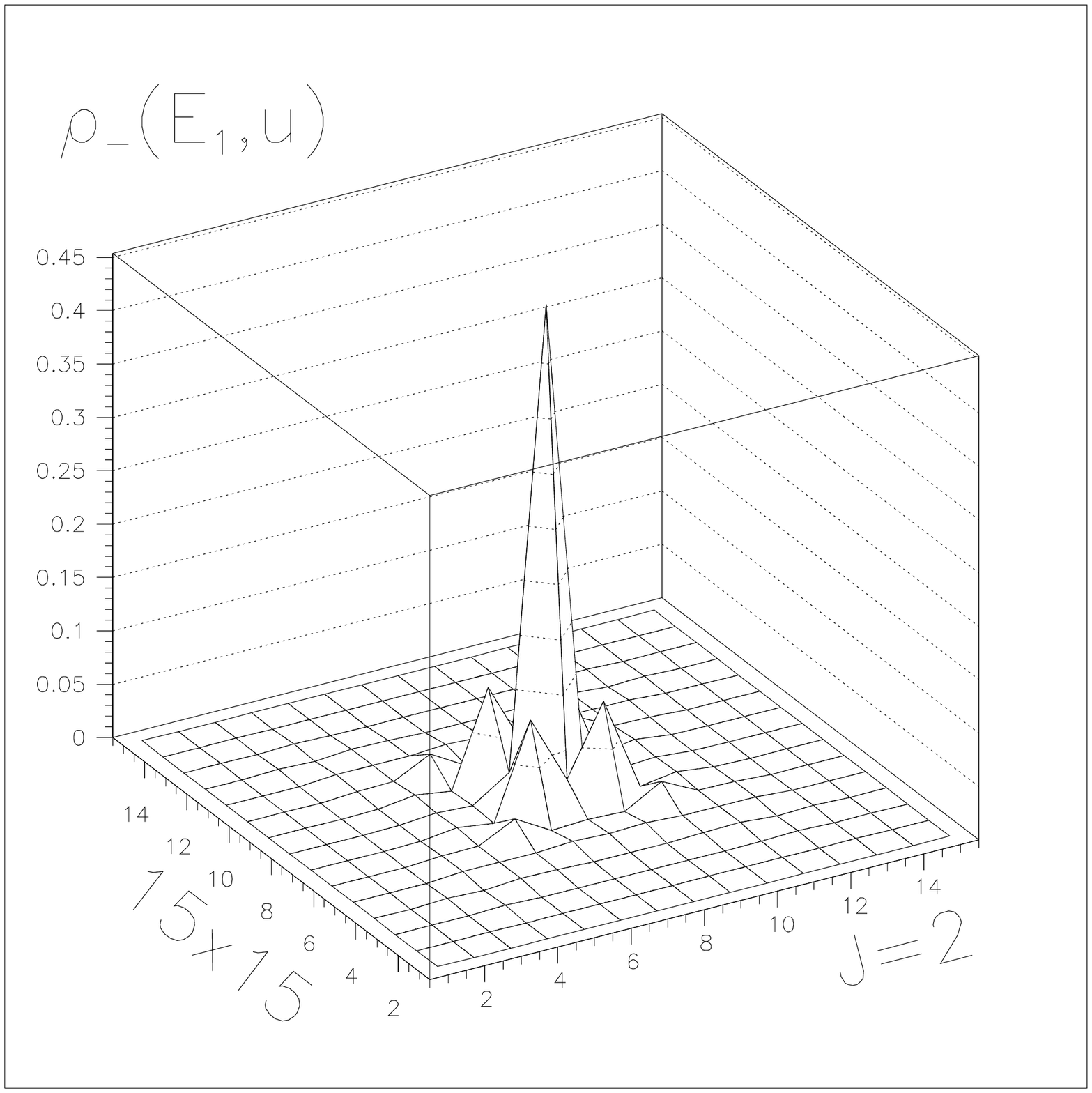}
\caption{\label{fig19}
LDOS for a single impurity. a) $\rho_+(\epsilon_1,i,\uparrow)$ for $J=1$,
b) $\rho_-(\epsilon_1,i,\downarrow)$ for $J=1$,
c) $\rho_+(\epsilon_1,i,\downarrow)$ for $J=2$,
d) $\rho_-(\epsilon_1,i,\uparrow)$ for $J=2$. (Note that in the vertical axis
label $u= \uparrow$ and $d=\downarrow$).
}
\end{figure*}

\begin{figure*}
\includegraphics[width=0.33\textwidth]{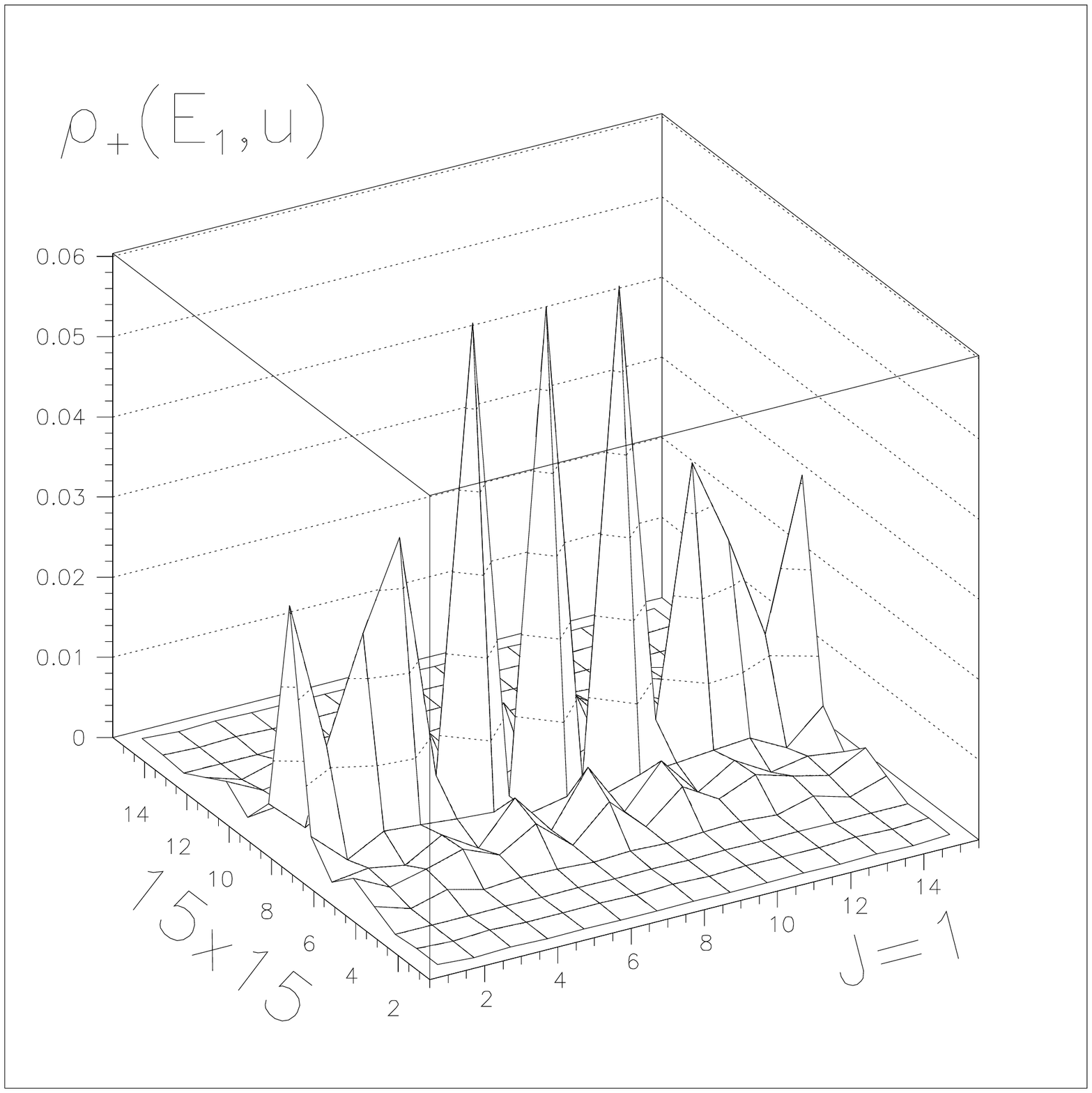}
\includegraphics[width=0.33\textwidth]{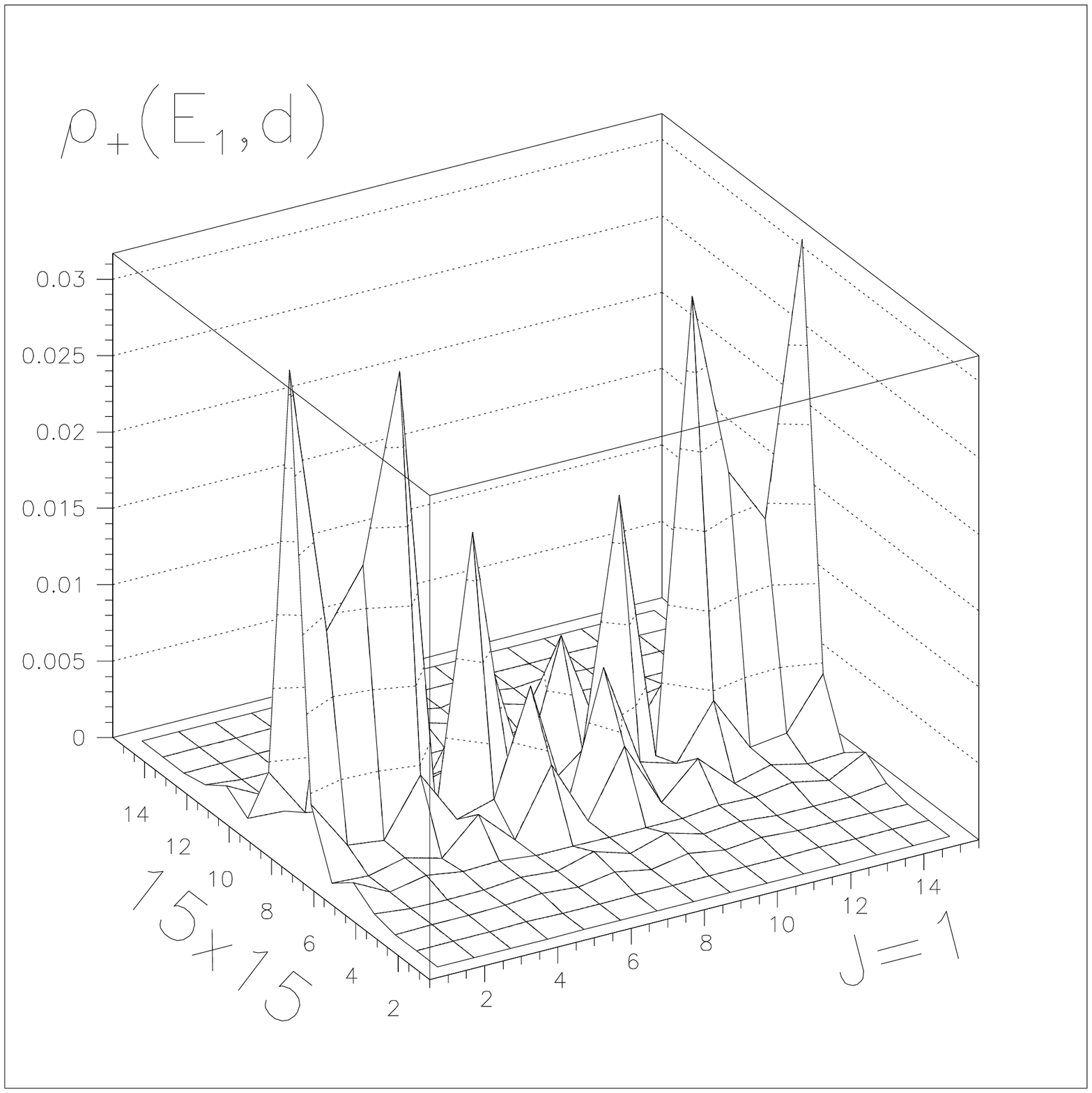}
\includegraphics[width=0.33\textwidth]{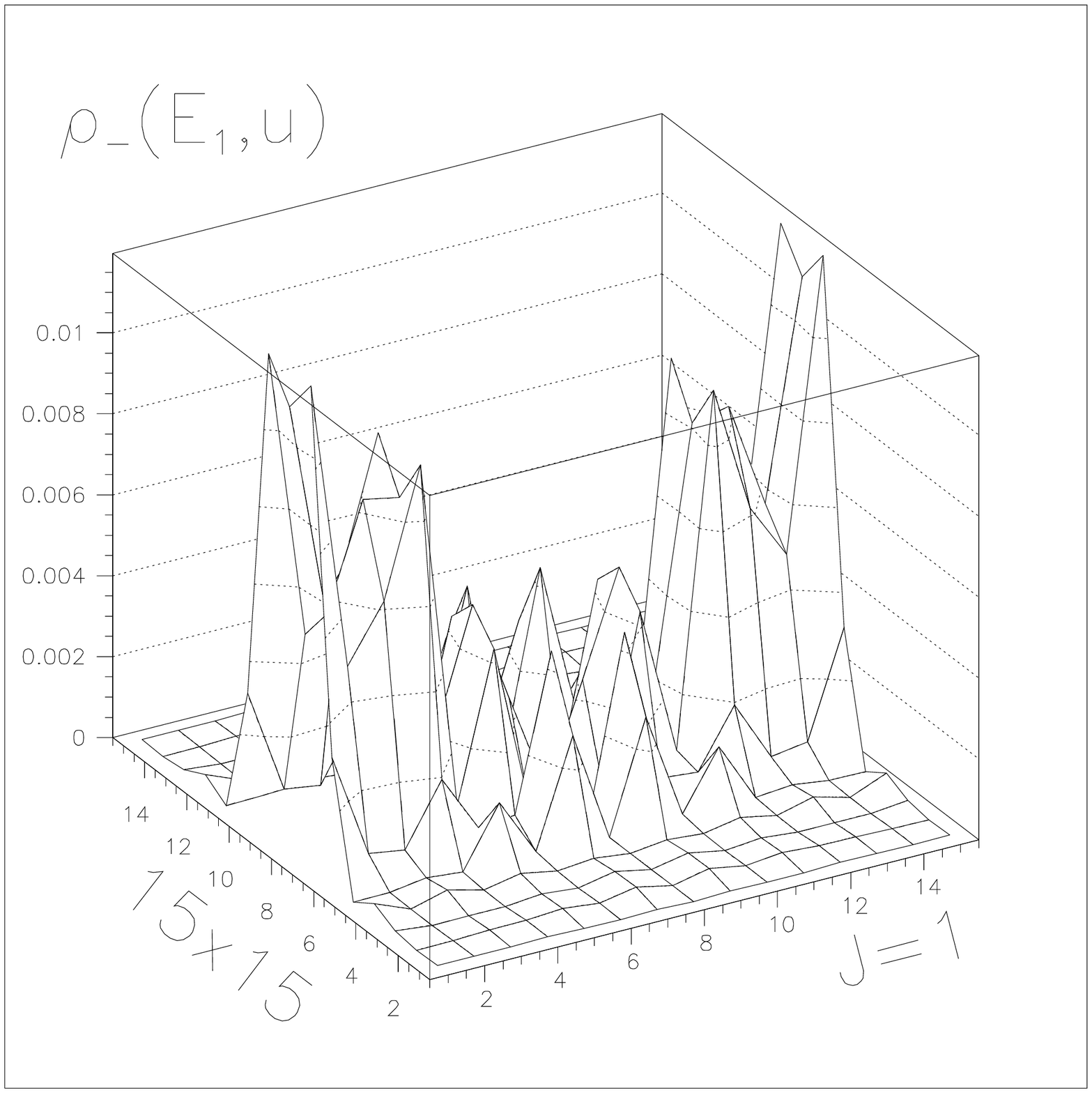}
\includegraphics[width=0.33\textwidth]{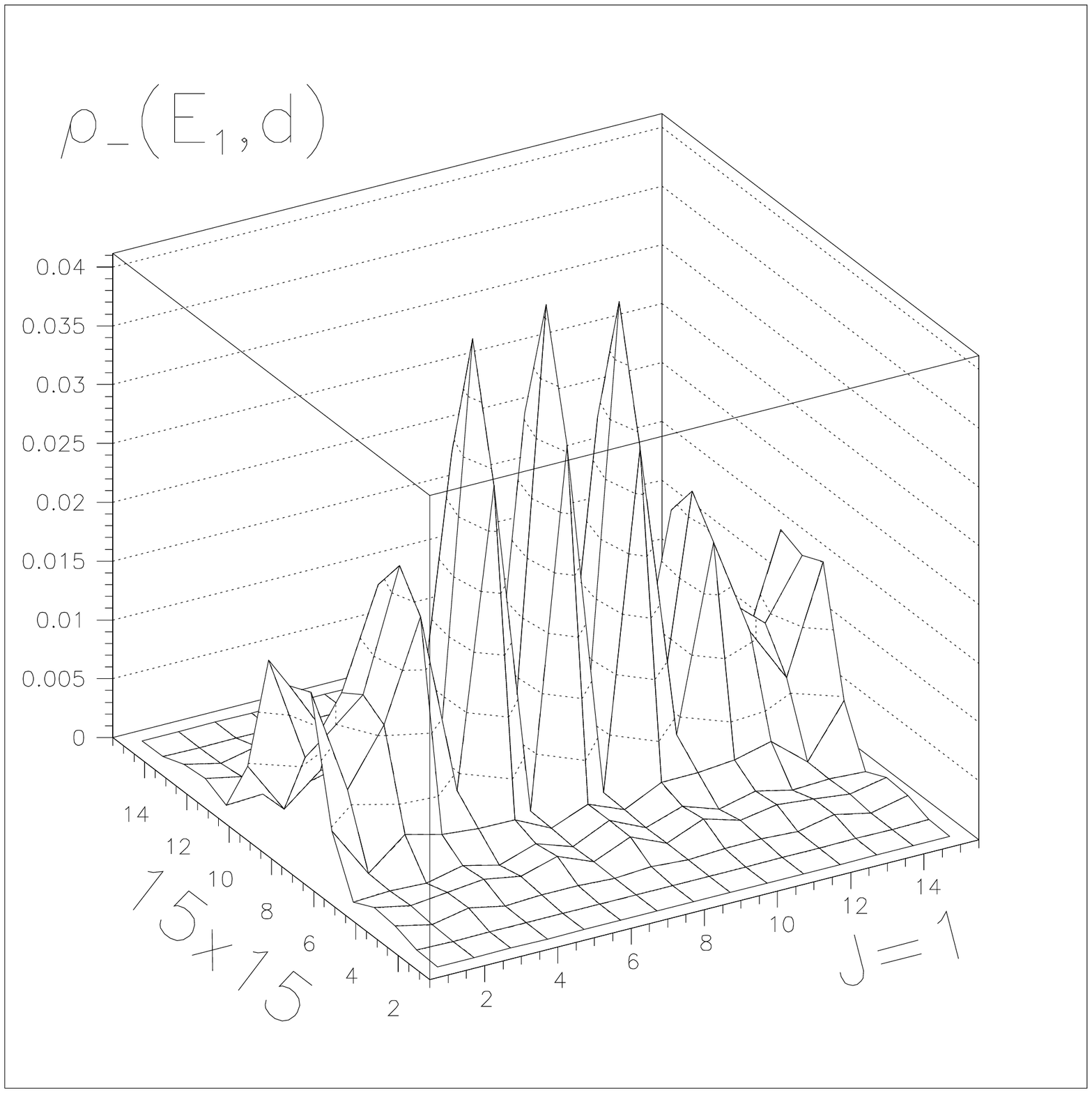}
\caption{\label{fig20}
LDOS for the domain wall DW1. a) $\rho_+(\epsilon_1,i,\uparrow)$ for $J=1$,
b) $\rho_+(\epsilon_1,i,\downarrow)$ for $J=1$,
c) $\rho_-(\epsilon_1,i,\uparrow)$ for $J=1$,
d) $\rho_-(\epsilon_1,i,\downarrow)$ for $J=1$. (Note that in the vertical axis
label $u= \uparrow$ and $d=\downarrow$).
}
\end{figure*}

\begin{figure*}
\includegraphics[width=0.33\textwidth]{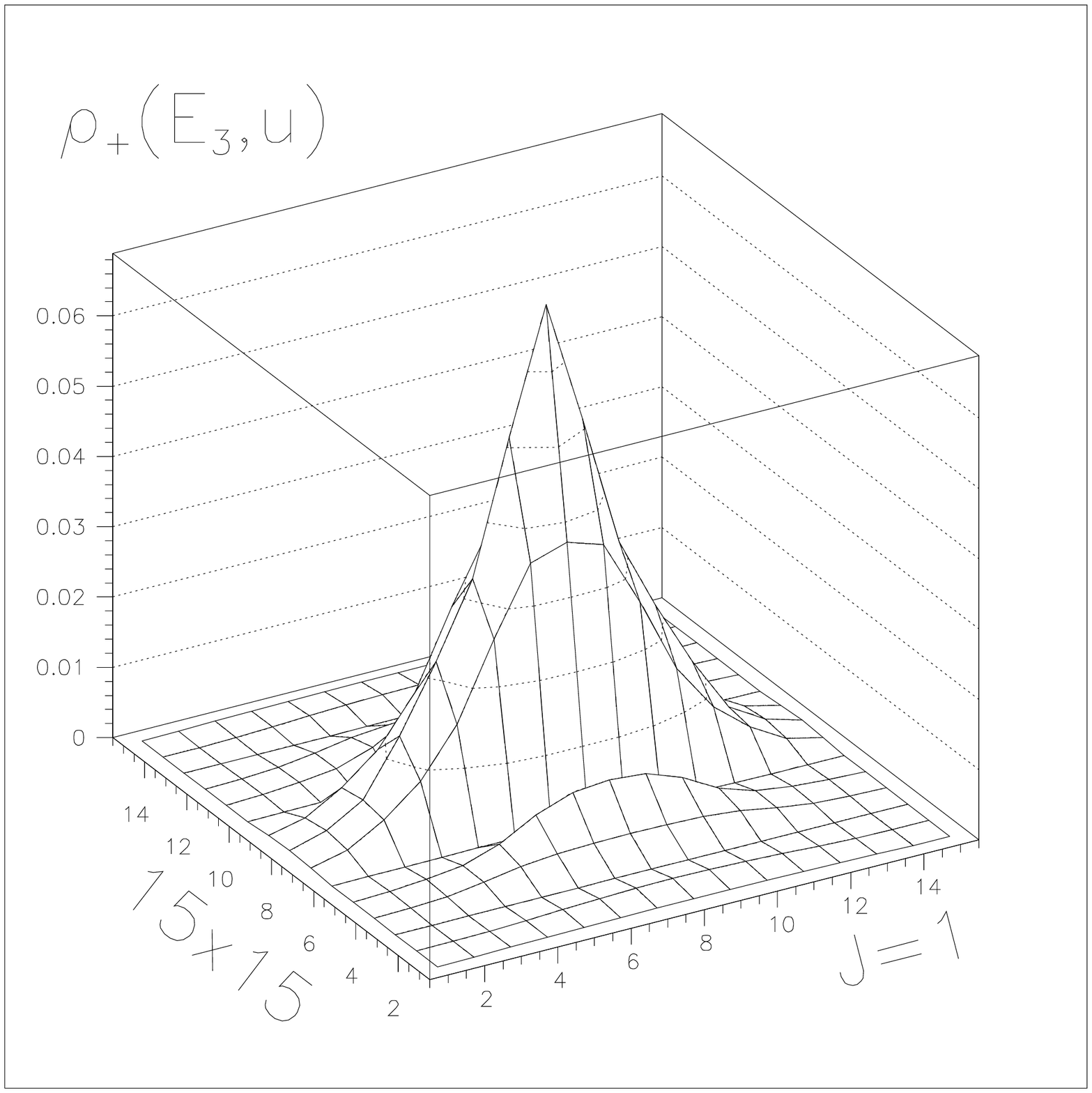}
\includegraphics[width=0.33\textwidth]{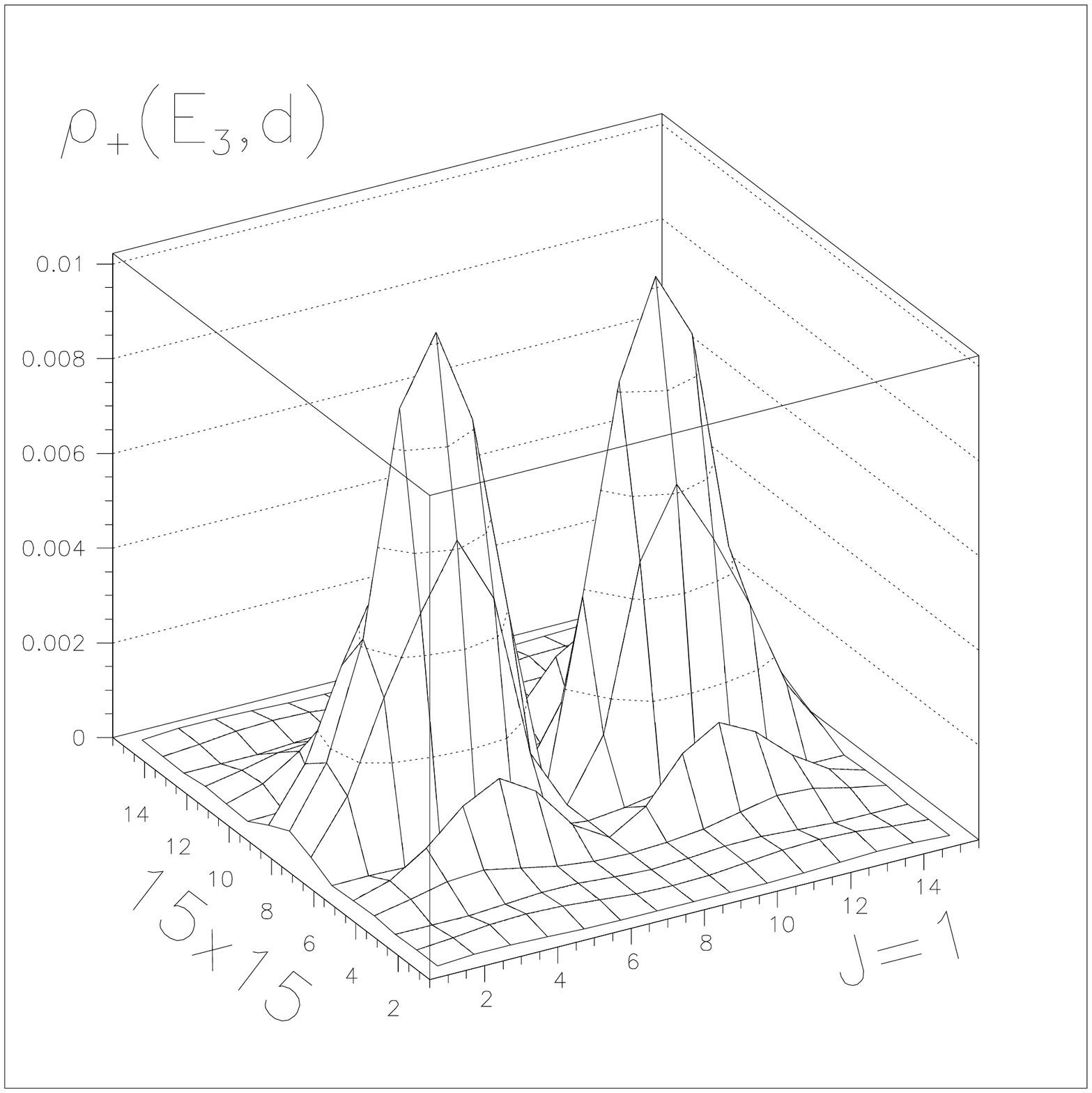}
\includegraphics[width=0.33\textwidth]{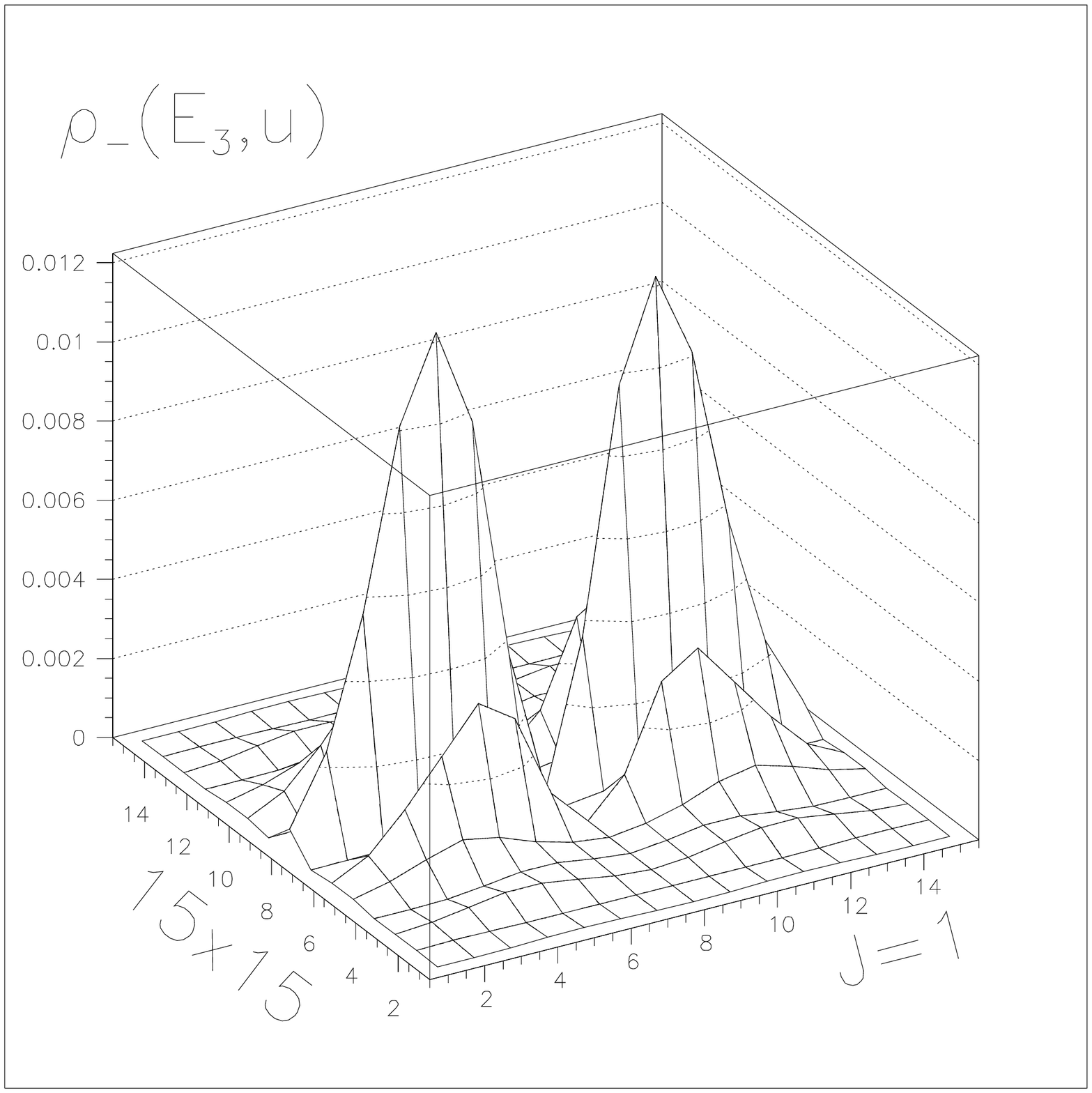}
\includegraphics[width=0.33\textwidth]{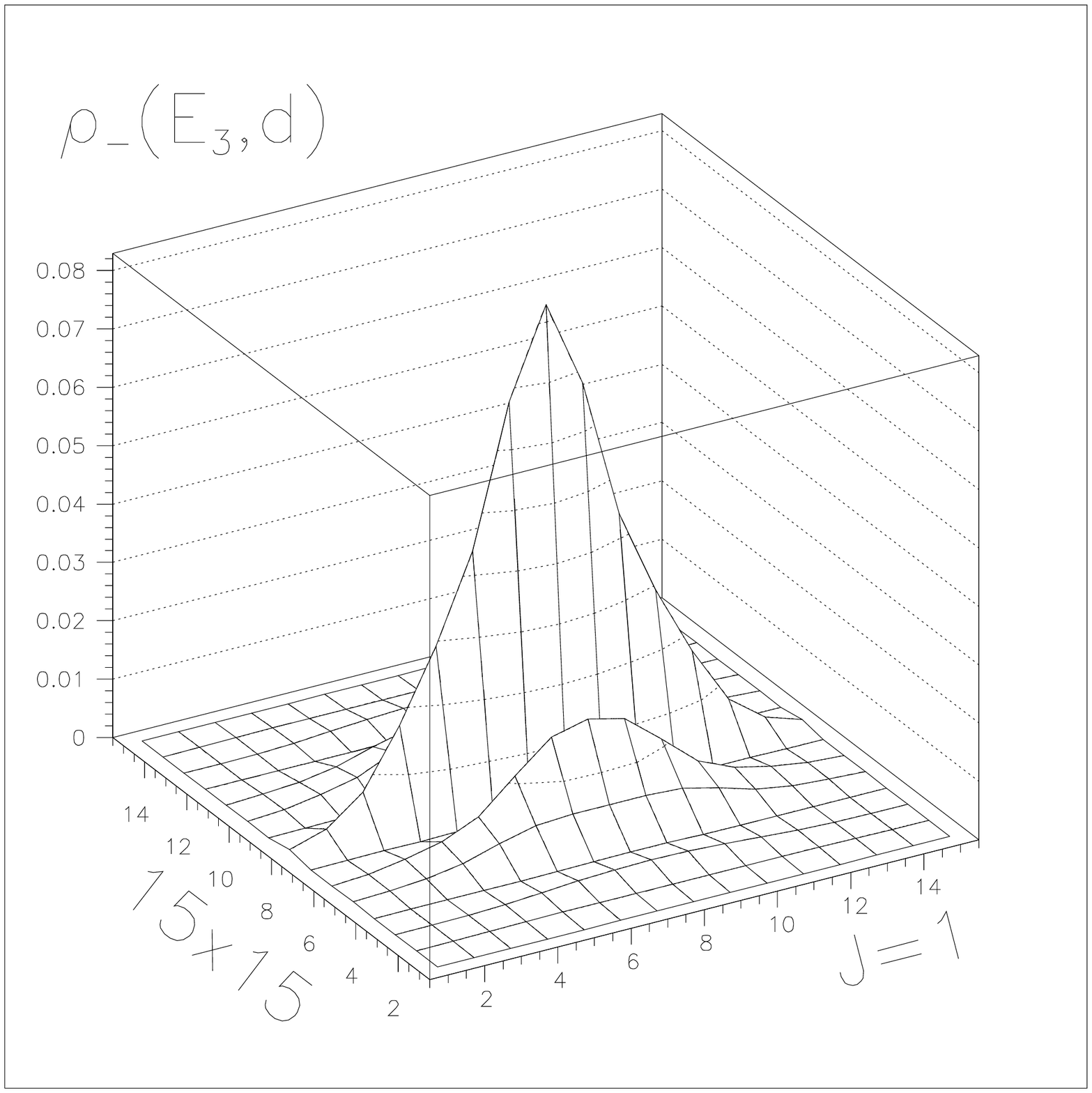}
\caption{\label{fig21}
LDOS for the domain wall DW1. a) $\rho_+(\epsilon_3,i,\uparrow)$ for $J=1$,
b) $\rho_+(\epsilon_3,i,\downarrow)$ for $J=1$,
c) $\rho_-(\epsilon_3,i,\uparrow)$ for $J=1$,
d) $\rho_-(\epsilon_3,i,\downarrow)$ for $J=1$. (Note that in the vertical axis
label $u= \uparrow$ and $d=\downarrow$).
}
\end{figure*}

\begin{figure*}
\includegraphics[width=0.33\textwidth]{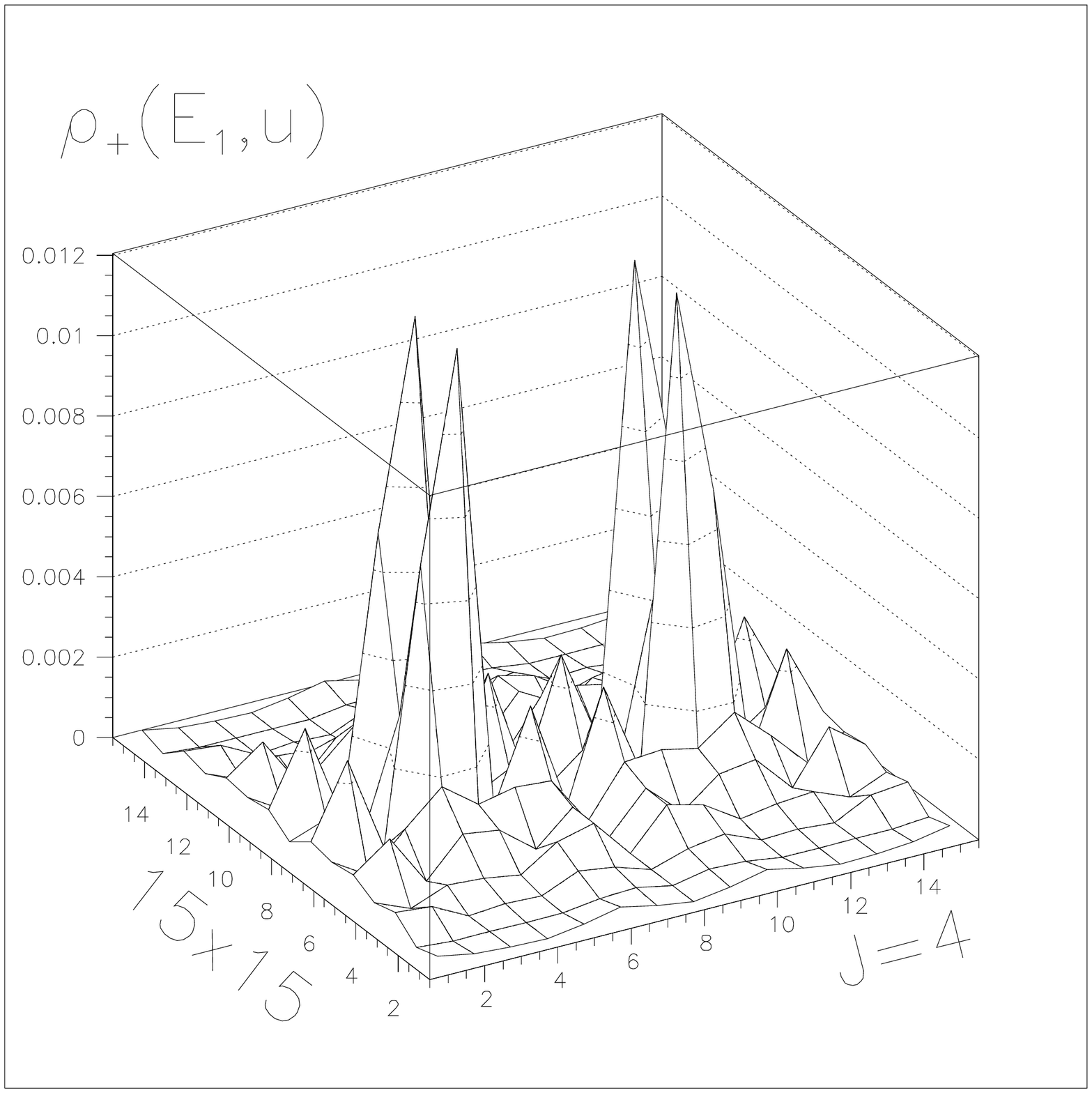}
\includegraphics[width=0.33\textwidth]{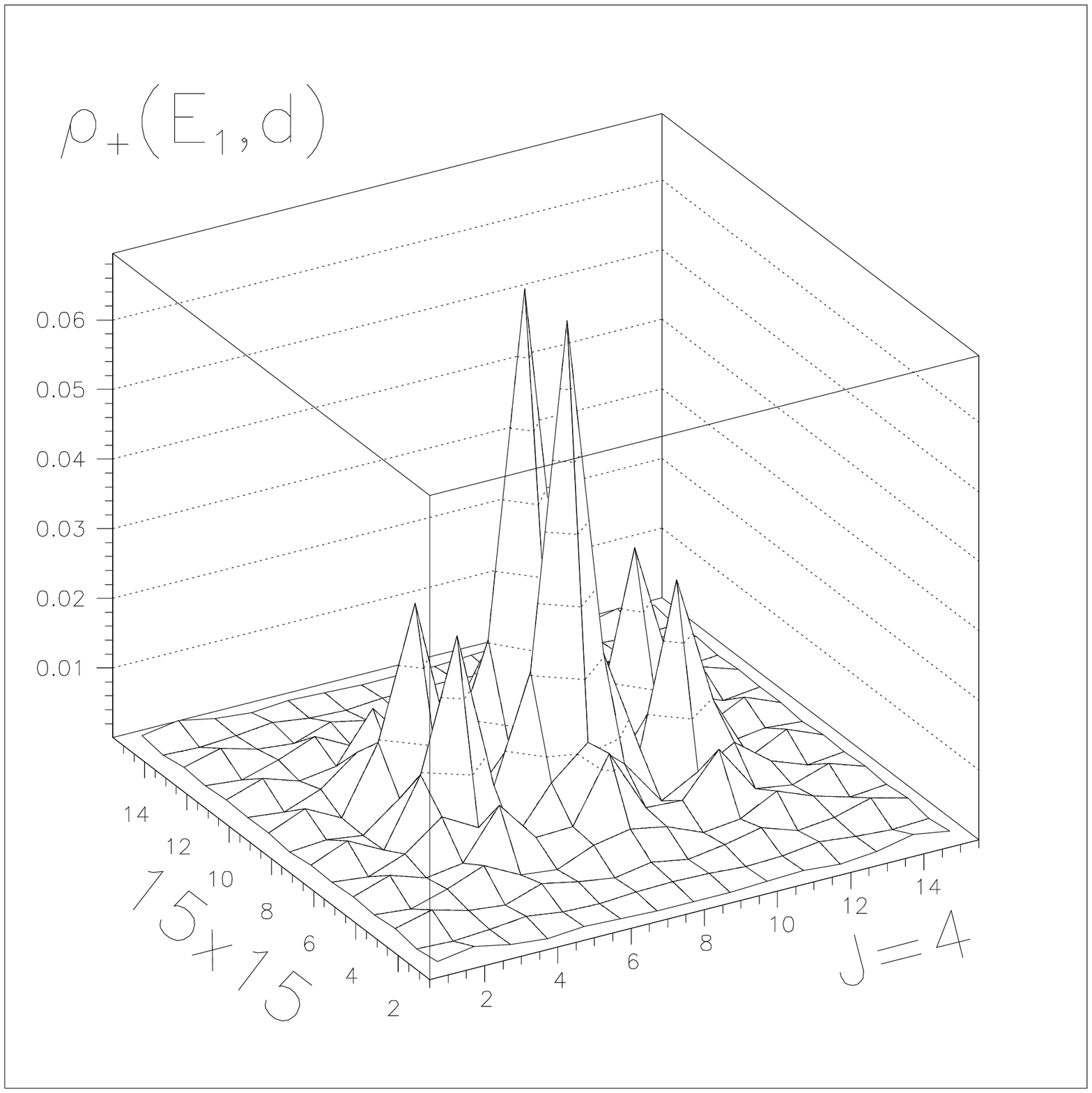}
\includegraphics[width=0.33\textwidth]{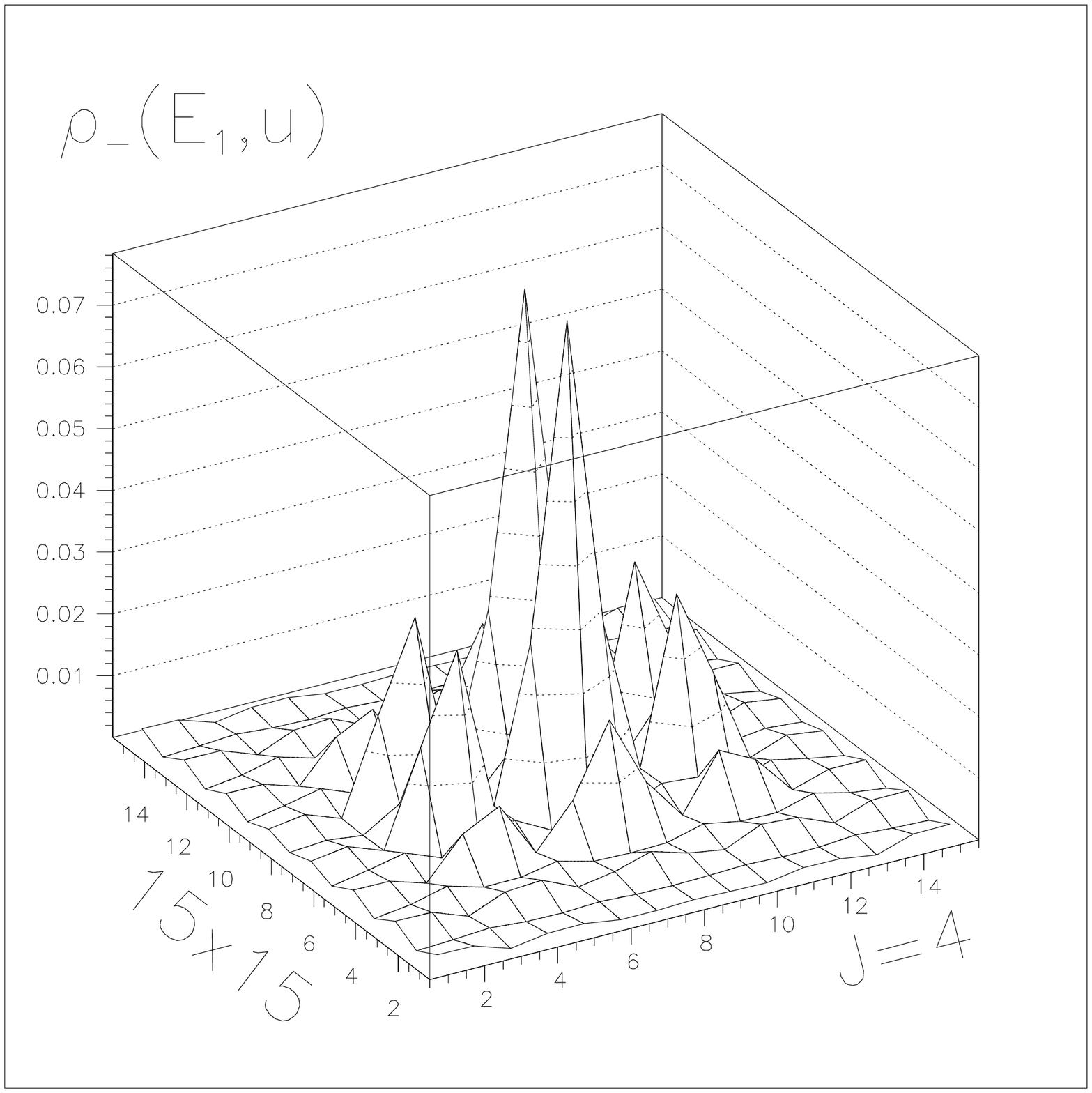}
\includegraphics[width=0.33\textwidth]{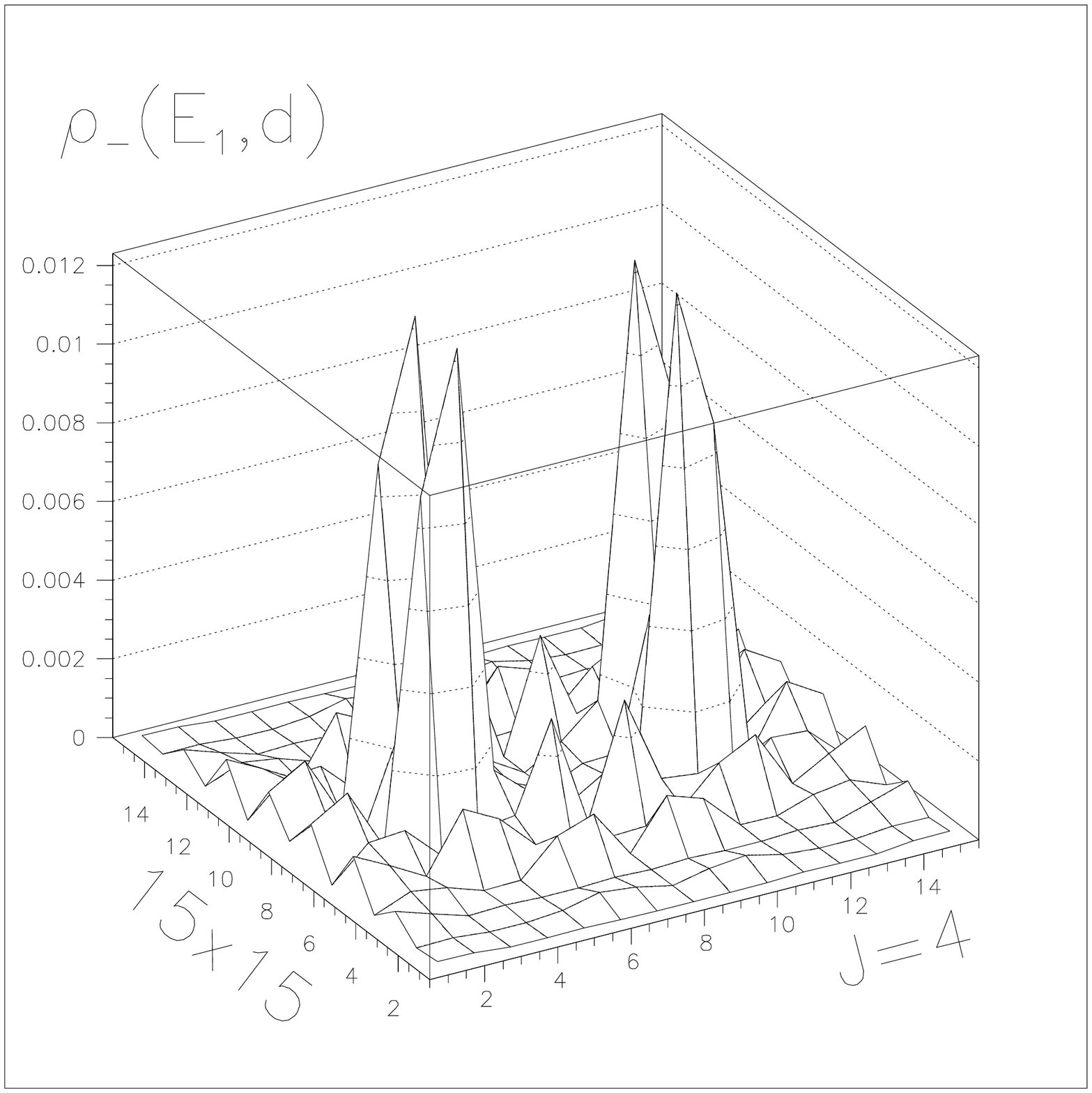}
\caption{\label{fig22}
LDOS for the domain wall DW1. a) $\rho_+(\epsilon_1,i,\uparrow)$ for $J=4$,
b) $\rho_+(\epsilon_1,i,\downarrow)$ for $J=4$,
c) $\rho_-(\epsilon_1,i,\uparrow)$ for $J=4$,
d) $\rho_-(\epsilon_1,i,\downarrow)$ for $J=4$. (Note that in the vertical axis
label $u= \uparrow$ and $d=\downarrow$).
}
\end{figure*}

\subsection{LDOS: $\rho_{\alpha}(\epsilon,i,\sigma)$}

It is perhaps clearer if we look into greater detail into the LDOS, separating
the spin components and trying to understand better the difference between the
positive and the negative energy states.

Consider once again the case of a single impurity.
Let us focus our attention on the first two levels, the first localized and the
next in the continuum. Actually these constitute a set
of 4 states due to the positive and negative energies. 
The analysis of the LDOS shows that for $J=1$, considering first the positive energies, the
first level has only a contribution from spin $\uparrow$ and the first level
with negative energy (symmetric to the other level) has only contribution from
spin $\downarrow$. The magnitude of the spectral weight at the impurity site is
different for the two states, as already noticed before \cite{morr2}. In Fig. \ref{fig19}
we show the results for the boundstates for both $J=1$ and $J=2$. 
Considering now the second level located in the continuum, both at positive and negative
energies, we obtain that there is a mixture of both spin components, even though
in the case of the state at positive energy the magnitude of the peak is larger for
the case of spin $\uparrow$ than for the case of spin $\downarrow$ and in the
case of the state in the continuum at negative energy the relative magnitudes of the
two spin components are reversed. Considering now the case of $J=2$, where the level
crossing has occurred, the nature of the states changes. The positive energy boundstate
has now only a contribution from the spin $\downarrow$ component and vice-versa,
the first negative energy state has only contribution from the spin component $\uparrow$.
As the level crossing occurred the spin content has changed. 
In Fig. \ref{fig19} we only show the non-vanishing contributions. The other contributions
vanish.
On the other hand, 
in the first state in the continuum, where the two spin components contribute, 
the magnitude of the $\uparrow$ component is now much larger than the $\downarrow$ component
while in the case of $J=1$ the magnitudes were of similar size. Also, the $\downarrow$
component of the second state of negative energy is now much smaller than the $\uparrow$
component. The second state has to compensate for the spin flip of the
lowest state by increasing the weight of the spin component aligned with the
external impurity spin.

In the case of two impurities the behavior is similar to the case of one impurity,
except that now the second state is also localized. Therefore, for $J=1$ there are
now two levels at positive energies with only spin $\uparrow$ contribution and
two states at negative energies with only spin $\downarrow$ contribution. As one crosses
to higher values of the coupling, for instance $J=2$, the spin contributions for the
boundstates change in a similar way to the single impurity case. For the set
of parameters we consider here the two states are nearly degenerate and therefore
the states change their nature basically simultaneously. Otherwise they would
change their nature in succession \cite{morr2}.

One may also consider a ferromagnetic chain where all the spins point in the $z$ direction.
This case is very similar to the single impurity case, as expected. All the positive
energy states have the same spin content at low values of the coupling. If the coupling
is large enough a similar situation will occur where all the boundstates have reversed
their spin content.

The case of the $DW1$ is however different. Even at a small value of the coupling, in the
sense that the first level crossing has not yet occurred, such as $J=1$, the various
states have a mixture of the two spin components. This is shown in Fig. \ref{fig20}.
The boundstates at positive energy have a mixture of the two components, even though
the magnitude at the impurity site is larger for the $\uparrow$ component.
For the negative energies the $\downarrow$ component has a larger magnitude.
In Fig. \ref{fig20} we only show results for the lowest level but a similar trend
is found for the other localized states.  
Note that both components have symmetric wavefunctions.
This does not happen for instance for the case of the third level shown in Fig. \ref{fig21}.
Even though the relative magnitudes of the spin components are the same, note that
while the $\uparrow$ component has a symmetric wave function the spin $\downarrow$
has an anti-symmetrical wave function. For the third level at negative energy
this is reversed. 

Increasing the coupling to $J=2$ does not change the nature of the first two level
states. However, increasing further to $J=4$ we see in Fig. \ref{fig22} that
there is once again a reversal of the magnitudes of the two spin components
of the first level. Even though both spin components contribute the magnitude
of the $\downarrow$ component is now larger than the magnitude of the $\uparrow$ 
component. This also occurs for the second level. For the third level however,
the two components have very similar magnitudes. 

As we have seen in some cases
the discrete states in the gap correspond to well defined spin polarized
states. Then the QPT changes the spin polarization of these states
leading to a sort of magnetic phase transition. However, in the case of
DW1 domain wall, each discrete level corresponds to a mixture of spin up
and down components because the spins in DW are oriented along axis x.
In this case the QPT does not show explicitly the transition between
spin up and down states. Nevertheless, the different spin components of
the composition are also interchanged during the QPT. Thus, the
analysis demonstrates that the QPT can be seen in the variation of
magnetic state corresponding to the discrete levels.
As for the origin of the QPT, we show that it is related only to the
level crossing because at each crossing point the ground state is
reconstructed.

\begin{figure}
\includegraphics[width=0.6\textwidth]{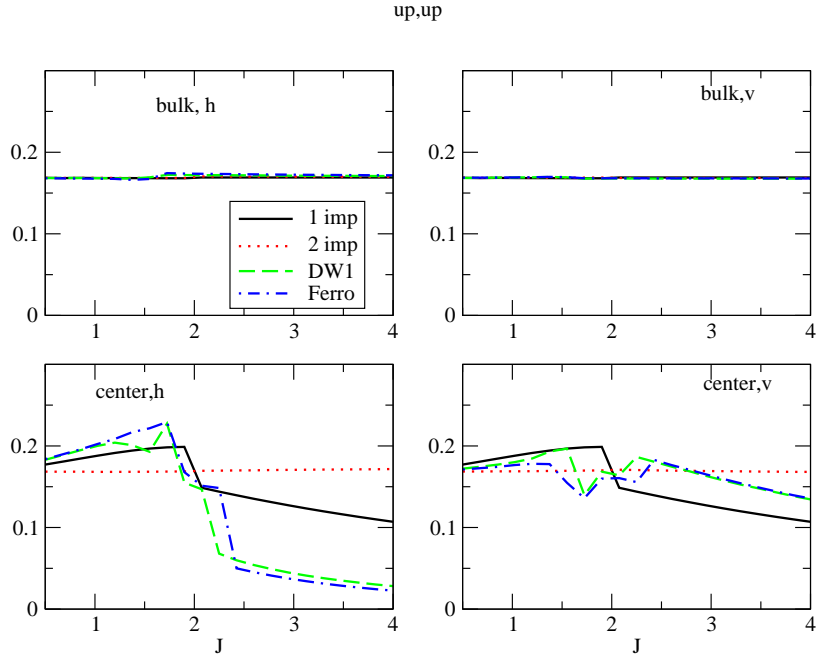}
\caption{\label{fig23}
Hopping matrix elements for up spin electrons from a bulk point and the
center point, in the horizontal (along $x$) and vertical (along $z$) directions.
}
\end{figure}

\begin{figure}
\includegraphics[width=0.6\textwidth]{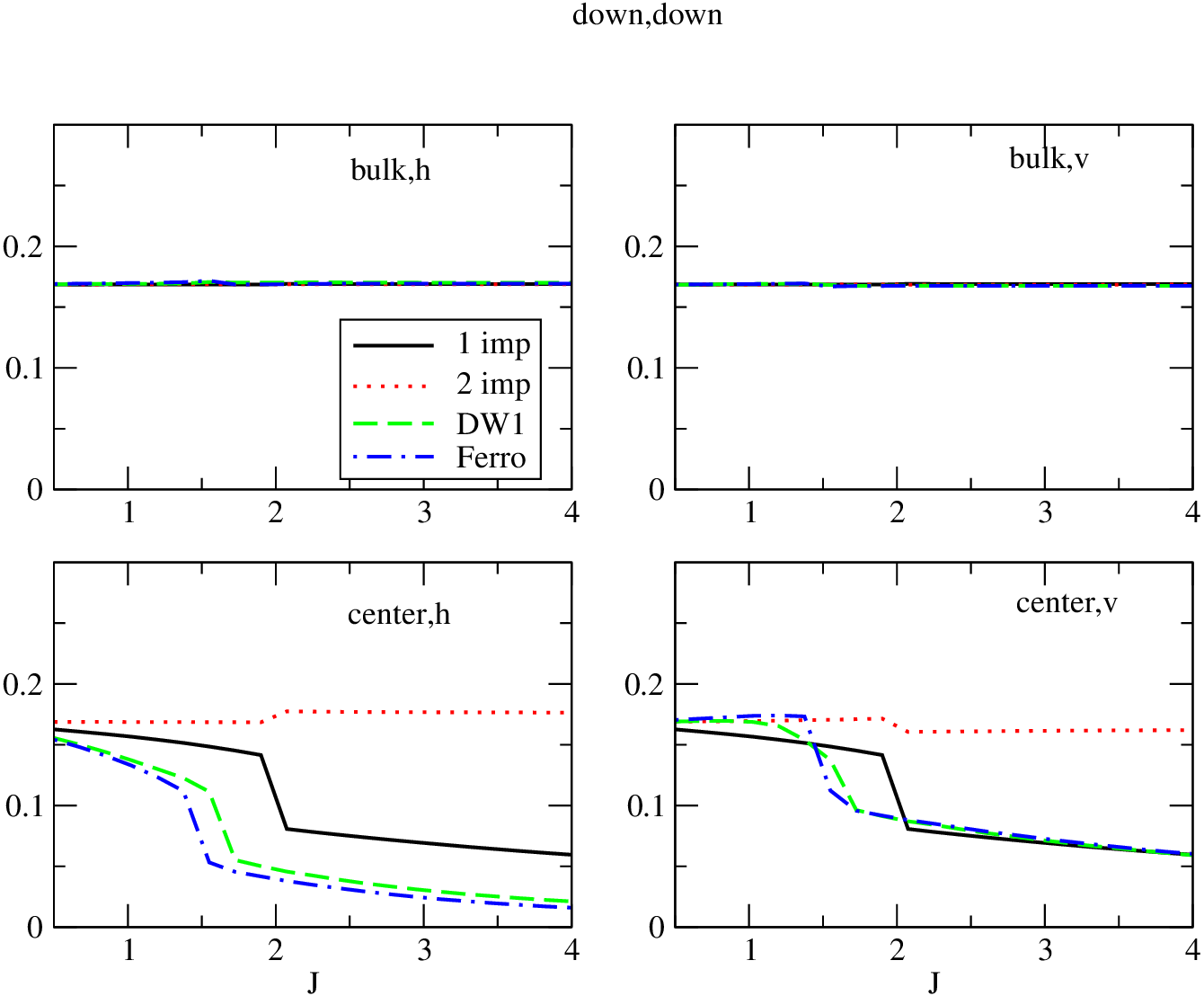}
\caption{\label{fig24}
Hopping matrix elements for down spin electrons from a bulk point and the
center point, in the horizontal (along $x$) and vertical (along $z$) directions.
}
\end{figure}

\subsection{Kinetic energy}

To further confirm the nature of the states we have calculated at $T=0$
\be
\langle c_{i\sigma}^{\dagger} c_{j,\sigma'} \rangle = \sum_n \sigma \sigma' v_n(i,\sigma)
v_n(j,\sigma')
\ee
We have considered two typical points in the system. The central point and a
point far from the central line (in the bulk). Also we have considered both
vertical (v) (along $z$) and horizontal (h) (along $x$) 
displacements of the electrons. In Figs. \ref{fig23},\ref{fig24}
we plot the hoppings for the cases of $\sigma,\sigma'=\uparrow$ and
$\sigma,\sigma'=\downarrow$ as a function of the coupling. We have considered the cases
of a single impurity, two impurities (four lattice sites apart), the domain wall $DW1$
and a ferromagnetic chain. In the bulk we expect the states to be extended
independently of the coupling. This is so and the hoppings are basically independent
of $J$ both along the vertical and horizontal directions and spin directions.
The central point is different however and when various impurity spins are considered
the horizontal and vertical directions are expected to be different. As shown in the
Figs. \ref{fig23},\ref{fig24} we see that as the coupling grows in general the hopping
decreases, consistently with the ``localized" nature of the states. 
This is particularly so along the chain direction, for the lines of spins.
In the case of two impurities, and due to the rather local nature of the effect of the
impurity spins, the hoppings at the central point do not change much as a function of the coupling.
As the $\uparrow$ spins are in majority the hopping term for the cases of $\uparrow,\uparrow$
are typically larger. Interestingly, and even though the hopping term of the 
Hamiltonian is diagonal in the spin, in the case of the domain wall $DW1$ there is a
non-vanishing spin flip term, even though quite small.

\section{Stability of domain wall}

\begin{figure}
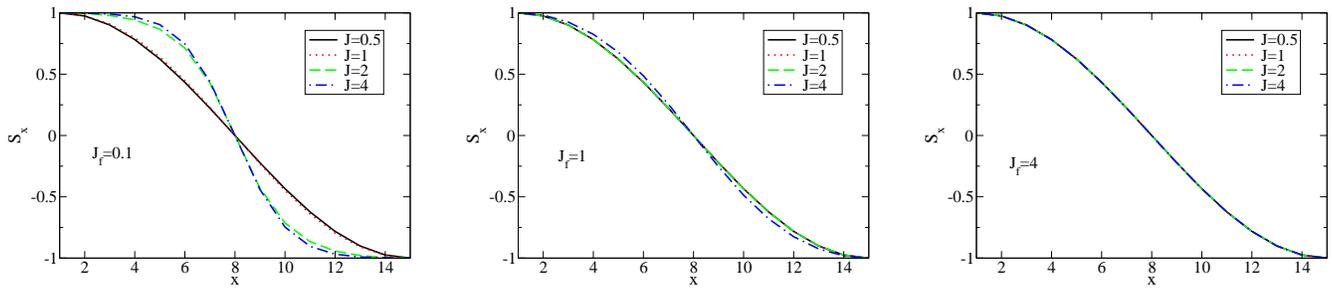

\includegraphics[width=0.3\textwidth]{fig24a.eps}
\hspace{0.5cm}
\includegraphics[width=0.3\textwidth]{fig24b.eps}
\hspace{0.5cm}
\includegraphics[width=0.3\textwidth]{fig24c.eps}
\caption{\label{fig25}
Profile of $S_x$ of the impurity spins for DW1 and for
for $J_f=0.1,1,4$ as a function of $J$. 
}
\end{figure}

So far we have assumed that the domain wall is stable. We may solve this stability issue
self-consistently and study the stability of the domain wall. This
may be achieved introducing effective interactions between the impurity spins,
possibly mediated by the quasiparticles. Let us approximate these interactions
by an Heisenberg like term. The part of the Hamiltonian involving the classical
spins may be written in mean field as
\be
-\sum_i J \left( S_i^x <\sigma^x>_i + S_i^z <\sigma^z>_i \right)
-\frac{1}{2} \sum_{<i,j>} J_f \left( S_i^x S_j^x + S_i^z S_j^z \right)
\ee
where we consider only the coupling between nearest neighbors, assumed to
be ferromagnetic for simplicity. In the mean field approximation for both
electrons and spins, 
the external impurity spins are determined from their mutual
interactions and the interaction with the average spin density of the electrons.
We look for the equilibrium impurity spin configuration minimizing the energy
with respect to the impurity spin components. 
This equilibrium distribution is then inserted in
the Bogoliubov-de Gennes equations and solved self-consistently. 
The impurity spins satisfy the constraint that
$(S_i^x)^2 + (S_i^z)^2 = (\vec{S}_i)^2$. An equivalent way is to
minimize the Hamiltonian with respect to the angles themselves, since in this
way the constraint is automatically enforced. Minimizing with respect to the
angles $\varphi_l$ we get the set of equations
\bea
&-& J \langle \sigma^z \rangle_i \cos \varphi_i + J \langle \sigma^x \rangle_i \sin \varphi_i
\nonumber \\
&+& J_f \sin \varphi_l \cos \varphi_{i+\delta} -J_f \cos \varphi_i \sin \varphi_{i+\delta} =0
\eea
where $\delta = \pm 1$. These equations hold at each impurity site.
The solution leads to a stable domain wall whose shape depends on $\frac{J}{J_f}$.
The results are shown in Fig. \ref{fig25}.
We see that the profiles are stable. In the case of a small coupling between the
impurity spins ($J_f=0.1$), for small $J$ the
profile is tending to a large $\lambda$, defined in Eqs. (\ref{lambda1},\ref{lambda2}),
which implies an almost
linear profile between the end spins. As J increases the value of
$\lambda$ decreases and there seems to be a rapid variation
between $J=1$ and $J=2$. 
Increasing the value of $J_f$, the stability is improved.
For large $J_j$ the profile is almost independent of $J$. Also note
that increasing $J_f$ the value of $\lambda$ increases and the profile becomes
almost linear for any value of $J$.

We should note that one strictly does not have  to introduce an effective Heisenberg interaction
between the impurity spins, if we were to treat them as quantum mechanical.
The Kondo interaction would couple the impurity spins to the conduction electron
spin density, which in turn would give rise dynamically to an effective long-range
interaction between the impurity spins. In our mean-field approach this interaction
is introduced phenomenologically, much in the same way as the attractive interaction
between the electrons to give rise to pairing. While a mean-field approach (BCS)
is quite good for the superconductivity order, the correct description of the
Kondo effect is much more involved as well as a proper treatment of the 
coupling between the impurity spins.
However, as shown before in a similar context \cite{schlottmann2} the results
are quite similar.

\section{Effect of temperature}

In this section we briefly study the effect of temperature on the results. Clearly the
quantum phase transitions are smeared out but the same trends prevail.

\begin{figure*}
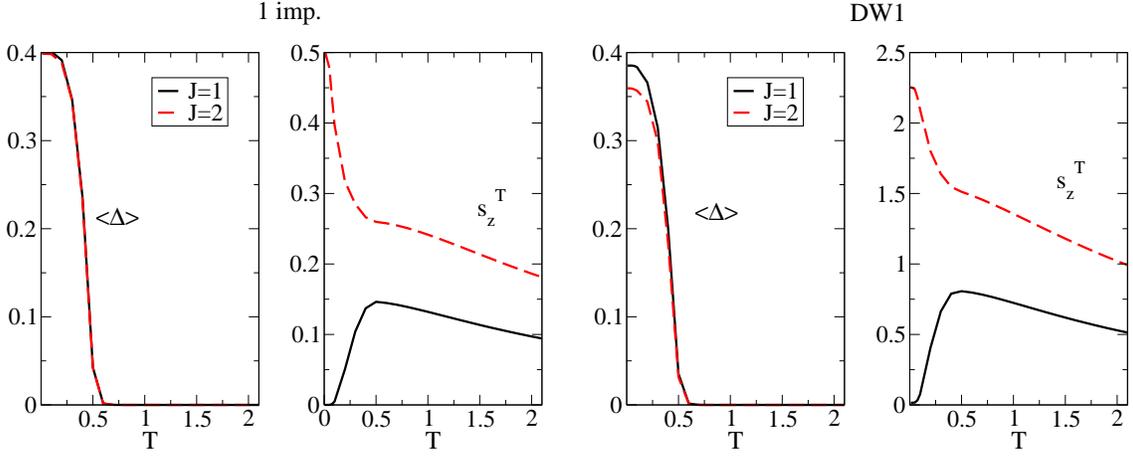

\includegraphics[width=0.4\textwidth]{fig25a.eps}
\hspace{0.4cm}
\includegraphics[width=0.4\textwidth]{fig25b.eps}
\caption{\label{fig28}
$<\Delta>$ and $s_z^T$ for one impurity and DW1
as a function of temperature for various values of $J$.
}
\end{figure*}

In Fig. \ref{fig28} we show the order parameter and the spin density for a single
impurity and the DW1 as
a function of temperature for the two typical cases of $J=1,2$. We see clearly
that the critical temperature is basically the same when we increase the
number of spins. 
Note the critical temperature for one impurity and the DW1 at about
$T \sim 0.5-0.6$. 
However, the behavior of the spin density is quite different
as a result of the underlying phase transition. In the case of $J=2$ the
spin density is finite and then decreases while for $J=1$ the increase in temperature
increases the density due to excitation to higher levels.
Note for both cases the different starting point
of $s_z^T$ due to the quantum phase transition when we go from $J=1$
to $J=2$. 

\begin{figure}
\includegraphics[width=0.4\textwidth]{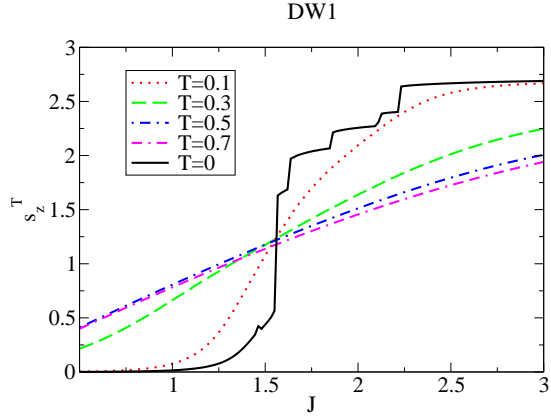}
\caption{\label{fig29}
Effect of temperature on the results
for the various magnetizations (total and left or right) for DW1. 
}
\end{figure}

Clearly, as shown in Fig. \ref{fig29}, the quantum transitions between the plateaus of the spin
densities as a function of the coupling are now smeared, but the same overall trend persists.
Even though the temperature smoothens
the curves and the quantum phase transitions disappear (they only occur
at $T=0$) the behaviors are robust to temperature. In particular the
regimes where at $T=0$ there is a finite value for one of the average
magnetizations it persists at small finite temperatures. 

\section{Summary}

In this work we considered the effect of correlated magnetic impurities
inserted in a conventional superconductor. 
Previous results on few impurities revealed the existence of
sequences of quantum phase transitions, associated with level crossings,
that lead to discontinuous changes in the properties of the system,
such as the magnetization. We extended these results to the case when
we have sets of impurity spins that are correlated through mutual
interactions, that may be originated via RKKY-like interactions mediated
by the electrons. In particular, we considered cases where the impurity
spins are organized in such a way that form domain walls which,
to simplify, we have limited in this work to one-dimensional arrays
of spins inserted in the superconductor. These domain wall structures may be
obtained imposing different boundary conditions at different sides of the
mesoscopic systems here considered. As in the case of a few impurities,
we found a series of quantum phase transitions that we have analyzed.
In general the introduction of foreign objects in the superconductor
originates interference effects. These are revealed in the LDOS. We have
presented detailed results which show the complex nature of the multiple
interference effects.

We have also shown that the domain wall structures considered here are
stable taking into account the mediated interactions between the impurity
spins. To simplify we considered in this work classical spins. The case of
a full quantum problem where the impurity spins are described by quantum
operators leads to Kondo like effects in the superconductor. This is
much more involved and the use of the Bogoliubov-de Gennes formalism
requires that we take this classical limit. Fortunately, at least if the
coupling between the electronic spin density and the impurity spins
is not large, the classical description is enough.

In this work we focused on the effect of the impurity spins on the
superconductor. The opposite problem of the effect of the superconductor 
on the impurity spins
may be interesting if, for instance by passing a current through the
superconductor, the spin torque created by a spin polarized current on
the impurity spins changes their relative orientations. This is a problem
that has received much attention in the context of spintronics, where spin
polarized currents passing through a magnetic semiconductor move the
position of the domain walls. The related problem in the context of the
magnetic correlated impurities spins in the superconductor will be
considered elsewhere.

\begin{acknowledgments}
This work is supported by FCT Grant
No.~POCI/FIS/58746/2004 in Portugal, 
the ESF Science Programme INSTANS 2005-2010,
Polish Ministry of Science and Higher Education as a research
project in years 2006-2009,
and by STCU Grant No. 3098 in Ukraine. 
\end{acknowledgments}


\end{document}